\documentclass{article}%
\usepackage{amssymb}
\usepackage{amsfonts}
\usepackage{amsmath}
\usepackage{geometry}
\usepackage{fancybox}
\usepackage{indentfirst}
\usepackage{layout}
\usepackage{makeidx}
\usepackage[final,hiresbb]{graphicx}
\usepackage{float}
\usepackage{amscd}%
\providecommand{\U}[1]{\protect\rule{.1in}{.1in}}

\ifx\pdfoutput\relax\let\pdfoutput=\undefined\fi
\newcount\msipdfoutput
\ifx\pdfoutput\undefined\else
\ifcase\pdfoutput\else
\ifx\paperwidth\undefined\else
\ifdim\paperheight=0pt\relax\else\pdfpageheight\paperheight\fi
\ifdim\paperwidth=0pt\relax\else\pdfpagewidth\paperwidth\fi
\fi\fi\fi

\begin{document}

\title{A Structurally Relativistic Quantum Theory\\Part 1. Foundations}
\author{Emile Grgin\thanks{Independent physicist, New York. \ eg2357@gmail.com}}
\maketitle

\begin{abstract}
The apparent impossibility of extending nonrelativistic quantum mechanics to a
relativistic quantum theory is shown to be due to the insufficient structural
richness of the field complex numbers over which quantum mechanics is built. A
new number system with the properties needed to support an inherently
relativistic quantum theory is brought to light, and investigated to a point
sufficient for applications.

\end{abstract}

\section{\label{S0}Introduction}

The finalization of quantum mechanics in the late 1920s gave rise to the still open
problem of its structural unification with relativity, meaning a merging built on a mathematical structure both quantum mechanical (like Hilbert space), and relativistic (like the locally Minkowskian Riemannian space).
We shall revisit the oldest and conceptually simplest approach to this
problem, for it has never been fully explored. The idea stems from the observation that while a final quantum theory
could not be built over the real numbers in the first quarter of the 20th
century, modern quanrum mechanics was finalized within a very short time once it became evident that a
transition from the real numbers to\emph{\ the structurally richer field of
complex numbers} was necessary. It is thus conceivable that the
unification of quantum mechanics with relativity might also require a
transition from the complex numbers to a structurally richer number system.

The only \emph{known} possibly relevant number systems richer than the complex numbers
are the higher division algebras: the quaternions and octonions --- but only
the former is an acceptable candidate for being associative. A quaternionic
quantum mechanics, thoroughly reviewed in Adler's monograph, \cite{Adler}, was
therefore developed over fifty years ago. As it did not prove useful to the
unification problem, the search for a unification based on a new number system
was not pursued. A reason for not seeking a \textbf{unifying number} system
outside the division algebras might have been the argument that for Born's
interpretation to be preserved, this system would have to be a normed algebra
--- in which case, by Hurwitz's theorem, the only option is the field of
quaternions.\footnote{All new concepts, and some standard ones that should be
emphasized, will be introduced in bold letters.} This argument is incorrect,
however, because a generalization of quantum mechanics would likely entail a
generalization of Born's interpretation \ --- as in the present
paper, where Born's interpretation generalizes to its relativistic version, referred to as
\textquotedblleft Zovko's interpretation\textquotedblright.

The option of seeking the hypothetical unifying number system outside the
division algebras is still open because no theorem precludes its existence. We
shall show that does exist by actually exhibiting it. It has been derived by two very different approaches:

\emph{Within physics:} In the author's previous works, \cite{GP} to \cite{G5},
briefly reviewed by Florin Moldoveanu in \cite{Mol}, the number system in
question emerged as one of the two concrete realizations of an abstract
version of quantum mechanics. The other realization is standard quantum
mechanics. This approach has the philosophical advantage of yielding the new
number system with an \textit{a priori} guarantee of its physical
significance. Its main disadvantage is its great length --- though substantial
simplifications seem possible and might be considered in the future.

\emph{Within mathematics:} In section \ref{S2} of the present self-contained
paper, the unifying number system is derived by removing a hitherto unnoticed degeneracy from the
field quaternions. The philosophical advantage of this approach is its
strictly mathematical nature: The unique number system which merges the
mathematical foundations of two apparently incompatible physical theories is
obtained \emph{without any reference to either theory.} It also has the
advantage of brevity.

From a broader perspective, the present work extends the work of the 19th
century algebraists --- from Hamilton in 1833 to Hurwitz in 1898. Mirroring
the often quoted adage about string theory (a piece of 21st century physics
that fell by chance into the 20th century), the unifying number system derived
in the sequel may be regarded as a piece of 19th century mathematics that was left on
the table to be investigated in the 21st century.

\vspace{10pt}%

The author most gratefully acknowledges the following contributions:%

Nikola Zovko, Institute Rudjer Bo\v{s}kovi\'{c} in Zagreb, has contributed the
idea, discussed in section \ref{s4.9}, which provides a bridge from the
\emph{algebra} of the unifying number system to the \emph{differential}
equations of motion (Schr\"{o}dinger's and Dirac's). This is duly recognized
as \textquotedblleft Zovko's interpretation\textquotedblright. All subsequent
physical interpretations are self-evident because the mathematical theorems in the new number system mirror known properties in modern physics.

Darko \v{Z}ubrini\'{c}, Department of Applied Mathematics at the Faculty of
Electrical Engineering and Computing, University of Zagreb, has made several very good suggestions concerning mathematical terminology, and found a few errors what was to be the final draft of this paper. Dr. \v{Z}ubrini\'{c} has also pointed out that the procedure referred
to as \textquotedblleft structural quantization\textquotedblright\ is not a
logically neutral separation of variables, but a postulate which can be
weakened to yield non-linear generalizations of Shr\"{o}dinger's equation.
This observation was taken into account without affecting the objective of the
paper, which is structural unification.

Tomislav Lozi\'{c}, independent designer of control systems, has read the
now-obsolete `final draft' with a specific objective: To identify the
conceptual difficulties a new reader might encounter in the heuristic steps
which generalized the field of complex numbers to the unifying number system.
To take his suggestions and objections into account, the entire approach was
rewritten from a different point of view. In the new version (the present
paper), the unifying number system comes out in section \ref{S2} as a unique
single-step generalization of the quaternions (from a Hurwitz structure to a
non-Hurwitz one). The original approach is reproduced in the Appendix.

Florin Moldoveanu, Committee for Philosophy and the Sciences, University of
Maryland, who is very familiar with the new number system for having followed its development for many years and discussed it by e-mail with the author, has kindly offered to go `with a very fine comb' through the mathematics of the final manuscript. Checking for typographical errors and re-doing all calculations, Dr. Moldoveanu has discovered several errors that had escaped the author's notice for being in formulae not subsequently used in the present paper.%

\section{\label{S2}The quantionic number system}

The unifying number system, hypothetical at this point, will be denoted by
$\mathcal{Q}$ and referred to as the \textbf{algebra of quantions} ---
pronounced `quant-ion', rhyming with `bastion', not with `fraction'. The
adjective, \textquotedblleft quantionic\textquotedblright, follows the model
of \textquotedblleft quaternionic\textquotedblright.

We shall arrive at this algebra by a comparative analysis of the division
algebras. The discussions will refer to the following graph, where the
standard symbols $\mathbb{R}\ $and $\mathbb{C}$ represent the fields of real
and complex numbers (both commutative and associative), $\mathbb{H},\ $ the
field of quaternions (not commutative but associative), and $\mathbb{O},\ $the
algebra of octonions (neither commutative nor associative). The doubly framed
structures, $\mathbb{R},\ \mathbb{C},$ and $\mathcal{Q},$ are the Leibnizian
number systems, (defined in subsection \ref{s2.3}).

\begin{center}%
\setlength{\unitlength}{0.01in}
\begin{picture}(580,250)
\put(510,147){\oval(140,36)}
\put(510,155){\makebox(0,0){pairs of }}
\put(510,139){\makebox(0,0){real numbers}}
\put(70,87){\oval(140,36)}
\put(70,95){\makebox(0,0){Quadruples of}}
\put(70,79){\makebox(0,0){complex numbers}}
\put(510,87){\oval(140,36)}
\put(510,95){\makebox(0,0){pairs of }}
\put(510,79){\makebox(0,0){complex numbers}}
\put(510,26){\oval(140,36)}
\put(510,35){\makebox(0,0){pairs of }}
\put(510,19){\makebox(0,0){quaternions}}
\put(460,195){\makebox(0,0){Division algebras}}
\put(145,230){\makebox(0,0){NON-HURWITZ STRUCTURES}}
\put(460,230){\makebox(0,0){HURWITZ STRUCTURES}}
\put(222,20){\makebox(0,0){Non-associative structures}}
\put(145,130){\makebox(0,0){The quantionic number system}}
\put(320,207){\makebox(0,0){$\mathbb{R}$}}
\put(320,147){\makebox(0,0){$\mathbb{C}$}}
\put(260,87){\makebox(0,0){$\mathcal{Q}$}}
\put(320,87){\makebox(0,0){$\mathbb{H}$}}
\put(320,26){\makebox(0,0){$\mathbb{O}$}}
\put(320,207){\circle{32}}
\put(320,207){\circle{30}}
\put(320,147){\circle{32}}
\put(320,147){\circle{30}}
\put(260,87){\circle{32}}
\put(260,87){\circle{30}}
\put(320,87){\circle{30}}
\put(320,27){\circle{30}}
\put(308,135){\vector(-1,-1){36}}
\put(320,190){\vector(0,-1){26}}
\put(320,130){\vector(0,-1){28}}
\put(320,72){\vector(0,-1){30}}
\put(337,147){\line(1,0){103}}
\put(335,87){\line(1,0){105}}
\put(335,27){\line(1,0){105}}
\put(140,87){\line(1,0){103}}
\put(195,98){\makebox(0,0){Level 2}}
\put(365,207){\makebox(0,0){Level 0}}
\put(380,158){\makebox(0,0){Level 1}}
\put(380,98){\makebox(0,0){Level 2}}
\put(380,38){\makebox(0,0){Level 3}}
\multiput(290,0)(26,0){12}{\line(1,0){9}}
\multiput(290,243)(26,0){12}{\line(1,0){9}}
\multiput(290,0)(0,26){10}{\line(0,1){9}}
\multiput(585,0)(0,26){10}{\line(0,1){9}}
\multiput(-5,0)(26,0){12}{\line(1,0){9}}
\multiput(-5,243)(26,0){12}{\line(1,0){9}}
\multiput(-5,0)(0,26){10}{\line(0,1){9}}
\thicklines\multiput(15,55)(35,0){16}{\line(1,0){20}}
\end{picture}%
\vspace{3pt}%

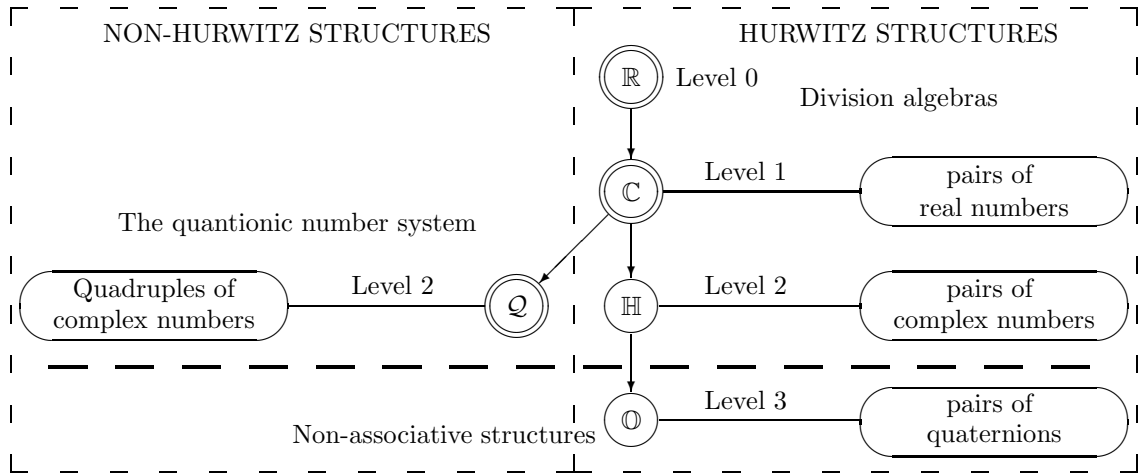
Figure \ref{S2}.1: The generalizations of the complex numbers.
\end{center}

The complex numbers were introduced by Cardano in 1545, but mathematicians
remained uncomfortable about the square root of minus one for almost three
centuries. This philosophical difficulty (for it is not a mathematical one)
was eliminated in 1833 by Hamilton, who redefined the field of complex numbers
as an algebra of pairs of real numbers. Their soundness thus no longer
depended on the apparently questionable object $\sqrt{-1}.$ It was based
instead on an ontologically neutral multiplication rule for pairs of real
numbers. The discovery of quaternions (Hamilton, 1843) and of octonions
(Cayley, 1845) followed the same approach, known as the \textquotedblleft
Cayley-Dickson construction\textquotedblright.%

\vspace{6pt}%

\textbf{The Cayley-Dickson construction: }\emph{Let} $a,b,u,$ and $v$ \emph{be
any elements of an algebra with a unit }$e,$\emph{\ an involution }$a^{\ast},$
\emph{and a norm} $a^{\ast}a.$\emph{ Pairs of these elements,} $\left\{
a,b\right\}  ,$ $\left\{  u,v\right\}  ,$ \emph{etc., are taken to represent}
\emph{the elements of a new algebra constructed by the following rules:}%
\begin{equation}
\left.
\begin{tabular}
[c]{rll}%
\emph{The new algebraic unit:} &  & $\left\{  e,0\right\}  $\\
\emph{The new involution:} &  & $\left\{  a,b\right\}  ^{\ast}=\left\{
a^{\ast},-b\right\}  $\\
\emph{The new product:} &  & $\left\{  a,b\right\}  \left\{  u,v\right\}
=\left\{  au-bv^{\ast},av+u^{\ast}b\right\}  $\\
\emph{The new norm:} &  & $\left\{  a,b\right\}  ^{\ast}\left\{  a,b\right\}
$%
\end{tabular}
\ \right\}  \label{2k1}%
\end{equation}

\vspace{6pt}%

Beginning with the field of real numbers, where $e=1$ and $a^{\ast}\equiv a, $
the Cayley-Dickson construction may be iteratively applied forever, but with
little interest past the octonions.

\subsubsection*{The normed algebras}

We shall now discuss each concept in the conditions (\ref{2k1}), taking into
account that the existence of a norm presupposes the existence of an involution.

The \emph{unit} $\left\{  1,0\right\}  $ is defined by the property $\left\{
1,0\right\}  \left\{  a,b\right\}  \equiv\left\{  a,b\right\}  \left\{
1,0\right\}  \equiv\left\{  a,b\right\}  ,$ and is denoted by
\textquotedblleft 1\textquotedblright\ in every algebra.

The \textbf{involution} is the identity in $\mathbb{R};$ complex conjugation
in $\mathbb{C};$ and is iteratively defined by (\ref{2k1}) in the algebras
$\mathbb{H}$ and $\mathbb{O}.$

An algebra is said to be \textbf{normed} if the norm of every element is
positive definite, that is, a non-negative real number which vanishes if and
only if the element in question is zero. The mapping of the algebra onto  $\mathbb{R}^+,\ $  referred to a the \textbf{norm function} is to be constructed only out of the operations which already exist in the algebra.

If the algebra has a unit and if the norm function is symmetric --- which is the case in the
division algebras --- every element other than zero has a unique inverse of the form,%

\begin{equation}
\left\{  a,b\right\}  ^{-1}=\frac{\left\{  a,b\right\}  ^{\ast}}{\left\{
a,b\right\}  ^{\ast}\left\{  a,b\right\}  } \label{2k3}%
\end{equation}

The four division algebras are therefore \textquotedblleft normed algebras with a
unit\textquotedblright .
The inverse of this statement is the subject of
Hurwitz's theorem (1898):%

\vspace{7pt}%

\textbf{Hurwitz's theorem:}\emph{\ Every normed algebra with a unit is
isomorphic with one of the four division algebras. \label{H-T}}%

\vspace{7pt}%

This is why the Cayley-Dickson construction is not very interesting past the octonions. 

In Figure \ref{S2}.1, only the central triangle at levels 1 and 2 need be considered because the field of real numbers has no algebraic structure of any relevance to our purposes, while the algebra of octonions is not
associative and is also much too distant from the complex numbers.
The search for the algebra of quantions (assuming that such an algebra exists) may therefore be limited to an analysis of the central triangle:

\begin{center}%
\setlength{\unitlength}{0.01in}
\begin{picture}(500,165)
\put(250,140){\circle{30}}
\put(140,30){\circle{30}}
\put(360,30){\circle{30}}
\put(250,140){\makebox(0,0){$\mathbb{C}$}}
\put(140,30){\makebox(0,0){$\mathcal{Q}$}}
\put(360,30){\makebox(0,0){$\mathbb{H}$}}
\put(230,120){\vector(-1,-1){70}}
\put(270,120){\vector(1,-1){70}}
\put(340,30){\vector(-1,0){180}}
\put(385,100){\makebox(0,0){$Inter-level$ $generalization$}}
\put(415,85){\makebox(0,0){The Cayley-Dickson construction}}
\put(195,95){\makebox(0,0)[r]{$Inter-level$ $generalization$}}
\put(250,40){\makebox(0,0){$Intra-level$ $generalization$}}
\put(250,20){\makebox(0,0){Takes us out of the reach}}
\put(250,5){\makebox(0,0){of Hurwitz's theorem}}
\put(120,36){\makebox(0,0)[r]{Structurally very rich}}
\put(120,24){\makebox(0,0)[r]{number system}}
\put(380,36){\makebox(0,0)[l]{Structurally rich}}
\put(380,24){\makebox(0,0)[l]{number system}}
\put(270,140){\makebox(0,0)[l]{Structurally poor number system}}
\end{picture}%

\vspace{10pt} %

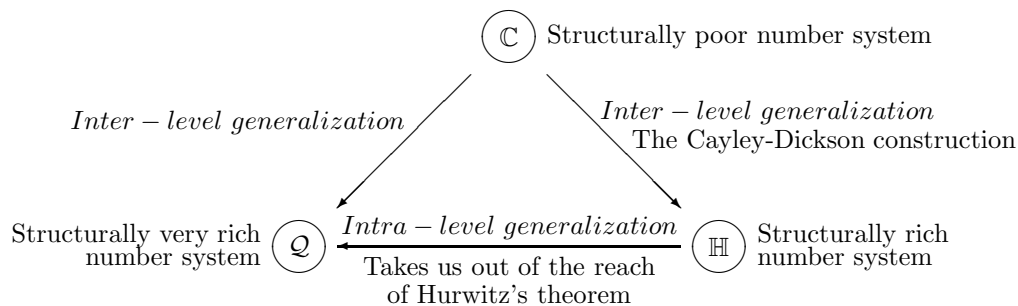
Figure \ref{S2}.2: Two paths to the unifying number system.\label{pTriangle}
\end{center}

\vspace{3pt}%

Of the two paths from $\mathbb{C}$ to $\mathcal{Q},$ the indirect one by way
of quaternions is the most transparent because the inter-level generalization
(the transition from level 1 to level 2) has already been completed by
Hamilton. 
The field $\mathbb{H}$ of quaternions will therefore play a key role
in deriving the algebra $\mathcal{Q}$ (but none whatsoever subsequently).

\subsection{\label{s2.1}The quaternions}

Referring to Figure \ref{S2}.2, we shall begin the search for $\mathcal{Q}$ at
the structurally rich algebra $\mathbb{H},$ with the expectation that some specific
property will suggest itself as a special case of a more general one ---
this being more likely to happen in the rich structure $\mathbb{H}$ than in
the simple structure $\mathbb{C}.$ To this end, the following conventions will
prove convenient:

1. The involution of complex numbers (complex conjugation) will be denoted by
a star, while the involution of quaternions will be denoted by a dagger.

2. The norm defined in (\ref{2k1}) will be denoted by $A\left(  x\right)  $
and referred to as the \textbf{algebraic norm.} Thus, for complex numbers,
$A\left(  z\right)  =z^{\ast}z,$ and for quaternions,
\begin{equation}
A\left(  Q\right)  =Q^{\dag}Q \label{2k4}%
\end{equation}
Since $Q^{\dag}Q=QQ^{\dag},$ the algebraic norm is symmetric:%
\begin{equation}
A\left(  Q^{\dag}\right)  =A\left(  Q\right)  \label{2k4a}%
\end{equation}

3. The inverse defined by (\ref{2k3}) will be referred to as the
\textbf{algebraic inverse.} For the complex numbers, the algebraic inverse is
$z^{-1}=z^{\ast}/A\left(  z\right)  .$ For the quaternions it is
\begin{equation}
Q^{-1}=Q^{\dag}/A\left(  Q\right)  \label{2k5}%
\end{equation}

The qualifier \textquotedblleft algebraic\textquotedblright\ is needed because
alternative definitions of the norm and of the inverse will soon be introduced.

We shall now discuss several representations of quaternions for two purposes:
To identify a property that could be generalized (different formalisms tend to
suggest different ideas), and as a model for studying the properties of quantions.

These representations (or formalisms) have no standard names; the names in the
subsection titles which follow have been selected for easy recall --- an
important consideration when several new concepts have to be introduced.

\subsubsection*{The Cayley-Dickson formalism}

The specialization of (\ref{2k1}) to quaternions yields
\begin{equation}
\left.
\begin{array}
[c]{c}%
Q=\left\{  a,b\right\}  \in\mathbb{H}\\
Q^{\dag}=\left\{  a^{\ast},-b\right\}  \in\mathbb{H}\\
\left\{  a,b\right\}  \left\{  u,v\right\}  =\left\{  au-bv^{\ast},bu^{\ast
}+av\right\}
\end{array}
\right\}  \label{2k6}%
\end{equation}
The algebraic norm is thus%
\begin{equation}
A\left(  Q\right)  =\left\{  a,b\right\}  \left\{  a,b\right\}  ^{\dag
}=\left\{  a,b\right\}  \left\{  a^{\ast},-b\right\}  =\left\{  a^{\ast
}a+b^{\ast}b,0\right\}  =\left(  a^{\ast}a+b^{\ast}b\right)  \left\{
1,0\right\}  \label{2k7}%
\end{equation}
and the algebraic inverse is%
\begin{equation}
\left\{  a,b\right\}  ^{-1}=\frac{1}{a^{\ast}a+b^{\ast}b}\left\{  a^{\ast
},-b\right\}  \label{2k8}%
\end{equation}

\subsubsection*{The linear formalism}

The underlying real linear space of the algebra of quaternions has four real
dimensions. Denoting the four basis elements by $\left\{  \mathbf{e}%
_{0},\mathbf{e}_{1},\mathbf{e}_{2},\mathbf{e}_{3}\right\}  ,$ a quaternion is thus
of the form%
\begin{equation}
Q=w\mathbf{e}_{0}+x\mathbf{e}_{1}+y\mathbf{e}_{2}+z\mathbf{e}_{3} \label{2k10}%
\end{equation}

The individuality of a particular quaternion is encoded in the quadruple
$\left\{  w,x,y,z\right\}  $ of real numbers, while the structure of the field
of quaternions is encoded in the multiplication table for the basis elements
--- which are therefore both basis vectors and algebraic objects.

Just as the pairs $\left\{  x,y\right\}  $ of real numbers defining a complex
number are commonly interpreted as points in the \textbf{Gaussian plane,} the
quadruples $\left\{  w,x,y,z\right\}  $ will be interpreted as points in a
four-dimensional real linear space. We shall refer to this space as the
\textbf{quaternionic Gaussian space.} This terminology is not standard but it
suits out purpose.

Instead of $\left\{  \mathbf{e}_{0},\mathbf{e}_{1},\mathbf{e}_{2}%
,\mathbf{e}_{3}\right\}  ,$ we shall use the standard symbols $\left\{
1,i,j,k\right\}  ,$%
\begin{equation}
Q=w+xi+yj+zk \label{2kb}%
\end{equation}
introduced by Hamilton along with the multiplication rules%
\begin{equation}
\left.
\begin{array}
[c]{l}%
i^{2}=j^{2}=k^{2}=-1\\
ij=k\ \cdots\ \text{cyclically}\\
i^{\dag}=-i,\ \text{etc.}%
\end{array}
\right\}  \label{2k10a}%
\end{equation}

To verify that these rules are consistent with the product (\ref{2k6}) in the
Cayley-Dickson formalism, let us expand the pairs of complex numbers in the
basis $\left\{  1,i,j,k\right\}  ,$
\begin{align*}
\left\{  a,b\right\}   &  =a+bj=\left(  w+xi\right)  +\left(  y+zi\right)
j=w+xi+yj+zk\\
\left\{  u,v\right\}   &  =u+vj=\left(  m+ni\right)  +\left(  r+si\right)
j=m+ni+rj+sk
\end{align*}
and multiply them according to the rules (\ref{2k6}):%
\[
\left\{  a,b\right\}  \left\{  u,v\right\}  =\left(  a+bj\right)  \left(
u+vj\right)  =\left(  au+bjvj\right)  +\left(  avj+bju\right)
\]
Computing the terms $bjvj$ and $bju,$%
\begin{align*}
bjvj  &  =\left(  y+zi\right)  j\left(  r+si\right)  j=\left(  y+zi\right)
\left(  rj+sji\right)  j\\
&  =\left(  y+zi\right)  \left(  -r+sjij\right)  =\left(  y+zi\right)  \left(
-r+si\right)  =-bv^{\ast}\\
bju  &  =\left(  y+zi\right)  \left(  mj+nji\right)  =\left(  y+zi\right)
\left(  m-ni\right)  j=bu^{\ast}j
\end{align*}
and substituting them into the \ expression for the product yields%
\[
\left\{  a,b\right\}  \left\{  u,v\right\}  =\left\{  au-bv^{\ast},bu^{\ast
}+av\right\}
\]
which is indeed the same as in (\ref{2k6}).

The involution%
\begin{equation}
Q^{\dag}=\left\{  a^{\ast},-b\right\}  =\left(  w+xi\right)  ^{\ast}-\left(
y+zi\right)  j=w-xi-yj-zk \label{2k11}%
\end{equation}
is consistent with (\ref{2k10a}). The algebraic norm (\ref{2k7}) assumes the
positive-definite form%
\begin{equation}
A\left(  Q\right)  =\left(  a^{\ast}a+b^{\ast}b\right)  \left\{  1,0\right\}
=\left(  w^{2}+x^{2}+y^{2}+z^{2}\right)  \left\{  1,0\right\}  \label{2k12}%
\end{equation}
The quaternionic Gaussian space is therefore a four-dimensional Euclidean
space $E^{4}.$

\subsubsection*{The matrix formalism}

Since their product is associative, the quaternions may be represented by
matrices whose elements are selected so as to make the algebra of these
matrices isomorphic with the algebra (\ref{2k6}). The subalgebra of
quaternions of the form $Q=\left\{  a,0\right\}  $ being the field of complex
numbers, the elements $a$ and $a^{\ast}$ must be on the main diagonal, and only on the main diagonal, in order
to ensure the stability of the product. Two solutions are stable:%
\begin{align*}
\text{Case I}\text{:\ }  &  Q=\left\{  a,b\right\}  =%
\begin{pmatrix}
a & B\\
C & a^{\ast}%
\end{pmatrix}
\\
\text{Case II}\text{:\ }  &  Q=\left\{  a,b\right\}  =%
\begin{pmatrix}
a^{\ast} & B\\
C & a
\end{pmatrix}
\end{align*}
where $B$ and $C$ are to be determined as linear functions of $b$ and
$b^{\ast} alone.$

\textbf{Case I:} The condition
\[
\left\{  a,b\right\}  \left\{  u,v\right\}  =%
\begin{pmatrix}
a & B\\
C & a^{\ast}%
\end{pmatrix}%
\begin{pmatrix}
u & V\\
W & u^{\ast}%
\end{pmatrix}
=%
\begin{pmatrix}
au-bv^{\ast} & X\\
Y & \left(  au-bv^{\ast}\right)  ^{\ast}%
\end{pmatrix}
\]
reduces to%
\[%
\begin{pmatrix}
BW+bv^{\ast} & Bu^{\ast}+Va-X\\
Wa^{\ast}+Cu-Y & vb^{\ast}+CV
\end{pmatrix}
\allowbreak=0
\]
The equations on the main diagonal admit two solutions, characterized by the
sign indicator $\varepsilon=\pm1:$
\begin{align*}
B  &  =-\varepsilon b,\ \ W=\varepsilon v^{\ast}\\
V  &  =-\varepsilon v,\ \ C=\varepsilon b^{\ast}%
\end{align*}
This implies,%
\begin{align*}
\ C  &  =-B^{\ast}\\
W  &  =-V^{\ast}%
\end{align*}
Substitution of these results into the secondary diagonal yields%
\begin{align*}
X  &  =-\varepsilon\left(  bu^{\ast}+av\right) \\
Y  &  =\varepsilon\left(  b^{\ast}u+a^{\ast}v^{\ast}\right)
\end{align*}
Since these terms satisfy the consistency relation%
\[
Y=-X^{\ast}%
\]
solutions exist, the matrices representing a quaternion being%
\begin{equation}
Q=\left\{  a,b\right\}  =%
\begin{pmatrix}
a & -\varepsilon b\\
\varepsilon b^{\ast} & a^{\ast}%
\end{pmatrix}
\label{2k19}%
\end{equation}

Verification:%
\[%
\begin{pmatrix}
a & -\varepsilon b\\
\varepsilon b^{\ast} & a^{\ast}%
\end{pmatrix}%
\begin{pmatrix}
u & -\varepsilon v\\
\varepsilon v^{\ast} & u^{\ast}%
\end{pmatrix}
=%
\begin{pmatrix}
au-bv^{\ast} & -\varepsilon\left(  av+bu^{\ast}\right) \\
\varepsilon\left(  av+bu^{\ast}\right)  ^{\ast} & a^{\ast}u^{\ast}-vb^{\ast}%
\end{pmatrix}
\allowbreak
\]

\textbf{Case II:} The condition%

\[%
\begin{pmatrix}
a^{\ast} & B\\
C & a
\end{pmatrix}%
\begin{pmatrix}
u^{\ast} & V\\
W & u
\end{pmatrix}
=%
\begin{pmatrix}
au-bv^{\ast} & X\\
Y & \left(  au-bv^{\ast}\right)  ^{\ast}%
\end{pmatrix}
\]
reduces to%
\[%
\begin{pmatrix}
BW+bv^{\ast}+a^{\ast}u^{\ast}-au & Va^{\ast}-X+Bu\\
aW-Y+Cu^{\ast} & vb^{\ast}-a^{\ast}u^{\ast}+CV+au
\end{pmatrix}
\allowbreak\allowbreak=0
\]
which has no solution because, in general, $a^{\ast}u^{\ast}-au\neq0.$

The only two matrix representations of the quaternions are therefore (\ref{2k19}).
To understand their mutual relationship, let us consider the special case of
$a$ and $b$ real. For $\varepsilon=1,$ the matrix (\ref{2k19}) represents a
complex number; for $\varepsilon=-1,$ it represents its complex conjugate. We
may therefore drop $\varepsilon$ as redundant, for it merely encodes complex
conjugation. Tthe unique matrix representation of a quaternion is therefore
\begin{equation}
Q=%
\begin{pmatrix}
a & -b\\
b^{\ast} & a^{\ast}%
\end{pmatrix}
\label{2k20}%
\end{equation}

The matrix formation rule (\ref{2k20}) applied to the involution yields
\begin{equation}
Q^{\dag}=\left\{  a,b\right\}  ^{\dag}=\left\{  a^{\ast},-b\right\}  =%
\begin{pmatrix}
a^{\ast} & b\\
-b^{\ast} & a
\end{pmatrix}
\equiv%
\begin{pmatrix}
a & -b\\
b^{\ast} & a^{\ast}%
\end{pmatrix}
^{\dag} \label{2k22}%
\end{equation}
which says that \emph{the involution of a quaternion is equivalent to the Hermitian
conjugate of its matrix representation.}

The algebraic norm is%
\begin{equation}
A\left(  Q\right)  =Q^{\dag}Q=%
\begin{pmatrix}
a & -b\\
b^{\ast} & a^{\ast}%
\end{pmatrix}
^{\dag}%
\begin{pmatrix}
a & -b\\
b^{\ast} & a^{\ast}%
\end{pmatrix}
=\left(  bb^{\ast}+aa^{\ast}\right)  I\allowbreak\label{2k24}%
\end{equation}

While the results (\ref{2k22}) and (\ref{2k24}) merely confirm what should have been
expectated, the next one opens a door to the world outside the division algebras
--- which is what we need.

To obtain the algebraic inverse (\ref{2k5}), as defined in division algebras,
we substitute the expressions (\ref{2k22}) and (\ref{2k24}) for $Q^{\dag}$ and
$A\left(  Q\right)  :$
\begin{equation}
Q^{-1}=\frac{1}{A\left(  Q\right)  }Q^{\dag}=\frac{1}{bb^{\ast}+aa^{\ast}}%
\begin{pmatrix}
a^{\ast} & b\\
-b^{\ast} & a
\end{pmatrix}
\label{2k26}%
\end{equation}
But this is not the only way the inverse can be computed: Since a quantion $Q$ has a matrix representation, $Q^{-1}$ may be computed as a matrix
inverse:
\begin{equation}
Q^{-1}=%
\begin{pmatrix}
a & -b\\
b^{\ast} & a^{\ast}%
\end{pmatrix}
^{-1}=\frac{1}{bb^{\ast}+aa^{\ast}}%
\begin{pmatrix}
a^{\ast} & b\\
-b^{\ast} & a
\end{pmatrix}
\allowbreak\allowbreak\label{2k27}%
\end{equation}

The two solutions coincide numerically, but they are conceptually different:

The second version, (\ref{2k27}), exists in all matrix algebras --- even those
that do not have an involution.

The first version, (\ref{2k26}), exists in all division algebras --- including
the octonions, which do not have a matrix representation for not being associative.

To formalize this conceptual difference we shall refer to the denominator in
(\ref{2k27}) as the \textbf{metric norm,} and denote it by $M\left(  Q\right)
:$%
\begin{equation}
M\left(  Q\right)  =bb^{\ast}+aa^{\ast} \label{2k28}%
\end{equation}
Comparison with (\ref{2k7}) yields the relation%
\begin{equation}
A\left(  Q\right)  =M\left(  Q\right)  \left\{  1,0\right\}  \label{2k29}%
\end{equation}

The algebraic norm is thus conceptually a quaternion (owing to the presence of
the quaternionic unit $\left\{  1,0\right\}  $), while the metric norm is
conceptually a complex number (for its being the determinant of a matrix) ---
though both coalesce to the same positive real number.

Similarly, the matrix on the right-hand side of (\ref{2k27}) is formally the
Hermitian conjugate of $Q,$ but this is also to be regarded as a coincidence.
To formalize the conceptual difference between (\ref{2k26}) and (\ref{2k27}),
we shall refer to the second matrix in (\ref{2k27}) as the \textbf{metric
dual} of $Q,$ and denote it by a \textbf{sharp operator:}%
\begin{equation}
Q^{\#}=%
\begin{pmatrix}
a^{\ast} & b\\
-b^{\ast} & a
\end{pmatrix}
\label{2k30}%
\end{equation}
The `true' inverse of $Q$ --- which does not depend on the involution --- is
therefore%
\begin{equation}
Q^{-1}=Q^{\#}/M\left(  Q\right)  \label{2k31}%
\end{equation}

\subsubsection*{The vector formalism}

In the matrix representation (\ref{2k20}) of a quaternion, the complex numbers
$a$ and $b$ appear redundantly: as themselves and as their complex conjugates.
No information is therefore lost by reducing the matrix representing a
quaternion to the vector defined as its first column.

Using Dirac's very convenient formalism of bras and kets, we shall write the
column vector in question as a ket $\left\vert q\right)  :$%
\begin{equation}
\left\vert q\right)  =%
\begin{pmatrix}
a\\
b^{\ast}%
\end{pmatrix}
\in\mathcal{H} \label{2k36}%
\end{equation}
where $\mathcal{H}$ is a complex two-dimensional linear space.\footnote{In
order to avoid confusion with state vectors $\left\vert q\right\rangle $ in
quantionic Hilbert spaces, we shall use the symbol $\left\vert q\right)  $ in
vector representations of number systems.}\label{pUnitary}

Clearly, $\mathbb{H}$ and $\mathcal{H}$ are linearly isomorphic, but while the
former is an algebra, the latter is only a linear space.

The matrices (\ref{2k20}) may now be interpreted as linear operators in
$\mathcal{H}:$
\begin{equation}
P\left\vert q\right)  =%
\begin{pmatrix}
u & -v\\
v^{\ast} & u^{\ast}%
\end{pmatrix}%
\begin{pmatrix}
a\\
b^{\ast}%
\end{pmatrix}
=%
\begin{pmatrix}
ua-vb^{\ast}\\
\left(  va^{\ast}+ub\right)  ^{\ast}%
\end{pmatrix}
\allowbreak\label{2k37}%
\end{equation}

Since the norm $\left(  \ast|\ast\right)  $ of the vectors in $\mathcal{H}$ is
positive definite,%
\begin{equation}
\left(  q|q\right)  =%
\begin{pmatrix}
a^{\ast} & b
\end{pmatrix}%
\begin{pmatrix}
a\\
b^{\ast}%
\end{pmatrix}
=a^{\ast}a+b^{\ast}b\in\mathbb{R}^{+} \label{2k38}%
\end{equation}
$\mathcal{H}$ is a two-dimensional inner product space. We shall refer to it
as the \textbf{representation space} associated to the field of quaternions.
Owing to relation (\ref{2k37}), it is a left regular representation space for
this field.

The same quaternion, $\mathcal{Q},$ may therefore be represented by two
conceptually different objects: In the \textquotedblleft matrix
representation\textquotedblright, $\mathcal{Q}$ is given by the matrix
(\ref{2k20}); in the \textquotedblleft vector representation\textquotedblright%
, $\mathcal{Q}$ is given by the ket (\ref{2k36}). Both concepts will play
equally important roles in quantionic differential field equations (Part 2 of
the present work).

The norm $\left(  q|q\right)  $ in $\mathcal{H}$ will be denoted by $H\left(
Q\right)  $ and referred to as the \textbf{Hermitian norm.} Numerically, the
norm
\[
H\left(  Q\right)  =\left(  q|q\right)
\]
coincides with the algebraic and metric norms,%
\begin{equation}
A\left(  Q\right)  =M\left(  Q\right)  =H\left(  Q\right)  \label{2k39}%
\end{equation}
but this is a coincidence, as can be concluded from the following observations:%
\begin{equation}
\left.
\begin{tabular}
[c]{rl}%
$A\left(  Q\right)  $ & is a quaternion which happens to be real\\
$M\left(  Q\right)  $ & is a complex number which happens to be real\\
$H\left(  Q\right)  $ & is a real number
\end{tabular}
\ \right\}  \label{2k40}%
\end{equation}

\subsection{\label{s2.2}Generalizing the quaternions}

In some cases, a \emph{loss of generality}\ may be regarded as a
\textbf{degeneracy}\ (for example, in the special case of $m=0,$ the hyperboloid
$E^{2}-(\vec{p}) ^{2}=m^{2}$ degenerates to a nul- cone). In
the opposite direction, it is sometimes justified to refer to the
\emph{elimination of a degeneracy} as a \textbf{generalization.} The
transition from quaternions to quantions, shown in Figure \ref{S2}.2 on page
\pageref{pTriangle}, is such a generalization. Being intra-level, it is much
milder than the Cayley-Dickson construction, which is an inter-level generalization.

The degeneracy is formally expressed in relations (\ref{2k39}) and in the
numerical equation
\begin{equation}
Q^{\#}=Q^{\ast} \label{2n1}%
\end{equation}
But the observations (\ref{2k40}) suggest that (\ref{2k39}) is a special case of something more
general. Similarly, the definitions (\ref{2k31}) of $Q^{\#}$ by way of the
inverse of a matrix, and of $Q^{\ast}$ as a postulated involution, suggest the
existence of a new algebra --- to be called the algebra of quantions --- in
which these objects are not only conceptually different but numerically
different as well.

Since $\mathcal{Q}$ is to be an algebra which structurally differs from the
algebra $\mathbb{H}$ of quaternions only in not being degenerate, the matrix
representation (\ref{2k20}) of quaternions generalizes to a matrix
representation of quantions,%
\begin{equation}
Q=\left[  \text{a}\ \text{complex }m\times m\text{ matrix}\right]
\in\mathcal{Q} \label{2n2}%
\end{equation}
where $m$ is greater than two but initially unknown. The number of independent complex variables being two in the case of quaternions, ($a$ and $b$), let it be $n$ for quantions, where it is also greater than two but initially unknown. We shall refer to the number $n$ of independent variables, whether they
are real or complex, as the number of (real or complex) \textbf{degrees of
freedom}. Clearly, $\mathcal{Q}$ is a subalgebra of the general algebra $\mathcal{M}$ which has $m^2 $ complex degrees of freedom.

We shall now ensure the existence in $\mathcal{Q}$ of all properties of
quaternions other than degeneracy: the existence of a\emph{\ unit,} of an
\emph{involution} (and hence of an \emph{algebraic norm}), of a \emph{metric
dual} (and hence of a \emph{metric norm}), and of a \emph{vector representation.}

\textbf{The unit:} The $m\times m$ unit matrix $I$ exists in $\mathcal{M}.$ We
include it in $\mathcal{Q}:$%
\begin{equation}
I\in\mathcal{Q}\subset\mathcal{M} \label{2n3}%
\end{equation}

\textbf{The involution:} The Hermitian conjugate $Q^{\dag}$ exists for every
$Q\in\mathcal{M}.$ We include it in $\mathcal{Q}:$%
\begin{equation}
Q\in\mathcal{Q}\Longrightarrow Q^{\dag}\in\mathcal{Q} \label{2n4}%
\end{equation}

\textbf{The algebraic norm:} Since, in general, $Q^{\dag}Q\neq QQ^{\dag},$ (otherwise
$\mathcal{Q}$ would be a division algebra), either product may be
conventionally defined as the algebraic norm. We take it to be
\begin{equation}
A\left(  Q\right)  \overset{def}{=}Q^{\dag}Q \label{2n5}%
\end{equation}

\textbf{The metric norm and the metric dual:} In general, the inverse of a
matrix is%
\begin{equation}
Q^{-1}=\frac{1}{\det\left(  Q\right)  }\tilde{Q} \label{2n6}%
\end{equation}
where the determinant $\det Q$ is a homogeneous polynomial of order $m$ in the
$n$ independent variables, and $\tilde{Q}$ denotes the matrix of cofactors of $Q.$
The elements of $\tilde{Q}$ are homogeneous polynomials of order $n-1$ in the
same variables.

Note: \emph{We are interested only in functional relationships.} For a
numerically given matrix $Q,$ the denominator $\det\left(  Q\right)  $ may
vanish, but it does not matter because functional relationships are not
defined over numbers but over variables --- which are just place holders for arbitrary numbers.

We now require that the inverse matrix $Q^{-1}$ of a matrix $Q\in\mathcal{Q} $
be of the form%
\begin{equation}
Q^{-1}=\frac{1}{M\left(  Q\right)  }Q^{\#} \label{2n7}%
\end{equation}
where $M\left(  Q\right)  $ is to be interpreted as a metric norm. After all
cancellations in (\ref{2n6}), the denominator must therefore be a second order homogeneous
polynomial in the $n$ independent complex variables. If this requirement is
satisfied, the elements of the matrix $Q^{\#}$ are necessarily linear
functions of the $n$ variables, so that $Q^{\#}$ may be interpreted as the
metric dual of $Q.$

For the expressions (\ref{2n6}) and (\ref{2n7}) to be equal, all matrices
$Q\in\mathcal{Q}$ must be simultaneously reducible, meaning reducible by the
same similarity transformation (possibly combined with a mirror reflection),
to matrices of the form%
\begin{equation}
Q=%
\begin{pmatrix}
N_{1} & 0 & 0 & 0\\
0 & N_{2} & 0 & 0\\
0 & 0 & \ddots & 0\\
0 & 0 & 0 & N_{k}%
\end{pmatrix}
\label{2n8}%
\end{equation}
in which the blocks $N_{i}$ are irreducible $2\times2$ matrices in the same $n$
independent variables.

These conditions immediately imply
\begin{equation}
\left.
\begin{array}
[c]{c}%
n\,=2^{2}=4\\
m=2k
\end{array}
\right\}  \label{2n9}%
\end{equation}
while the sub-matrices $N_{i}$ must satisfy the conditions%
\begin{equation}
\det N_{1}=\det N_{2}=\cdots\det N_{k} \label{2n10}%
\end{equation}
Denoting the common numerical value of these determinants by $M\left(
Q\right)  ,$ the inverse of $Q$ is given by (\ref{2n6}), and its metric dual is
\begin{equation}
Q^{\#}=%
\begin{pmatrix}
\tilde{N}_{1} & 0 & 0 & 0\\
0 & \tilde{N}_{2} & 0 & 0\\
0 & 0 & \ddots & 0\\
0 & 0 & 0 & \tilde{N}_{k}%
\end{pmatrix}
\label{2n11}%
\end{equation}

\textbf{The vector representation:} By its definition, the dimension of the
vector representation space $\mathcal{H}$ is $m=2k$ (because the matrices $Q$
are $m\times m=\left(  2k\right)  \times\left(  2k\right)  $), but it is also
$n=4$ (which is the number of independent variables). This implies $k=2,$
which simplifies (\ref{2n8}) to%
\begin{equation}
Q=%
\begin{pmatrix}
N & 0\\
0 & M
\end{pmatrix}
\label{2n12}%
\end{equation}
where the four elements of the matrix $M$ are some linear combinations of the
four independent elements of the matrix $N,$ and satisfy the condition
(\ref{2n10}), that is
\begin{equation}
\det M=\det N \label{2n13}%
\end{equation}

Owing to this condition, a unimodular $2\times2$ matrix $S$ exists, such that%
\begin{equation}
M=SNS^{-1} \label{2n14}%
\end{equation}
and therefore%
\begin{equation}
Q=%
\begin{pmatrix}
I & 0\\
0 & S
\end{pmatrix}%
\begin{pmatrix}
N & 0\\
0 & N
\end{pmatrix}%
\begin{pmatrix}
I & 0\\
0 & S^{-1}%
\end{pmatrix}
\label{2n15}%
\end{equation}

It follows from this observation that every algebra $\mathcal{Q}$ is
isomorphic to the algebra of block-diagonal matrices of the type%
\begin{equation}
Q=%
\begin{pmatrix}
N & 0\\
0 & N
\end{pmatrix}
\label{2n16}%
\end{equation}
where $N$ is an arbitrary complex $2\times2$ matrix.

We shall prove in the next subsection that the quantions are exclusively of
the form (\ref{2n16}), implying that the group of similarity transformations of the type
(\ref{2n15}) is not allowed.

\subsection{\label{s2.3}The Leibnizian number systems}

Being meant to generalize quantum mechanics, the algebra of quantions is to be
regarded as a generalization of the field of complex numbers, as on the
left-hand side of Figure \ref{S2}.2 on page \pageref{pTriangle}, though it was
simpler and more instructive to derive it indirectly by way of the field of
quaternions, as on the right-hand side of the diagram. But this indirect
approach might conceal a trap: If some property which is essential in the field of
complex numbers happens to be lost in the transition to the field of quaternions, the solution (\ref{2n15}) might be too general.

The matrix (\ref{2n15}) does contain an $\infty^{6}$ of cases, encoded in the
arbitrary complex unimodular matrix $S,$ but this is not necessarily final
because the differential structure of the complex numbers, which is not present in
quaternions, has not yet been imposed on the algebra of quantions. The
differential structure in question concerns the properties of derivation
operators. Since this issue does not arise in workaday mathematics, let us
introduce it carefully.

For smooth real functions $f\left(  x\right)  ,$ the derivation operator
$\frac{d}{dx},$ defined by the standard limiting procedure, assigns a tangent
to every point of the curve defined by $f.$ A self-consistent real
differential calculus has been developed on this idea.

For complex functions, $f\left(  z\right)  =u\left(  x,y\right)  +iv\left(
x,y\right)  ,$ the derivation operator%
\begin{equation}
\frac{d}{dz}=\frac{1}{2}\left(  \partial_{x}-i\partial_{y}\right)  \label{2v1}%
\end{equation}
is well defined when the operators $\partial_{x}$ and $\partial_{y}$ are
applicable to the real functions $u$ and $v,$ but this is merely a
transformation of variables (from two real to one complex). But if we limit
the domain of the operator $\frac{d}{dz}$ to the class of functions $f\left(
z\right)  $ which satisfy the Cauchy-Riemann condition%
\begin{equation}
\frac{d}{dz^{\ast}}f=\frac{1}{2}\left(  \partial_{x}+i\partial_{y}\right)  f=0
\label{2v2}%
\end{equation}
we enter\ the theory of analytic function, which is not a trivially
complexified real analysis but a new world --- one of the most beautiful in
all of mathematics.

Summarizing these observations, we see that only some classes of functions
over the fields of real and complex numbers admit a structurally sound
derivation operator. The question is whether the same is true for the other
number systems considered so far.

This question cannot be answered, as above, by constructing (or failing to
construct) a derivation operator for each particular algebra of interest. A
general existence condition for a derivation operator in an algebra
$\mathcal{A}$ is to be based on an abstract and structurally necessary
definition of such an operator.

To arrive at such a definition, let us begin with the basic algebraic concept of automorphism. An automorphism
$T$  of an algebra $\mathcal{A}$ is a linear mapping of the algebra
$\mathcal{A}$ onto itself,%
\[
T:\mathcal{A}\rightarrow\mathcal{A}%
\]
which commutes with the algebraic product,%
\begin{equation}
T\left(  FG\right)  =\left(  TF\right)  \left(  TG\right)  \label{2v3}%
\end{equation}
for all $F,G\in\mathcal{A}.$ It is a continuous automorphism if $T$ is a
function of a (real) parameter, $T=T_{t},$ such that the totality of these
transformations is an Abelian group:%
\begin{align*}
T_{0}  &  =I\\
T_{t}T_{u}  &  =T_{u}T_{t}=T_{\left(  t+u\right)  }%
\end{align*}
For an infinitesimal automorphism,%
\begin{equation}
T=I+\varepsilon\mathcal{D} \label{2v4}%
\end{equation}
the condition (\ref{2v3}) becomes%
\begin{equation}
\text{\textbf{\ \ \ }}\mathcal{D}\left(  FG\right)  =\left(  \mathcal{D}%
F\right)  G+F\left(  \mathcal{D}G\right)  \label{2v5}%
\end{equation}
where $\mathcal{D}$ is referred to as a \textbf{derivation operator,} and the
condition (\ref{2v5}) as the \textbf{Leibniz condition,} or \textbf{Leibniz
identity.}

If $\mathcal{A}$ is an algebra of $n\times n$ matrices for some $n,$ the
derivation operator $\mathcal{D}$ is also an $n\times n$ matrix with the same
number of degrees of freedom (because every independent variable
$a\in\mathcal{A}$ gives rise to a partial derivation operator $\partial_{a}$
in the matrix $\mathcal{D}$).

The Leibniz condition (\ref{2v5}) has therefore two aspects: It is a
\emph{differential} condition (which is its primary role), and an
\emph{algebraic} condition (which cannot be avoided if the matrix
$\mathcal{D}$ does not commute with all matrices in $\mathcal{A}$).

To separate these two conditions, let us temporarily denote by $D$ the algebraic form of
the matrix $\mathcal{D}.$ (One obtains $D$ from $\mathcal{D}$ by substituting
a new complex variable for each of the $n$ partial derivation operator
$\partial_{i}.$) The purely algebraic aspect of the Leibniz condition is
therefore%
\begin{equation}
D\left(  FG\right)  =\left(  DF\right)  G+F\left(  DG\right)  \label{2v6}%
\end{equation}
The totality of matrices $D$ is an $n-$dimensional linear space over the same
field as the algebra $\mathcal{A}.$ Let us denote it by $\mathcal{P}.$

The condition (\ref{2v6}) can be satisfied even if the algebra $\mathcal{A}$
is non-commutative because the ordering of $F$ and $G$ is the same in all
three terms. On the other hand, this condition is not satisfied if $F$ and $D
$ do not commute because their ordering is reversed in the third term. Since
$F$ is any element of $\mathcal{A}$ while $D$ is any element of $\mathcal{P},
$ the Leibniz condition (\ref{2v6}) can be satisfied only if $\mathcal{P}$ is
the \textbf{commutant} of $\mathcal{A}$ --- meaning that all matrices
$D\in\mathcal{P}$ commute with all matrices $F\in\mathcal{A}.$

This conclusion yields the abstract criterion for separating the number
systems that admit a derivation operator of the same number of degrees of
freedom from those that do not. We shall refer to the former as
\textbf{Leibnizian number systems}.

\textbf{Definition:} \emph{A number system is Leibnizian if it has a commutant
of the same number of degrees of freedom.}

Let us consider the relevant special cases:

(1) The commutative number systems $\mathbb{R}$ and $\mathbb{C}$ are trivially
Leibnizian for being their own commutants.

(2) The field of quaternions is not Leibnizian because the representative
matrices (\ref{2k20}) have two degrees of freedom and are irreducible. They
therefore commute only with the matrices $zI,$ which have one degree of freedom.

(3) \textbf{Theorem:}\ The algebra represented by the matrices (\ref{2n12}) is
Leibnizian if and only if $M=N.$

Proof: Let us write $D$ in block form,
\[
D=%
\begin{pmatrix}
W & Y\\
X & Z
\end{pmatrix}
\]
where $W$ to $Z$ are unknown complex $2\times2$ matrices that \emph{are not
functions} of the variables in $N$ and $M.$ The commutativity condition%
\[
\left[  D,Q\right]  =\left[
\begin{pmatrix}
W & Y\\
X & Z
\end{pmatrix}
,%
\begin{pmatrix}
N & 0\\
0 & M
\end{pmatrix}
\right]  =0
\]
is equivalent to%
\begin{align}
\left[  N,W\right]   &  =\left[  M,Z\right]  =0\label{2v10}\\
MX-XN  &  =NY-YM=0 \label{2v11}%
\end{align}
Since $N$ and $M$ are irreducible, (\ref{2v10}) implies%
\begin{align*}
W  &  =wI\\
Z  &  =zI
\end{align*}
where $w$ and $z$ are arbitrary complex numbers.

If $M\neq N,$ the conditions (\ref{2v11}) have no solutions as identities
(meaning the same solutions $X$ and $Y$ for all matrices $N$ and $M$).

If $M=N,$ the solutions are%
\begin{align*}
X  &  =xI\\
Y  &  =yI
\end{align*}

The most general solution for $D$ is therefore%
\begin{equation}
D=%
\begin{pmatrix}
wI & yI\\
xI & zI
\end{pmatrix}
\in\mathcal{P} \label{2v20}%
\end{equation}
for arbitrary complex numbers $w,x,y,z$ (four degrees of freedom). All
matrices (\ref{2v20}) indeed commute with all matrices
\begin{equation}
Q=%
\begin{pmatrix}
N & 0\\
0 & N
\end{pmatrix}
\in\mathcal{Q} \label{2v21}%
\end{equation}

Clearly, the linear space $\mathcal{P}$ of matrices $D$ is not only a linear
space but an algebra (it is stable under matrix multiplication).

\subsection{\label{2.4}Conclusions}

We have arrived at the conclusion that \emph{the only Leibnizian number
systems} are the doubly circled ones in Figure \ref{S2}.1 on page
3, namely $\mathbb{R},$ $\mathbb{C},$ and $\mathcal{Q}.$ They support,
respectively, classical mechanics, nonrelativistic quantum mechanics, and, by
the author's tentative contention, relativistic quantum mechanics.

Nonrelativistic quantum mechanics has brought to light many physical phenomena
that clashed with classical intuitions (stability of atoms, superposition,
entanglement, etc.). They all depend ultimately, and in an essential way, on
the theory's underlying number system being the field of complex numbers.

We are therefore to expect that a theory built on a new number system much
richer than the complex numbers will have consequences that clash with our
quantum mechanical intuitions. But a an important lesson must be learned from a century of
modern physics: Disagreements between intuitions based on a current theory
but related to questions outside the theory's domain of appplicability, and conclusions drawn from a
mathematically final generalization of this theory, tend to be settled in favor of the latter (unless
the new theory is physically `wrong' --- in the sense of not being the theory
selected by nature).\footnote{The only known major mathematically final
theories are nonrelativistic quantum mechanics and non-quantum general relativity. This is
because the Hilbert space and the locally Minkowskian Riemannian space are
mathematical structures that have no close neighbours. One says that they are rigid
structures.} We therefore take it as a rule to ignore possibly
no-longer-applicable physical intuitions, and to follow the mathematics
wherever it leads. It is the final results
that must be physically acceptable, not necessarily the initial intuitions.

Summarizing, the new Leibnizian number system whose properties remain to be
investigated is the algebra $\mathcal{Q}$ of quantions. Closely related to this
algebra is its commutant, the algebra $\mathcal{P}.$ To distinguish them, when
the distinction is not obvious from the context, we shall refer to the
elements of $\mathcal{Q}$ as \textbf{q-quantions,} and to the elements of
$\mathcal{P}$ as \textbf{p-quantions. }

These algebras and their corresponding sharp and dagger transforms are
explicitly listed for reference on this page and the next.
\begin{equation}
Q=%
\begin{pmatrix}
N & 0\\
0 & N
\end{pmatrix}
=%
\begin{pmatrix}
q_{1} & q_{3} & 0 & 0\\
q_{2} & q_{4} & 0 & 0\\
0 & 0 & q_{1} & q_{3}\\
0 & 0 & q_{2} & q_{4}%
\end{pmatrix}
\in\mathcal{Q} \label{2w1}%
\end{equation}%
\begin{equation}
P=%
\begin{pmatrix}
p_{1}I & p_{2}I\\
p_{3}I & p_{4}I
\end{pmatrix}
=%
\begin{pmatrix}
p_{1} & 0 & p_{2} & 0\\
0 & p_{1} & 0 & p_{2}\\
p_{3} & 0 & p_{4} & 0\\
0 & p_{3} & 0 & p_{4}%
\end{pmatrix}
\in\mathcal{P} \label{2w2}%
\end{equation}

The involutions are the Hermitian conjugates:%
\begin{equation}
Q^{\dag}=%
\begin{pmatrix}
N^{\dag} & 0\\
0 & N^{\dag}%
\end{pmatrix}
=%
\begin{pmatrix}
q_{1}^{\ast} & q_{2}^{\ast} & 0 & 0\\
q_{3}^{\ast} & q_{4}^{\ast} & 0 & 0\\
0 & 0 & q_{1}^{\ast} & q_{2}^{\ast}\\
0 & 0 & q_{3}^{\ast} & q_{4}^{\ast}%
\end{pmatrix}
\in\mathcal{Q} \label{2w3}%
\end{equation}%
\begin{equation}
P^{\dag}=%
\begin{pmatrix}
p_{1}^{\ast}I & p_{3}^{\ast}I\\
p_{2}^{\ast}I & p_{4}^{\ast}I
\end{pmatrix}
=%
\begin{pmatrix}
p_{1}^{\ast} & 0 & p_{3}^{\ast} & 0\\
0 & p_{1}^{\ast} & 0 & p_{3}^{\ast}\\
p_{2}^{\ast} & 0 & p_{4}^{\ast} & 0\\
0 & p_{2}^{\ast} & 0 & p_{4}^{\ast}%
\end{pmatrix}
\in\mathcal{P} \label{2w4}%
\end{equation}
The algebraic norms are therefore%
\begin{align}
A\left(  Q\right)   &  =Q^{\dag}Q=%
\begin{pmatrix}
N^{\dag}N & 0\\
0 & N^{\dag}N
\end{pmatrix}
\nonumber\\
&  =%
\begin{pmatrix}
q_{1}q_{1}^{\ast}+q_{2}q_{2}^{\ast} & q_{3}q_{1}^{\ast}+q_{4}q_{2}^{\ast} &
0 & 0\\
q_{1}q_{3}^{\ast}+q_{2}q_{4}^{\ast} & q_{3}q_{3}^{\ast}+q_{4}q_{4}^{\ast} &
0 & 0\\
0 & 0 & q_{1}q_{1}^{\ast}+q_{2}q_{2}^{\ast} & q_{3}q_{1}^{\ast}+q_{4}%
q_{2}^{\ast}\\
0 & 0 & q_{1}q_{3}^{\ast}+q_{2}q_{4}^{\ast} & q_{3}q_{3}^{\ast}+q_{4}%
q_{4}^{\ast}%
\end{pmatrix}
\allowbreak\label{2w5}%
\end{align}%
\begin{align}
A\left(  P\right)   &  =P^{\dag}P=%
\begin{pmatrix}
\left(  p_{1}p_{1}^{\ast}+p_{3}p_{3}^{\ast}\right)  I & \left(  p_{2}%
p_{1}^{\ast}+p_{4}p_{3}^{\ast}\right)  I\\
\left(  p_{1}p_{2}^{\ast}+p_{3}p_{4}^{\ast}\right)  I & \left(  p_{2}%
p_{2}^{\ast}+p_{4}p_{4}^{\ast}\right)  I
\end{pmatrix}
\nonumber\\
&  =\allowbreak%
\begin{pmatrix}
p_{1}p_{1}^{\ast}+p_{3}p_{3}^{\ast} & 0 & p_{2}p_{1}^{\ast}+p_{4}p_{3}^{\ast}
& 0\\
0 & p_{1}p_{1}^{\ast}+p_{3}p_{3}^{\ast} & 0 & p_{2}p_{1}^{\ast}+p_{4}%
p_{3}^{\ast}\\
p_{1}p_{2}^{\ast}+p_{3}p_{4}^{\ast} & 0 & p_{2}p_{2}^{\ast}+p_{4}p_{4}^{\ast}
& 0\\
0 & p_{1}p_{2}^{\ast}+p_{3}p_{4}^{\ast} & 0 & p_{2}p_{2}^{\ast}+p_{4}%
p_{4}^{\ast}%
\end{pmatrix}
\allowbreak\label{2w6}%
\end{align}

From the inverses%
\[
Q^{-1}=%
\begin{pmatrix}
N^{-1} & 0\\
0 & N^{-1}%
\end{pmatrix}
=\frac{1}{q_{1}q_{4}-q_{2}q_{3}}%
\begin{pmatrix}
q_{4} & -q_{3} & 0 & 0\\
-q_{2} & q_{1} & 0 & 0\\
0 & 0 & q_{4} & -q_{3}\\
0 & 0 & -q_{2} & q_{1}%
\end{pmatrix}
\allowbreak
\]%
\[
P^{-1}=\frac{1}{p_{1}p_{4}-p_{2}p_{3}}%
\begin{pmatrix}
p_{4}I & -p_{2}I\\
-p_{3}I & p_{1}I
\end{pmatrix}
\allowbreak=\frac{1}{p_{1}p_{4}-p_{2}p_{3}}%
\begin{pmatrix}
p_{4} & 0 & -p_{2} & 0\\
0 & p_{4} & 0 & -p_{2}\\
-p_{3} & 0 & p_{1} & 0\\
0 & -p_{3} & 0 & p_{1}%
\end{pmatrix}
\]
we extract the metric duals,%
\begin{equation}
Q^{\#}=%
\begin{pmatrix}
N^{\#} & 0\\
0 & N^{\#}%
\end{pmatrix}
=%
\begin{pmatrix}
q_{4} & -q_{3} & 0 & 0\\
-q_{2} & q_{1} & 0 & 0\\
0 & 0 & q_{4} & -q_{3}\\
0 & 0 & -q_{2} & q_{1}%
\end{pmatrix}
\label{2w7}%
\end{equation}%
\begin{equation}
P^{\#}=%
\begin{pmatrix}
p_{4}I & -p_{2}I\\
-p_{3}I & p_{1}I
\end{pmatrix}
=%
\begin{pmatrix}
p_{4} & 0 & -p_{2} & 0\\
0 & p_{4} & 0 & -p_{2}\\
-p_{3} & 0 & p_{1} & 0\\
0 & -p_{3} & 0 & p_{1}%
\end{pmatrix}
\label{2w8}%
\end{equation}
and the metric norms,%
\begin{equation}
M\left(  Q\right)  =q_{1}q_{4}-q_{2}q_{3} \label{2w9}%
\end{equation}%
\begin{equation}
M\left(  P\right)  =p_{1}p_{4}-p_{2}p_{3} \label{2w10}%
\end{equation}

The following relations are valid in both $\mathcal{Q}$ and $\mathcal{P}.$%
\begin{align}
Q^{\#}Q  &  =M\left(  Q\right)  \ I\label{2w11}\\
\left(  QR\right)  ^{\dag}  &  =R^{\dag}Q^{\dag}\label{2w12}\\
\left(  QR\right)  ^{\ast}  &  =R^{\ast}Q^{\ast} \label{2w13}%
\end{align}

Knowing that fundamental physical entities are always represented by
irreducible objects, it might seem strange that a number system expected to be
fundamental would be an algebra of reducible objects, namely of block-diagonal
matrices (\ref{2w1}). 

The reason is that the space of numerical variables
supports only the algebra of functions of these variables. A different space
is needed to support differential operators. The two may apparently coincide, as is the case
in elementary differential calculus, but their separation is already evident
in differential geometry, where displacement vectors belong to cotangent
spaces while the partial differential operators define the tangent spaces.

The conceptual difference between tangent and cotangent spaces is even more pronounced in
quantionic mathematics, where the counterparts of these linear spaces are
different algebras, namely the mutual commutants $\mathcal{Q}$ and
$\mathcal{P}.$

Let us also point out that the algebra $\mathcal{Q}$ of quantions is not an isolated number
system (like the fields $\mathbb{C}$ or $\mathbb{H}$): It uniquely defines the algebra
$\mathcal{P},$ and, as shown in the next section, both together provide a
tensorial factoring of the most general matrix algebra $\mathcal{M}:$
\begin{equation}
\mathcal{M} = \mathcal{Q} \otimes \mathcal{P}
\end{equation} %
 This might be a mere
curiosity if the algebra $\mathcal{M}$ were nothing more than a mathematical envelope of
$\mathcal{Q}$ and $\mathcal{P},$ but, as shown in Part 2 of this article (in
preparation), the real part of $\mathcal{M}$ has a physical interpretation: In
relation (\ref{3t11}), the four quantions $Q_{0}$ to $Q_{3}$ in the
Q-structuring of $\mathcal{M}$ (illustrated in Figure \ref{S3}.2 on page
\pageref{pTable2}) give rise to the four intermediate vector fields (mesons)
of the electroweak theory.%

\section{\label{S3}Linear bases in the algebra $\mathcal{M}$}

To develop a formalism in which the properties of the quantionic algebras
$\mathcal{Q}$ and $\mathcal{P}$ are most transparent, we shall select in the
algebra $\mathcal{M}$ of complex $4\times4$ matrices a basis of 16 linearly
independent Hermitian matrices structured by the subalgebras $\mathcal{Q}$ and
$\mathcal{P}$ themselves. These basis elements will be referred to as
\textbf{basis matrices} when their products are relevant, and as \textbf{basis
vectors} when the only relevant structure is linearity.

Viewed as a 16-dimensional \emph{complex} linear space, $\mathcal{M}$ is a
direct sum of two 16-dimensional \emph{real} linear spaces of Hermitian and
antihermitian matrices,%
\[
\mathcal{M}=\mathcal{M}_{h}\oplus i\mathcal{M}_{h}%
\]
where the label \textquotedblleft$h$\textquotedblright\ stands for
\textquotedblleft Hermitian\textquotedblright. While the algebra $\mathcal{M}$
is associative, $\mathcal{M}_{h}$ is a non-associative Jordan algebra (the
product in $\mathcal{M}_{h}$ is half the anticommutator of the product in
$\mathcal{M},$ but this is not relevant in the present article).

The linear space $\mathcal{M}_{h}$ decomposes into a direct sum of four real
linear subspaces (the label \textquotedblleft$0$\textquotedblright\ stands for
\textquotedblleft traceless\textquotedblright):%
\begin{equation}%
\begin{tabular}
[c]{|cccccccccc|}\hline
&  &  &  &  &  &  &  &  & \\
& $\mathcal{M}_{h}$ & $=$ & $\mathbb{R}I$ & $\oplus$ & $\mathcal{Q}_{0h}$ &
$\oplus$ & $\mathcal{P}_{0h}$ & $\oplus$ & $\mathcal{Q}_{0h}\otimes
\mathcal{P}_{0h}$\\
&  &  &  &  &  &  &  &  & \\
Dimensions: & $16$ & $=$ & $1$ & $+$ & $3$ & $+$ & $3$ & $+$ & $3\times3$\\
&  &  &  &  &  &  &  &  & \\\hline
\end{tabular}
\label{3s2}%
\end{equation}

The decomposition (\ref{3s2}) will be referred to as the \textbf{quantionic
structuring}\ of $\mathcal{M}_{h},$ and, by extension, of $\mathcal{M}.$ In a
Hermitian basis, the difference between $\mathcal{M}_{h}$ and \ $\mathcal{M}$
is only in the coefficients, which are complex numbers in general but real
numbers for Hermitian matrices.

The algebra $\mathcal{M}$ is also the Clifford algebra over the Dirac
matrices. So understood, it is structured as a lattice of Minkowski
multivectors whose elements are matrices, some Hermitian, some antihermitian.
Both decompositions of $\ \mathcal{M}$ are comparatively illustrated in the
following diagrams (two additional ones are shown in Figure \ref{S3}.2):

\begin{center}%
\setlength{\unitlength}{0.01in}
\begin{picture}(400,150)
\put(55,60){\oval(30,120)}
\put(100,105){\oval(120,30)}
\put(145,15){\oval(30,30)[br]}
\put(55,0){\line(1,0){90}}
\put(160,15){\line(0,1){90}}
\put(55,105){\oval(28,28)}
\put(86,45){\oval(30,88)[l]}
\put(144,45){\oval(30,88)[r]}
\put(86,89){\line(1,0){58}}
\put(86,1){\line(1,0){58}}
\put(300,108){\oval(140,24)}
\put(300,84){\oval(180,24)}
\put(300,60){\oval(220,24)}
\put(300,36){\oval(180,24)}
\put(300,12){\oval(140,24)}
\put(100,135){\makebox(0,0){\bf Quantionic structuring}}
\put(300,135){\makebox(0,0){\bf Spinorial structuring}}
\put(55,105){\makebox(0,0){$\mathbb{C}$}}
\put(55,60){\makebox(0,0){$\mathcal{Q}$}}
\put(100,105){\makebox(0,0){$\mathcal{P}$}}
\put(115,45){\makebox(0,0){$\mathcal{Q}_0 \bigotimes\mathcal{P}_0$}}
\put(300,108){\makebox(0,0){Scalars (1D)}}
\put(300,84){\makebox(0,0){Vectors (4D)}}
\put(300,60){\makebox(0,0){Bi-vectors (6D)}}
\put(300,36){\makebox(0,0){Pseudo-vectors (4D)}}
\put(300,12){\makebox(0,0){Pseudo-scalars (1D)}}
\end{picture}%

\label{pEnvelope}%
\vspace{4pt}%

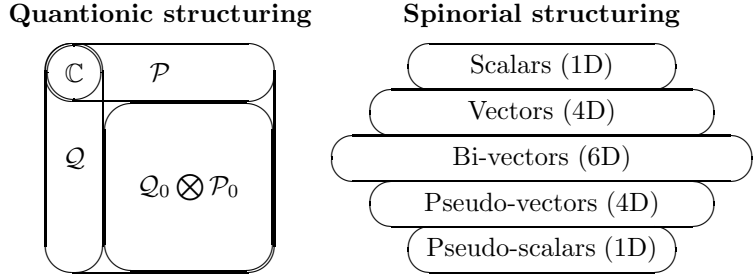
Figure \ref{S3}.1: The quantionic and spinorial decomposition of
$\mathcal{M}.$
\end{center}

The spinorial (or Clifford) structuring plays no role in deriving the
properties of quantions, but it will be needed in Part 2 to establish the
equivalence between the quantionic field equation (derived from Zovko's
interpretation) and the Dirac equation (derived as the square root of
Klein-Gordon's equation). We shall therefore need a convenient name for the
relationship between the quantionic and Clifford structuring of the same
algebra $\mathcal{M}.$ Let us refer to it by the self-descriptive term
\textbf{algebraic quantion-spinor equivalence}, shortened to \textbf{algebraic qs-equivalence.}

\subsection{\label{s3.1}The quantionic structuring}

Let us introduce the following four Hermitian matrices%
\begin{equation}
\Theta_{\mu}\overset{def}{=}%
\begin{pmatrix}
\sigma_{\mu} & 0\\
0 & \sigma_{\mu}%
\end{pmatrix}
\in\mathcal{Q} \label{3t2}%
\end{equation}
as the quantionic basis in the subalgebra $\mathcal{Q}$ of matrices of type
(\ref{2w1}). This choice establishes a formal link between quantions and
standard objects in physics. In compact form:
\begin{equation}
\Theta_{\mu}=I\times\sigma_{\mu} \label{3t2a}%
\end{equation}
where we use the following expansion rule for the direct product $A\times B$
of matrices:\ (1) Write down the matrix $A.$ (2) Multiply each of its
components by $B.$

By a standard convention, we select the Pauli matrices in their right-handed
orientation:%
\begin{equation}%
\begin{tabular}
[c]{lllllll}%
$\sigma_{0}=%
\begin{pmatrix}
1 & 0\\
0 & 1
\end{pmatrix}
,$ &  & $\sigma_{1}=%
\begin{pmatrix}
0 & 1\\
1 & 0
\end{pmatrix}
,$ &  & $\ \ \sigma_{2}=%
\begin{pmatrix}
0 & -i\\
i & 0
\end{pmatrix}
,$ &  & $\ \sigma_{3}=%
\begin{pmatrix}
1 & 0\\
0 & -1
\end{pmatrix}
$%
\end{tabular}
\ \ \label{3t3}%
\end{equation}
Two different bases may therefore be selected in the algebra $\mathcal{P}:$
\begin{align}
&  \Theta^{\mu}=\sigma_{\mu}\times I\ \ \cdots\ \ \text{Right-handed
orientation}\ \ \label{3t5}\\
&  \Theta^{\mu}=\sigma_{\mu}^{\ast}\times I\ \ \cdots\ \ \text{Left-handed
orientation} \label{3t6}%
\end{align}
These bases differ only in the sign of $\sigma_{2},$%
\begin{equation}
\left\{  \Theta^{1},\Theta^{2},\Theta^{3}\right\}  _{\left(  left\right)
}=\left\{  \Theta^{1},-\Theta^{2},\Theta^{3}\right\}  _{\left(  right\right)
} \label{3t6a}%
\end{equation}
and both commute with the basis (\ref{3t2}) in $\mathcal{Q}.$

The Greek indices attached to the kernel $\Theta$ are labels that specify
matrices. \emph{Thus, }$\Theta_{\mu}$ \emph{and }$\Theta^{\mu}$\emph{\ are not
vector components.} By contrast, $\mu$ has two interpretations in Dirac's
gamma matrices $\gamma^{\mu}:$ It is a vector index \emph{and} a matrix label.
In\emph{\ }$\Theta_{\mu}$ \emph{and }$\Theta^{\mu},$ the positions of the
labels\emph{\ }are fixed because these matrices belong to different algebras.
The standard summation rule is nevertheless admissible for being only a
short-hand convention. The choice of subscripts in $\Theta_{\mu}$ and of
superscripts in $\Theta^{\mu}$ is arbitrary at this point, but has been
selected to reflect the geometric interpretations introduced later.

We are to determine which of the options (\ref{3t5}) and (\ref{3t6}) is to be
chosen as a basis in $\mathcal{P}.$ We may not have a choice because the
right-handed orientation has already been selected for the basis in
$\mathcal{Q},$ and the two orientation are essentially different: A similarity
transformation
\begin{equation}
\Theta^{\mu}=S\Theta_{\mu}S^{-1} \label{3t7}%
\end{equation}
relates bases of the same orientation, but a reflection is also needed if the
orientations are different.

It is not evident \textit{a priori} which orientation of the basis in
$\mathcal{P}$ is structurally correct. A precedent can be found in \cite{Dir}
(pages 256-257), where Dirac uses the right-handed orientation in both cases:
His matrices $\sigma_{i}$ and $\rho_{i}$ are related to our theta matrices by
the relations $\sigma_{i}=\Theta_{i}$ and $\rho_{i}=\Theta_{right-handed}%
^{i},\ $but the matrices $\sigma_{i}$ and $\rho_{i}$ are only auxiliary. The
matrices that matter are $\alpha_{1},\alpha_{2},\alpha_{3},$ and $\alpha_{m},$
where $\alpha_{i}=\Theta_{i}\Theta^{1}$ and $\alpha_{m}=\Theta^{3}.$ Since the
matrix $\Theta^{2},$ which is the only one affected by the choice of
orientation in $\mathcal{P},$ plays no role in Dirac's work, this work cannot
help us to decide which orientation should be selected in the present case.

The only way to proceed is to chose an orientation at random and work with it
until the answer becomes evident. If the wrong orientation has been selected,
one easily switches to the other one by changing the sign of every occurrence
of $\Theta^{2}.$ This exercise (which need not be repeated) yielded an
unequivocal answer: \emph{It is the left-handed orientation which must be
selected.\footnote{The right-handed orientation was selected in \cite{G5}.
This unfortunate choice gave rise to several spurious concepts (like the
transformation $W$) which took much space and time to develop but ultimately
led nowhere. This is why Volume II could not be written. Readers of Volume I
are advised to ignore everything related to the basis matrices $\Omega$ (which
are now replaced by the matrices $\Theta$ to obviate confusion). Readers not
familiar with Volume I are advised to ignore it altogether.}} Thus,%
\begin{equation}%
\begin{tabular}
[c]{ccccccc}%
$\Theta^{0}=%
\begin{pmatrix}
I & 0\\
0 & I
\end{pmatrix}
,$ &  & $\Theta^{1}=%
\begin{pmatrix}
0 & I\\
I & 0
\end{pmatrix}
,$ &  & $\Theta^{2}=%
\begin{pmatrix}
0 & iI\\
-iI & 0
\end{pmatrix}
,$ &  & $\Theta^{3}=%
\begin{pmatrix}
I & 0\\
0 & -I
\end{pmatrix}
$%
\end{tabular}
\ \label{3t8}%
\end{equation}

The bases $\left\{  \Theta_{\mu}\right\}  $ and $\left\{  \Theta^{\mu
}\right\}  $ are therefore related by a mirror reflection in addition to a
similarity transformation (\ref{3t7}). They satisfy the following identities
\begin{equation}
\left.
\begin{array}
[c]{l}%
\Theta_{i}\Theta_{i}=\Theta^{i}\Theta^{i}=I\\
\Theta_{i}\Theta_{j}=i\Theta_{k}\ \text{... cyclically}\\
\Theta^{i}\Theta^{j}=-i~\Theta^{k}\ \text{... cyclically}\\
\left[  \Theta_{i},\Theta^{j}\right]  =0
\end{array}
\right\}  \label{3t8a}%
\end{equation}

The unit matrix $\Theta^{0}=\Theta_{0}=I$ and the six traceless matrices
$\Theta_{i}\in\mathcal{Q}$ and $\Theta^{i}\in\mathcal{P}$ account for seven of
the sixteen basis matrices in $\mathcal{M}.$ The other nine basis matrices are
the products $\Theta_{i}^{j}=\Theta_{i}\Theta^{j}\equiv\Theta^{j}\Theta_{i}.$
Collecting them, we shall use the notations%
\begin{equation}
\Theta_{\mu}^{\nu}=\Theta_{\mu}\Theta^{\nu} \label{3t9}%
\end{equation}

The matrices $\Theta_{\mu}^{\nu}$ are Hermitian for being products of mutually
commuting Hermitian matrices:%
\[%
\begin{tabular}
[c]{|c|cccc|}\hline
$\Theta_{\mu}^{\nu}$ & $I$ $\left(  =\Theta^{0}\right)  $ & $\Theta^{1}$ &
$\Theta^{2}$ & $\Theta^{3}$\\\hline
&  &  &  & \\
$I$ $\left(  =\Theta_{0}\right)  $ & \multicolumn{1}{|l}{$\Theta_{0}^{0}=%
\begin{pmatrix}
I & 0\\
0 & I
\end{pmatrix}
$} & \multicolumn{1}{l}{$\Theta_{0}^{1}=%
\begin{pmatrix}
0 & I\\
I & 0
\end{pmatrix}
$} & \multicolumn{1}{l}{$\Theta_{0}^{2}=%
\begin{pmatrix}
0 & iI\\
-iI & 0
\end{pmatrix}
$} & \multicolumn{1}{l|}{$\Theta_{0}^{3}=%
\begin{pmatrix}
I & 0\\
0 & -I
\end{pmatrix}
$}\\
& \multicolumn{1}{|l}{} & \multicolumn{1}{l}{} & \multicolumn{1}{l}{} &
\multicolumn{1}{l|}{}\\
$\Theta_{1}$ & \multicolumn{1}{|l}{$\Theta_{1}^{0}=%
\begin{pmatrix}
\sigma_{1} & 0\\
0 & \sigma_{1}%
\end{pmatrix}
$} & \multicolumn{1}{l}{$\Theta_{1}^{1}=%
\begin{pmatrix}
0 & \sigma_{1}\\
\sigma_{1} & 0
\end{pmatrix}
$} & \multicolumn{1}{l}{$\Theta_{1}^{2}=%
\begin{pmatrix}
0 & i\sigma_{1}\\
-i\sigma_{1} & 0
\end{pmatrix}
$} & \multicolumn{1}{l|}{$\Theta_{1}^{3}=%
\begin{pmatrix}
\sigma_{1} & 0\\
0 & -\sigma_{1}%
\end{pmatrix}
$}\\
& \multicolumn{1}{|l}{} & \multicolumn{1}{l}{} & \multicolumn{1}{l}{} &
\multicolumn{1}{l|}{}\\
$\Theta_{2}$ & \multicolumn{1}{|l}{$\Theta_{2}^{0}=%
\begin{pmatrix}
\sigma_{2} & 0\\
0 & \sigma_{2}%
\end{pmatrix}
$} & \multicolumn{1}{l}{$\Theta_{2}^{1}=%
\begin{pmatrix}
0 & \sigma_{2}\\
\sigma_{2} & 0
\end{pmatrix}
$} & \multicolumn{1}{l}{$\Theta_{2}^{2}=%
\begin{pmatrix}
0 & i\sigma_{2}\\
-i\sigma_{2} & 0
\end{pmatrix}
$} & \multicolumn{1}{l|}{$\Theta_{2}^{3}=%
\begin{pmatrix}
\sigma_{2} & 0\\
0 & -\sigma_{2}%
\end{pmatrix}
$}\\
& \multicolumn{1}{|l}{} & \multicolumn{1}{l}{} & \multicolumn{1}{l}{} &
\multicolumn{1}{l|}{}\\
$\Theta_{3}$ & \multicolumn{1}{|l}{$\Theta_{3}^{0}=%
\begin{pmatrix}
\sigma_{3} & 0\\
0 & \sigma_{3}%
\end{pmatrix}
$} & \multicolumn{1}{l}{$\Theta_{3}^{1}=%
\begin{pmatrix}
0 & \sigma_{3}\\
\sigma_{3} & 0
\end{pmatrix}
$} & \multicolumn{1}{l}{$\Theta_{3}^{2}=%
\begin{pmatrix}
0 & i\sigma_{3}\\
-i\sigma_{3} & 0
\end{pmatrix}
$} & \multicolumn{1}{l|}{$\Theta_{3}^{3}=%
\begin{pmatrix}
\sigma_{3} & 0\\
0 & -\sigma_{3}%
\end{pmatrix}
$}\\
&  &  &  & \\\hline
\end{tabular}
\]

\begin{center}
Table \ref{S3}.1: The matrices $\Theta_{\mu}^{\nu}$ in block form.
\label{pTable1}
\end{center}

Ten of the matrices $\Theta_{\mu}^{\nu}$ are symmetric (those that contain the
label "2" an even number of times), while six are antisymmetric (those that
contain the label "2" exactly once):%

\[
Antisymmetric\ \cdots\ \left\{
\begin{tabular}
[c]{lll}%
$\Theta_{2}$ & $\Theta_{2}^{1}$ & $\Theta_{2}^{3}$\\
$\Theta^{2}$ & $\Theta_{1}^{2}$ & $\Theta_{3}^{2}$%
\end{tabular}
\right.  \
\]

The products of theta matrices are easily computed from their definitions,
their commutation rules, and their products (\ref{3t8a}). For example:%
\begin{align*}
\Theta_{1}^{3}\Theta_{3}^{2}  &  =\left(  \Theta_{1}\Theta^{3}\right)  \left(
\Theta_{3}\Theta^{2}\right)  =\Theta_{1}\Theta^{3}\Theta_{3}\Theta^{2}%
=\Theta_{1}\Theta_{3}\Theta^{3}\Theta^{2}\\
&  =\left(  \Theta_{1}\Theta_{3}\right)  \left(  \Theta^{3}\Theta^{2}\right)
=\left(  -i\Theta_{2}\right)  \left(  i\Theta^{1}\right)  =\Theta_{2}%
\Theta^{1}=\Theta_{2}^{1}%
\end{align*}

Since the theta matrices have exactly one non-vanishing entry in each row and
in each column, all of unit absolute value, most computations are easily
performed by visual inspection alone once these matrices are displayed in
$4\times4$ format --- which is done in the following table:
\[%
\begin{tabular}
[c]{|c|c|c|c|}\hline
$%
\begin{tabular}
[c]{l}%
\\
$\Theta_{0}^{0}=I$%
\end{tabular}
$ & $%
\begin{tabular}
[c]{l}%
\\
$\Theta_{0}^{1}=\Theta^{1}$%
\end{tabular}
$ & $%
\begin{tabular}
[c]{l}%
\\
$\Theta_{0}^{2}=\Theta^{2}$%
\end{tabular}
$ & $%
\begin{tabular}
[c]{l}%
\\
$\Theta_{0}^{3}=\Theta^{3}$%
\end{tabular}
$\\
&  &  & \\
$%
\begin{pmatrix}
1 & 0 & 0 & 0\\
0 & 1 & 0 & 0\\
0 & 0 & 1 & 0\\
0 & 0 & 0 & 1
\end{pmatrix}
$ & $%
\begin{pmatrix}
0 & 0 & 1 & 0\\
0 & 0 & 0 & 1\\
1 & 0 & 0 & 0\\
0 & 1 & 0 & 0
\end{pmatrix}
$ & $%
\begin{pmatrix}
0 & 0 & i & 0\\
0 & 0 & 0 & i\\
-i & 0 & 0 & 0\\
0 & -i & 0 & 0
\end{pmatrix}
$ & $%
\begin{pmatrix}
1 & 0 & 0 & 0\\
0 & 1 & 0 & 0\\
0 & 0 & -1 & 0\\
0 & 0 & 0 & -1
\end{pmatrix}
$\\\hline
$%
\begin{tabular}
[c]{l}%
\\
$\Theta_{1}^{0}=\Theta_{1}$%
\end{tabular}
$ & $%
\begin{tabular}
[c]{l}%
\\
$\Theta_{1}^{1}$%
\end{tabular}
$ & $%
\begin{tabular}
[c]{l}%
\\
$\Theta_{1}^{2}$%
\end{tabular}
$ & $%
\begin{tabular}
[c]{l}%
\\
$\Theta_{1}^{3}$%
\end{tabular}
$\\
&  &  & \\
$%
\begin{pmatrix}
0 & 1 & 0 & 0\\
1 & 0 & 0 & 0\\
0 & 0 & 0 & 1\\
0 & 0 & 1 & 0
\end{pmatrix}
$ & $%
\begin{pmatrix}
0 & 0 & 0 & 1\\
0 & 0 & 1 & 0\\
0 & 1 & 0 & 0\\
1 & 0 & 0 & 0
\end{pmatrix}
$ & $%
\begin{pmatrix}
0 & 0 & 0 & i\\
0 & 0 & i & 0\\
0 & -i & 0 & 0\\
-i & 0 & 0 & 0
\end{pmatrix}
$ & $%
\begin{pmatrix}
0 & 1 & 0 & 0\\
1 & 0 & 0 & 0\\
0 & 0 & 0 & -1\\
0 & 0 & -1 & 0
\end{pmatrix}
$\\\hline
$%
\begin{tabular}
[c]{l}%
\\
$\Theta_{2}^{0}=\Theta_{2}$%
\end{tabular}
$ & $%
\begin{tabular}
[c]{l}%
\\
$\Theta_{2}^{1}$%
\end{tabular}
$ & $%
\begin{tabular}
[c]{l}%
\\
$\Theta_{2}^{2}$%
\end{tabular}
$ & $%
\begin{tabular}
[c]{l}%
\\
$\Theta_{2}^{3}$%
\end{tabular}
$\\
&  &  & \\
$%
\begin{pmatrix}
0 & -i & 0 & 0\\
i & 0 & 0 & 0\\
0 & 0 & 0 & -i\\
0 & 0 & i & 0
\end{pmatrix}
$ & $%
\begin{pmatrix}
0 & 0 & 0 & -i\\
0 & 0 & i & 0\\
0 & -i & 0 & 0\\
i & 0 & 0 & 0
\end{pmatrix}
$ & $%
\begin{pmatrix}
0 & 0 & 0 & 1\\
0 & 0 & -1 & 0\\
0 & -1 & 0 & 0\\
1 & 0 & 0 & 0
\end{pmatrix}
$ & $%
\begin{pmatrix}
0 & -i & 0 & 0\\
i & 0 & 0 & 0\\
0 & 0 & 0 & i\\
0 & 0 & -i & 0
\end{pmatrix}
$\\\hline
$%
\begin{tabular}
[c]{l}%
\\
$\Theta_{3}^{0}=\Theta_{3}$%
\end{tabular}
$ & $%
\begin{tabular}
[c]{l}%
\\
$\Theta_{3}^{1}$%
\end{tabular}
$ & $%
\begin{tabular}
[c]{l}%
\\
$\Theta_{3}^{2}$%
\end{tabular}
$ & $%
\begin{tabular}
[c]{l}%
\\
$\Theta_{3}^{3}$%
\end{tabular}
$\\
&  &  & \\
$%
\begin{pmatrix}
1 & 0 & 0 & 0\\
0 & -1 & 0 & 0\\
0 & 0 & 1 & 0\\
0 & 0 & 0 & -1
\end{pmatrix}
$ & $%
\begin{pmatrix}
0 & 0 & 1 & 0\\
0 & 0 & 0 & -1\\
1 & 0 & 0 & 0\\
0 & -1 & 0 & 0
\end{pmatrix}
$ & $%
\begin{pmatrix}
0 & 0 & i & 0\\
0 & 0 & 0 & -i\\
-i & 0 & 0 & 0\\
0 & i & 0 & 0
\end{pmatrix}
$ & $%
\begin{pmatrix}
1 & 0 & 0 & 0\\
0 & -1 & 0 & 0\\
0 & 0 & -1 & 0\\
0 & 0 & 0 & 1
\end{pmatrix}
$\\\hline
\end{tabular}
\]

\begin{center}
Table \ref{S3}.2: The $4\times4$ matrices $\Theta_{\mu}^{\nu}.$\label{pTable2}
\end{center}

\subsection{\label{s3.2}Reciprocity}

In the decomposition
\begin{equation}
M=k_{\nu}^{\mu}\Theta_{\mu}^{\nu} \label{3t10}%
\end{equation}
of an arbitrary matrix $M\in\mathcal{M},$ the 16 complex coefficients $k_{\nu
}^{\mu}$ form a new $4\times4$ matrix which we denote by $K,$
\[
K=\left(  k_{\nu}^{\mu}\right)
\]

Since the matrices $\Theta_{\nu}^{\mu}$ are Hermitian, the matrix $M$ is
Hermitian if and only if the matrix $K$ is real.

Relation (\ref{3t10}) establishes a one-to-one correspondence between the
matrices%
\begin{equation}
M=%
\begin{pmatrix}
m_{1}^{1} & m_{1}^{2} & m_{1}^{3} & m_{1}^{4}\\
m_{2}^{1} & m_{2}^{2} & m_{2}^{3} & m_{2}^{4}\\
m_{3}^{1} & m_{3}^{2} & m_{3}^{3} & m_{3}^{4}\\
m_{4}^{1} & m_{4}^{2} & m_{4}^{3} & m_{4}^{4}%
\end{pmatrix}
\in\mathcal{M} \label{4u25}%
\end{equation}
and the 16 coefficients $k_{\nu}^{\mu}.$ The matrix $K$ of these coefficients
belongs to a linear space of matrices, denoted by $\mathcal{K}:$
\begin{equation}
K=%
\begin{pmatrix}
k_{0}^{0} & k_{0}^{1} & k_{0}^{2} & k_{0}^{3}\\
k_{1}^{0} & k_{1}^{1} & k_{1}^{2} & k_{1}^{3}\\
k_{2}^{0} & k_{2}^{1} & k_{2}^{2} & k_{2}^{3}\\
k_{3}^{0} & k_{3}^{1} & k_{3}^{2} & k_{3}^{3}%
\end{pmatrix}
\in\mathcal{K} \label{4u27}%
\end{equation}
While $\mathcal{M}$ is an \emph{algebra of matrices}, $\mathcal{K}$ is only a
\emph{linear space of matrices.}

Substitution of the $\Theta$ matrices given in Table \ref{S3}.2 into relation
(\ref{3t10}) yields the 16 elements $m_{j}^{i}$ of the matrix $M$ as linear
combinations of the 16 elements $k_{\beta}^{\alpha}$ of the matrix $K:$%
\[
M=%
\begin{pmatrix}
k_{0}^{0}+k_{3}^{0}+k_{3}^{3}+k_{0}^{3} & -ik_{3}^{2}+k_{3}^{1}-ik_{0}%
^{2}+k_{0}^{1} & k_{1}^{0}+k_{1}^{3}+ik_{2}^{0}+ik_{2}^{3} & -ik_{1}^{2}%
+k_{1}^{1}+k_{2}^{2}+ik_{2}^{1}\\
ik_{3}^{2}+k_{3}^{1}+ik_{0}^{2}+k_{0}^{1} & k_{0}^{0}+k_{3}^{0}-k_{3}%
^{3}-k_{0}^{3} & ik_{1}^{2}+k_{1}^{1}-k_{2}^{2}+ik_{2}^{1} & k_{1}^{0}%
-k_{1}^{3}+ik_{2}^{0}-ik_{2}^{3}\\
k_{1}^{0}+k_{1}^{3}-ik_{2}^{0}-ik_{2}^{3} & -ik_{1}^{2}+k_{1}^{1}-k_{2}%
^{2}-ik_{2}^{1} & k_{0}^{0}-k_{3}^{0}-k_{3}^{3}+k_{0}^{3} & ik_{3}^{2}%
-k_{3}^{1}-ik_{0}^{2}+k_{0}^{1}\\
ik_{1}^{2}+k_{1}^{1}+k_{2}^{2}-ik_{2}^{1} & k_{1}^{0}-k_{1}^{3}-ik_{2}%
^{0}+ik_{2}^{3} & -ik_{3}^{2}-k_{3}^{1}+ik_{0}^{2}+k_{0}^{1} & k_{0}^{0}%
-k_{3}^{0}+k_{3}^{3}-k_{0}^{3}%
\end{pmatrix}
\allowbreak\allowbreak\allowbreak
\]

Let us define the matrix
\[
T=%
\begin{pmatrix}
1 & 1 & 1 & 1\\
1 & 1 & -1 & -1\\
1 & -1 & 1 & -1\\
1 & -1 & -1 & 1
\end{pmatrix}
\]
whose inverse is%
\[
T^{-1}=\frac{1}{4}T
\]
With its help, the one-to-one mapping%
\begin{equation}
\mathcal{K}\rightleftarrows\mathcal{M} \label{5c0}%
\end{equation}
relating the spaces $\mathcal{M}$ and $\mathcal{K}$ is given by the relations
\[%
\begin{tabular}
[c]{|ll|ll|}\hline
&  &  & \\
$\left(
\begin{array}
[c]{c}%
m_{1}^{1}\\
m_{2}^{2}\\
m_{3}^{3}\\
m_{4}^{4}%
\end{array}
\right)  =T\left(
\begin{array}
[c]{c}%
k_{0}^{0}\\
k_{3}^{0}\\
k_{0}^{3}\\
k_{3}^{3}%
\end{array}
\right)  $ &  &  & $\left(
\begin{array}
[c]{c}%
k_{0}^{0}\\
k_{3}^{0}\\
k_{0}^{3}\\
k_{3}^{3}%
\end{array}
\right)  =\frac{1}{4}T\left(
\begin{array}
[c]{c}%
m_{1}^{1}\\
m_{2}^{2}\\
m_{3}^{3}\\
m_{4}^{4}%
\end{array}
\right)  $\\
&  &  & \\
$\left(
\begin{array}
[c]{c}%
m_{1}^{2}\\
m_{2}^{1}\\
m_{3}^{4}\\
m_{4}^{3}%
\end{array}
\right)  =T\left(
\begin{array}
[c]{c}%
k_{0}^{1}\\
k_{3}^{1}\\
-ik_{0}^{2}\\
-ik_{3}^{2}%
\end{array}
\right)  $ &  &  & $\left(
\begin{array}
[c]{c}%
k_{0}^{1}\\
k_{3}^{1}\\
-ik_{0}^{2}\\
-ik_{3}^{2}%
\end{array}
\right)  =\frac{1}{4}T\left(
\begin{array}
[c]{c}%
m_{1}^{2}\\
m_{2}^{1}\\
m_{3}^{4}\\
m_{4}^{3}%
\end{array}
\right)  $\\
&  &  & \\
$\left(
\begin{array}
[c]{c}%
m_{1}^{4}\\
m_{2}^{3}\\
m_{3}^{2}\\
m_{4}^{1}%
\end{array}
\right)  =T\left(
\begin{array}
[c]{c}%
k_{1}^{1}\\
ik_{2}^{1}\\
-ik_{1}^{2}\\
k_{2}^{2}%
\end{array}
\right)  $ &  &  & $\left(
\begin{array}
[c]{c}%
k_{1}^{1}\\
ik_{2}^{1}\\
-ik_{1}^{2}\\
k_{2}^{2}%
\end{array}
\right)  =\frac{1}{4}T\left(
\begin{array}
[c]{c}%
m_{1}^{4}\\
m_{2}^{3}\\
m_{3}^{2}\\
m_{4}^{1}%
\end{array}
\right)  $\\
&  &  & \\
$\left(
\begin{array}
[c]{c}%
m_{1}^{3}\\
m_{2}^{4}\\
m_{3}^{1}\\
m_{4}^{2}%
\end{array}
\right)  =T\left(
\begin{array}
[c]{c}%
k_{1}^{0}\\
ik_{2}^{0}\\
k_{1}^{3}\\
ik_{2}^{3}%
\end{array}
\right)  $ &  &  & $\left(
\begin{array}
[c]{c}%
k_{1}^{0}\\
ik_{2}^{0}\\
k_{1}^{3}\\
ik_{2}^{3}%
\end{array}
\right)  =\frac{1}{4}T\left(
\begin{array}
[c]{c}%
m_{1}^{3}\\
m_{2}^{4}\\
m_{3}^{1}\\
m_{4}^{2}%
\end{array}
\right)  $\\
&  &  & \\\hline
\end{tabular}
\
\]

\begin{center}
Table \ref{S3}.3: The linear isomorphism $\mathcal{K}\longleftrightarrow
\mathcal{M}.$
\end{center}

While the matrices $M\in\mathcal{M}$ and $K\in\mathcal{K}$ carry the same
information,\ in the sense that no information is lost in passing from one to
the other, they bring to light different properties of the objects they represent.

As an analogy, let us consider the pairing of periodic crystallographic
lattices: the spacial lattice and its reciprocal. They represent,
respectively, the positions of the ions in Euclidean space, and the $\vec{k}$
vectors in the reciprocal space. The former is adapted to the local properties
of crystals, the latter to their global properties.

Similarly, the matrices $M$ are adapted to the investigation of the
\emph{algebraic properties} of quantions, the matrices $K$ are adapted to the
investigation of their \emph{geometric properties.} We may thus say that the
former are related to quantum mechanic, the latter to relativity. This is
reflected in the choice of indices:\ Latin indices, 1 to 4, in $\mathcal{M};$
Greek indices, 0 to 3, in $\mathcal{K}.$ We shall refer to the relationship
$\mathcal{M}\rightleftarrows\mathcal{K}$ displayed in Table \ref{S3}.3 as
\textbf{reciprocity.} Stated suggestively, \emph{reciprocity refers to to the
relationship between quantum and relativistic properties,} as it manifests
itself in the quantionic approach.

\subsection{\label{s3.3}The spinorial structuring of $\mathcal{M}$}

A standard linear basis in $\mathcal{M}$ consists of the multivectors built
over the Dirac gamma matrices. We take the latter to be in the Weyl gauge.
While this choice may seem arbitrary at this point,\ we shall see in Part 2
that it is implied by the quantionic derivation of the Dirac equation.

The complete spinorial basis in $\mathcal{M}$ is displayed in the following
table:%
\begin{equation}%
\begin{tabular}
[c]{|r|ll|}\hline
&  & \\
$\left.
\begin{tabular}
[c]{l}%
$\text{scalar}$\\
\multicolumn{1}{c}{$I$}%
\end{tabular}
\right\}  $ & $I=%
\begin{pmatrix}
\sigma_{0} & 0\\
0 & \sigma_{0}%
\end{pmatrix}
$ & \\
&  & \\
$\left.
\begin{tabular}
[c]{l}%
$\text{vector}$\\
\multicolumn{1}{c}{$\gamma^{\mu}$}%
\end{tabular}
\right\}  $ & $\gamma^{0}=%
\begin{pmatrix}
0 & \sigma_{0}\\
\sigma_{0} & 0
\end{pmatrix}
\ $ & $\gamma^{i}=%
\begin{pmatrix}
0 & \sigma_{i}\\
-\sigma_{i} & 0
\end{pmatrix}
$\\
&  & \\
$\left.
\begin{tabular}
[c]{c}%
bi-$\text{vector}$\\
$\gamma^{\mu\nu}=\gamma^{\mu}\gamma^{\nu}$%
\end{tabular}
\right\}  $ & $\gamma^{0i}=%
\begin{pmatrix}
-\sigma_{i} & 0\\
0 & \sigma_{i}%
\end{pmatrix}
\ $ & $\gamma^{ij}=-i%
\begin{pmatrix}
\sigma_{k} & 0\\
0 & \sigma_{k}%
\end{pmatrix}
$\\
&  & \\
$\left.
\begin{tabular}
[c]{c}%
$\text{pseudo-vector}$\\
\multicolumn{1}{l}{$\gamma^{\mu\nu\rho}=\gamma^{\mu}\gamma^{\nu}\gamma^{\rho}%
$}%
\end{tabular}
\right\}  $ & $\gamma^{123}=i%
\begin{pmatrix}
0 & -\sigma_{0}\\
\sigma_{0} & 0
\end{pmatrix}
$ & $\gamma^{0ij}=-i%
\begin{pmatrix}
0 & \sigma_{k}\\
\sigma_{k} & 0
\end{pmatrix}
$\\
&  & \\
\multicolumn{1}{|c|}{$\left.
\begin{tabular}
[c]{c}%
$\text{pseudo-scalar}$\\
\multicolumn{1}{l}{$\gamma^{5}=i\gamma^{0123}=i\gamma^{0}\gamma^{1}\gamma
^{2}\gamma^{3}$}%
\end{tabular}
\right\}  $} & $\gamma^{5}=%
\begin{pmatrix}
-\sigma_{0} & 0\\
0 & \sigma_{0}%
\end{pmatrix}
$ & \\
&  & \\\hline
\end{tabular}
\label{3u1}%
\end{equation}

\begin{center}
Table \ref{S3}.4: The spinorial basis in block form. \label{pTable3}
\end{center}

Comparison of these matrices with the theta matrices in Table \ref{S3}.1
yields the following correspondences between the quantionic and spinorial
bases:
\begin{equation}%
\begin{tabular}
[c]{|llcrc|llcll|}\hline
&  &  &  &  &  &  &  &  & \\
& $\gamma^{0}$ & $=$ & $\Theta^{1}$ &  &  & $\Theta_{i}$ & $=$ & $i\gamma
^{jk}$ & \\
&  &  &  &  &  &  &  &  & \\
& $\gamma^{i}$ & $=$ & $-i\Theta_{i}^{2}$ &  &  & $\Theta^{1}$ & $=$ &
$\gamma^{0}$ & \\
&  &  &  &  &  &  &  &  & \\
& $\gamma^{5}$ & $=$ & $-\Theta^{3}$ &  &  & $\Theta^{2}$ & $=$ &
$-\gamma^{123}$ & \\
&  &  &  &  &  &  &  &  & \\
& $\gamma^{0i}$ & $=$ & $-\Theta_{i}^{3}$ &  &  & $\Theta^{3}$ & $=$ &
$-\gamma^{5}$ & \\
&  &  &  &  &  &  &  &  & \\
& $\gamma^{ij}$ & $=$ & $-i\Theta_{k}$ &  &  & $\Theta_{i}^{1}$ & $=$ &
$i\gamma^{0jk}$ & \\
&  &  &  &  &  &  &  &  & \\
& $\gamma^{123}$ & $=$ & $-\Theta^{2}$ &  &  & $\Theta_{i}^{2}$ & $=$ &
$i\gamma^{i}$ & \\
&  &  &  &  &  &  &  &  & \\
& $\gamma^{0ij}$ & $=$ & $-i\Theta_{k}^{1}$ &  &  & $\Theta_{i}^{3}$ & $=$ &
$-\gamma^{0i}$ & \\
&  &  &  &  &  &  &  &  & \\\hline
\end{tabular}
\label{3u4}%
\end{equation}

\begin{center}
Table \ref{S3}.5: The qs-equivalence.\label{pTable4}
\end{center}

\subsection{\label{s3.4}The p-q structuring of $\mathcal{M}$}

Owing to the associativity of the matrix product and the commutativity of the
q-quantions with p-quantions, the expression (\ref{3t10}) may be interpreted
in two different ways: as%
\begin{equation}
M=\left(  k_{\nu}^{\mu}\Theta_{\mu}\right)  \Theta^{\nu}=Q_{\nu}\Theta^{\nu
}=Q_{0}\Theta^{0}+Q_{1}\Theta^{1}+Q_{2}\Theta^{2}+Q_{3}\Theta^{3} \label{3t11}%
\end{equation}
or as%
\begin{equation}
M=\left(  k_{\nu}^{\mu}\Theta^{\nu}\right)  \Theta_{\mu}=P^{\mu}\Theta_{\mu
}=P^{0}\Theta_{0}+P^{1}\Theta_{1}+P^{2}\Theta_{2}+P^{3}\Theta_{3} \label{3t12}%
\end{equation}
The coefficients $Q_{\nu}$ are q-quantions and the coefficients $P^{\mu}$ are p-quantions.

We shall refer to these two expressions as the q-decomposition and
p-decomposition of $\mathcal{M}$ respectively. They are illustrated in the
following diagrams, which bring out different properties of quantions than
those illustrated in Figure \ref{S3}.1.

\begin{center}%
\setlength{\unitlength}{0.01in}
\begin{picture}(560,230)
\thicklines\put(50,15){\line(1,0){160}}
\put(50,15){\line(0,1){160}}
\put(210,175){\line(-1,0){160}}
\put(210,175){\line(0,-1){160}}
\put(340,15){\line(1,0){160}}
\put(340,15){\line(0,1){160}}
\put(500,175){\line(-1,0){160}}
\put(500,175){\line(0,-1){160}}
\thinlines\put(90,15){\line(0,1){160}}
\put(130,15){\line(0,1){160}}
\put(170,15){\line(0,1){160}}
\put(340,55){\line(1,0){160}}
\put(340,95){\line(1,0){160}}
\put(340,135){\line(1,0){160}}
\put(70,95){\makebox(0,0){$P^0 I$}}
\put(110,95){\makebox(0,0){$P^1\Theta_1$}}
\put(150,95){\makebox(0,0){$P^2\Theta_2$}}
\put(190,95){\makebox(0,0){$P^3\Theta_3$}}
\put(420,155){\makebox(0,0){$Q_0 I $}}
\put(420,115){\makebox(0,0){$Q_1 \Theta^1$}}
\put(420,75){\makebox(0,0){$Q_2 \Theta^2$}}
\put(420,35){\makebox(0,0){$Q_3 \Theta^3$}}
\put(125,200){\makebox(0,0){\bf{P-structuring}}}
\put(420,200){\makebox(0,0){\bf{Q-structuring}}}
\end{picture}%

\vspace{4pt}%

Figure \ref{S3}.2: The two decomposition of $\mathcal{M}.$
\end{center}

\section{\label{S4}The mathematics of quantions}

The properties of quantions which follow immediately from the definitions
(\ref{2w1}) and (\ref{2w2}) are already listed on pages \pageref{2w1} to
\pageref{2w13}. In the present section we develop their less obvious
properties and many related concepts.

\subsection{\label{s4.1}The vector representation of quantions}

The vector formalism for quaternions is based on a linear mapping of the field
of quaternions onto the two-dimensional representation space $\mathcal{H}:$
\[
\omega:\mathbb{H}\rightarrow\mathcal{H}%
\]
As this mapping consists in truncating the $2\times2$ matrix (\ref{2k20}) to
its first column, (\ref{2k36}), it may be represented by the distinguished
vector
\[
\left\vert \omega\right)  =%
\begin{pmatrix}
1\\
0
\end{pmatrix}
\in\mathcal{H}%
\]
which yields%
\[
\left\vert q\right)  =Q\left\vert \omega\right)  \in\mathcal{H}%
\]
for every $Q\in\mathbb{H}.$ Since the first column of a quaternion uniquely
defines the second, the mapping $\omega$ is one-to-one.%

\vspace{8pt}%

To extend the vector formalism to quantions, we introduce two new objects:%
\[%
\begin{tabular}
[c]{rll}%
$\mathcal{H}$ & $\cdots$ & A four-dimensional inner product space (the
\textbf{representation space})\\
$\left\vert \omega\right)  \in\mathcal{H}$ & $\cdots$ & A distinguished ket
(the \textbf{linking vector})
\end{tabular}
\
\]

The linking vector is \textit{a priori} arbitrary, but it will soon be
restricted to a unique one.

For a given $\left\vert \omega\right)  ,$ two new objects are associated with
every matrix $M\in\mathcal{M}:$

(1)\ A unique ket $\left\vert M\right)  \in$ $\mathcal{H}:$%
\begin{equation}
\left\vert M\right)  \overset{def}{=}M\left\vert \omega\right)  \in\mathcal{H}
\label{4a4}%
\end{equation}
referred to as the \textbf{vector representation} of $M.$ For an arbitrary
matrix $M,$ this is obviously not a faithful representation.

(2) A unique complex number,%
\begin{equation}
\left\langle M\right\rangle \overset{def}{=}\left(  \omega|M|\omega\right)
\in\mathbb{C} \label{4a5}%
\end{equation}
referred to as the \textbf{expectation value} of $M.$

While \textquotedblleft vector representation\textquotedblright\ and
\textquotedblleft linking vector\textquotedblright\ are descriptive terms, the
name \textquotedblleft expectation value\textquotedblright\ is only mnemonic
(by analogy with the formally similar expressions in quantum mechanics, but
without an \textit{a priori} probabilistic meaning).

Taking $M=I$ in relation (\ref{4a4}) yields
\begin{equation}
I\left\vert \omega\right)  =\left\vert I\right)  =\left\vert \omega\right)
\label{4a6a}%
\end{equation}
while by taking $M=AB$ for arbitrary $A,B\in\mathcal{M},$ relation (\ref{4a4})
implies%
\begin{equation}
\left\vert AB\right)  =A\left\vert B\right)  =A\left\vert BI\right)
=AB\left\vert \omega\right)  \label{4a7}%
\end{equation}

The unit matrix being real, $I^{\ast}=I,$ it follows from (\ref{4a6a}) that
the linking vector is real,
\[
\left\vert \omega^{\ast}\right)  =\left\vert \omega\right)
\]
Thus,%
\begin{equation}
\left\vert \omega\right)  =\left(
\begin{array}
[c]{c}%
a\\
b\\
c\\
d
\end{array}
\right)  \in\mathbb{R}^{4} \label{4a8}%
\end{equation}

At this point, the four real numbers $a,b,c,d$ are still arbitrary. We may
therefore define them so as to ensure the structural consistency of the vector
representation of quantions with their other properties. Since these
properties are implicitly encoded in Table \ref{S3}.2, it suffices to consider
the expectation values of the basis matrices,
\begin{equation}
\left\langle \Theta_{\nu}^{\mu}\right\rangle =\left(  \omega|\Theta_{\nu}%
^{\mu}|\omega\right)  \label{4a9}%
\end{equation}

For an arbitrarily selected linking vector $\left\vert \omega\right)  ,$ the
matrix $\left\langle \Theta_{\nu}^{\mu}\right\rangle $ is a new object that
has no intrinsic meaning. Turning this observation around, we may select
$\left\langle \Theta_{\nu}^{\mu}\right\rangle $ as a structurally meaningful
matrix, and then compute $\left\vert \omega\right)  $ from the equations
(\ref{4a9}). Since only one structurally distinguished matrix exists in
$\mathcal{M},$ namely the unit matrix%
\[
I=%
\begin{pmatrix}
1 & 0 & 0 & 0\\
0 & 1 & 0 & 0\\
0 & 0 & 1 & 0\\
0 & 0 & 0 & 1
\end{pmatrix}
=\left(  \delta_{\nu}^{\mu}\right)
\]
the defining equation for $\left\vert \omega\right)  $ imposes itself as
\begin{equation}
\left\langle \Theta_{\nu}^{\mu}\right\rangle =k^{2}\delta_{\nu}^{\mu}
\label{4a10}%
\end{equation}
where $k$ is a real coefficient which may be freely selected. To solve this
matrix equation, we first solve four of its 16 scalar equations. A convenient
choice is
\begin{align*}
\left\langle \Theta_{0}^{0}\right\rangle  &  =k^{2}=a^{2}+b^{2}+c^{2}+d^{2}\\
\left\langle \Theta_{1}^{1}\right\rangle  &  =k^{2}=2\left(  ad+bc\right) \\
\left\langle \Theta_{3}^{3}\right\rangle  &  =k^{2}=a^{2}-b^{2}-c^{2}+d^{2}\\
\left\langle \Theta_{0}^{3}\right\rangle  &  =0=a^{2}+b^{2}-c^{2}-d^{2}%
\end{align*}
The difference $\left\langle \Theta_{0}^{0}\right\rangle -\left\langle
\Theta_{3}^{3}\right\rangle =1-1=0=b^{2}+c^{2}$ yields $b=c=0.$ The solutions of
the two remaining equations are:%
\[
\left\vert \omega\right)  =\pm\frac{\left\vert k\right\vert }{\sqrt{2}}\left(
\begin{array}
[c]{c}%
1\\
0\\
0\\
1
\end{array}
\right)
\]
One easily verifies that they satisfy the 12 other equations (\ref{4a10}).

Since reversing the sign of $\left\vert \omega\right)  $ is equivalent to
acting on $\mathcal{H}$ with the operator $-I,$ and since $\det\left(
-I\right)  =1,$ the negative solution for $\left\vert \omega\right)  $ may be
ignored for not being essentially different.

To select the numerical value of $k,$ let us consider the expectation values
of q-quantions:%
\[
\left(  \omega|Q|\omega\right)  =\frac{k^{2}}{2}%
\begin{pmatrix}
1 & 0 & 0 & 1
\end{pmatrix}%
\begin{pmatrix}
q_{1} & q_{3} & 0 & 0\\
q_{2} & q_{4} & 0 & 0\\
0 & 0 & q_{1} & q_{3}\\
0 & 0 & q_{2} & q_{4}%
\end{pmatrix}
\left(
\begin{array}
[c]{c}%
1\\
0\\
0\\
1
\end{array}
\right)  =\frac{k^{2}}{2}\left(  q_{1}+q_{4}\right)
\]

Three options suggest themselves for being distinguished by some special property:

(1) The case of \ $k=1:\ \ \left\langle Q\Theta^{\mu}\right\rangle $ is the
coefficient of $\Theta_{\mu}$ in the linear expansion of $Q\in\mathcal{Q}:$
\[
Q=\left\langle Q\right\rangle I+\left\langle Q\Theta^{1}\right\rangle
\Theta_{1}+\left\langle Q\Theta^{2}\right\rangle \Theta_{2}+\left\langle
Q\Theta^{3}\right\rangle \Theta_{3}%
\]

(2) The case of \ $k=\sqrt{2}:\ \ \left\langle Q\right\rangle $ is the trace
of the $2\times2$ sub-matrix $N$ in (\ref{2w1}).
\[
\left\langle Q\right\rangle =Tr\left(  N\right)
\]

(3) The case of \ $k=2:\ \ \left\langle Q\right\rangle $ is the trace of the
$4\times4$ matrix $Q:$
\[
\left\langle Q\right\rangle =Tr\left(  Q\right)
\]

The first option is conceptually appealing because (as shown later) the
quantionic Gaussian space of coefficients has a physical interpretation as a
linear Minkowski space. The second option is computationally appealing because
there is no square root of two to carry along in calculations. The third
option has no clear advantage.

Having worked with the first two options, the author came to favor the first.
We therefore select the following form of the linking vector:%
\begin{equation}
\left\vert \omega\right)  =\frac{1}{\sqrt{2}}\left(
\begin{array}
[c]{c}%
1\\
0\\
0\\
1
\end{array}
\right)  \label{4a15}%
\end{equation}
The relation (\ref{4a10}) simplifies to%
\begin{equation}
\left\langle \Theta_{\nu}^{\mu}\right\rangle =\left(  \omega|\Theta_{\nu}%
^{\mu}|\omega\right)  =\delta_{\nu}^{\mu} \label{4a19}%
\end{equation}

\subsection{\label{s4.2}Algebraic duality}

The algebras $\mathcal{Q}$ and $\mathcal{P}$ are globally related as mutual
commutants (every $Q\in\mathcal{Q}$ commutes with every $P\in\mathcal{P}$),
but they are not locally related --- meaning that no point-to-point
relationship $\mathcal{Q}\rightleftarrows\mathcal{P}$ exists \textit{a
priori.} The linking vector (\ref{4a15}) introduces such a relationship.

The linear mapping
\begin{equation}
\omega:\mathcal{M}\longrightarrow\mathcal{H} \label{4a6}%
\end{equation}
defined by (\ref{4a4}) maps the 16-dimensional space $\mathcal{M}$ onto the
4-dimensional space $\mathcal{H}.$ Looking at this mapping from the opposite
direction, let $\left\vert z\right)  \in\mathcal{H}$ be an arbitrary vector,
and let us consider its pre-image, $\omega^{-1}\left(  \left\vert z\right)
\right)  .$ It is a 12-dimensional subset of $\mathcal{M},$ denoted by
$\mathcal{M}_{\left\vert z\right)  }$ in Figure \ref{S4}.1.%
\[
\mathcal{M}_{\left\vert z\right)  }=\omega^{-1}\left(  \left\vert z\right)
\right)  \subset\mathcal{M}%
\]

\begin{center}%
\setlength{\unitlength}{0.01in}
\begin{picture}(500,300)
\put(150,220){\circle*{8}}
\put(350,220){\circle*{8}}
\put(250,20){\circle*{8}}
\put(250,220){\vector(1,0){90}}
\put(250,220){\vector(-1,0){90}}
\put(200,120){\vector(1,-2){45}}
\put(200,120){\vector(-1,2){45}}
\put(300,120){\vector(-1,-2){45}}
\put(300,120){\vector(1,2){45}}
\put(250,30){\vector(0,1){162}}
\put(250,20){\oval(300,40)}
\put(200,20){\makebox(0,0){$\mathcal{H}$}}
\put(190,100){\makebox(0,0){$\omega$}}
\put(310,100){\makebox(0,0){$\omega$}}
\put(230,100){\makebox(0,0){$\Omega_q$}}
\put(270,100){\makebox(0,0){$\Omega_p$}}
\put(135,220){\oval(40,80)[l]}
\put(365,220){\oval(40,80)[r]}
\put(135,180){\line(1,0){230}}
\put(135,260){\line(1,0){230}}
\put(40,230){\oval(40,120)[l]}
\put(480,230){\oval(40,120)[r]}
\put(40,170){\line(1,0){440}}
\put(40,290){\line(1,0){440}}
\put(120,220){\oval(120,50)}
\put(380,220){\oval(120,50)}
\put(250,200){\makebox(0,0){$\mathcal{M}_z$}}
\put(250,278){\makebox(0,0){$\mathcal{M}$}}
\put(100,230){\makebox(0,0){$\mathcal{Q}$}}
\put(400,230){\makebox(0,0){$\mathcal{P}$}}
\put(135,220){\makebox(0,0){$Q$}}
\put(365,220){\makebox(0,0){$P$}}
\put(263,20){\makebox(0,0){$|z)$}}
\put(265,150){\makebox(0,0){$\omega^{-1}$}}
\thicklines\put(195,120){\vector(1,-2){20}}
\put(305,120){\vector(-1,-2){20}}
\put(225,80){\vector(-1,2){20}}
\put(275,80){\vector(1,2){20}}
\thinlines\put(250,230){\makebox(0,0){Algebraic duality}}
\end{picture}%

\vspace{4pt}%

Figure \ref{S4}.1: The definition of algebraic duality.
\end{center}

The intersections of $\mathcal{M}_{\left\vert z\right)  }$ with the
subalgebras $\mathcal{Q}$ and $\mathcal{P}$ are the well-defined single points
$Q$ and $P.$ As this cannot be visualized in the two-dimensional Figure
\ref{S4}.1, $Q$ and $P$ are to be imagined as the only points at which the
algebras $\mathcal{Q}$ and $\mathcal{P}$ intersect the set $\mathcal{M}%
_{\left\vert z\right)  }:$
\begin{align*}
\mathcal{Q}\cap\mathcal{M}_{\left\vert z\right)  }  &  =Q\in\mathcal{Q}\\
\mathcal{P}\cap\mathcal{M}_{\left\vert z\right)  }  &  =P\in\mathcal{P}%
\end{align*}
The uniqueness of $Q$ and $P$ has two consequences:

(1) While the mapping $\omega$ is not invertible in all of $\mathcal{M},$ it
is invertible in each of the subalgebra $\mathcal{Q}$ and $\mathcal{P}$
separately. The inverse mappings are denoted by $\Omega_{q}$ and $\Omega_{p}$
in Figure \ref{S4}.1%
\begin{equation}
\left.
\begin{array}
[c]{c}%
\Omega_{q}:\mathcal{H}\rightarrow\mathcal{Q}\\
\Omega_{p}:\mathcal{H}\rightarrow\mathcal{P}%
\end{array}
\right\}  \label{4a16}%
\end{equation}

(2) The two quantionic algebras are in a one-to-one correspondence implied by
the mapping $\omega.$ We refer to this relationship as \textbf{algebraic
duality. }Thus, two quantions, $Q\in\mathcal{Q}$ and $P\in\mathcal{P},$ are
dual to each other if there exists a ket $\left\vert z\right)  \in\mathcal{H}$
such that%
\begin{equation}
Q\left\vert \omega\right)  =\left\vert z\right)  =P\left\vert \omega\right)
\label{4a17}%
\end{equation}

Given an arbitrary ket
\[
\left\vert z\right)  =\left(
\begin{array}
[c]{c}%
u\\
v\\
w\\
z
\end{array}
\right)  \in\mathcal{H}%
\]
it is easy to see that the unique solutions for the dual quantions $Q$ and $P$
are
\begin{align*}
Q  &  =\sqrt{2}%
\begin{pmatrix}
u & w & 0 & 0\\
v & z & 0 & 0\\
0 & 0 & u & w\\
0 & 0 & v & z
\end{pmatrix}
\in\mathcal{Q}\\
P  &  =\sqrt{2}%
\begin{pmatrix}
u & 0 & v & 0\\
0 & u & 0 & v\\
w & 0 & z & 0\\
0 & w & 0 & z
\end{pmatrix}
\in\mathcal{P}%
\end{align*}
The duality involves a transposition $u\leftrightarrows v,$ most obvious in
block form:%
\[
Q=%
\begin{pmatrix}
N & 0\\
0 & N
\end{pmatrix}
\]%
\[
N=\sqrt{2}%
\begin{pmatrix}
u & w\\
v & z
\end{pmatrix}
,\ \ \ P=\sqrt{2}%
\begin{pmatrix}
uI & vI\\
wI & zI
\end{pmatrix}
\]

The inverse mappings, $\Omega_{q}$ and $\Omega_{p},$ can be represented by
\textbf{inverse linking vectors}, which are Dirac bras:%
\begin{equation}
\left(  \Omega_{q}\right\vert =%
\begin{pmatrix}
\lambda_{1} & \lambda_{2} & \lambda_{3} & \lambda_{4}%
\end{pmatrix}
\label{4a20}%
\end{equation}%
\begin{equation}
\left(  \Omega_{p}\right\vert =%
\begin{pmatrix}
\pi_{1} & \pi_{2} & \pi_{3} & \pi_{4}%
\end{pmatrix}
\label{4a21}%
\end{equation}
\ \ They are defined by the relations%
\begin{equation}
\left(  \Omega_{q}\right\vert Q\left\vert \omega\right)  =Q\ \ \text{\ \ for
every }Q\in\mathcal{Q} \label{4a22}%
\end{equation}%
\begin{equation}
\left(  \Omega_{p}\right\vert P\left\vert \omega\right)  =P\ \ \text{\ \ for
every }P\in\mathcal{P} \label{4a23}%
\end{equation}
The components $\lambda_{1}$ to $\lambda_{4}$ and $\pi_{1}$ to $\pi_{4},$ are
necessarily matrices. They are displayed in the following tables.%

\[%
\begin{tabular}
[c]{|lll|}\hline
&  & \\
$\lambda_{1}=\frac{1}{\sqrt{2}}\left(  I+\Theta_{3}\right)  =\sqrt{2}%
\begin{pmatrix}
1 & 0 & 0 & 0\\
0 & 0 & 0 & 0\\
0 & 0 & 1 & 0\\
0 & 0 & 0 & 0
\end{pmatrix}
$ &  & $\lambda_{2}=\frac{1}{\sqrt{2}}\left(  \Theta_{1}-i\Theta_{2}\right)
=\sqrt{2}%
\begin{pmatrix}
0 & 0 & 0 & 0\\
1 & 0 & 0 & 0\\
0 & 0 & 0 & 0\\
0 & 0 & 1 & 0
\end{pmatrix}
$\\
&  & \\
$\lambda_{3}=\frac{1}{\sqrt{2}}\left(  \Theta_{1}+i\Theta_{2}\right)
=\sqrt{2}%
\begin{pmatrix}
0 & 1 & 0 & 0\\
0 & 0 & 0 & 0\\
0 & 0 & 0 & 1\\
0 & 0 & 0 & 0
\end{pmatrix}
$ &  & $\lambda_{4}=\frac{1}{\sqrt{2}}\left(  I-\Theta_{3}\right)  =\sqrt{2}%
\begin{pmatrix}
0 & 0 & 0 & 0\\
0 & 1 & 0 & 0\\
0 & 0 & 0 & 0\\
0 & 0 & 0 & 1
\end{pmatrix}
$\\
&  & \\\hline
\end{tabular}
\]

\begin{center}
Table \ref{S4}.1: The elements of $\left(  \Omega_{q}\right\vert .$
\end{center}

\[%
\begin{tabular}
[c]{|lll|}\hline
&  & \\
$\pi_{1}=\frac{1}{\sqrt{2}}\left(  I+\Theta^{3}\right)  =\sqrt{2}%
\begin{pmatrix}
1 & 0 & 0 & 0\\
0 & 1 & 0 & 0\\
0 & 0 & 0 & 0\\
0 & 0 & 0 & 0
\end{pmatrix}
$ &  & $\pi_{2}=\frac{1}{\sqrt{2}}\left(  \Theta^{1}-i\Theta^{2}\right)
=\sqrt{2}%
\begin{pmatrix}
0 & 0 & 1 & 0\\
0 & 0 & 0 & 1\\
0 & 0 & 0 & 0\\
0 & 0 & 0 & 0
\end{pmatrix}
$\\
&  & \\
$\pi_{3}=\frac{1}{\sqrt{2}}\left(  \Theta^{1}+i\Theta^{2}\right)  =\sqrt{2}%
\begin{pmatrix}
0 & 0 & 0 & 0\\
0 & 0 & 0 & 0\\
1 & 0 & 0 & 0\\
0 & 1 & 0 & 0
\end{pmatrix}
$ &  & $\pi_{4}=\frac{1}{\sqrt{2}}\left(  I-\Theta^{3}\right)  =\sqrt{2}%
\begin{pmatrix}
0 & 0 & 0 & 0\\
0 & 0 & 0 & 0\\
0 & 0 & 1 & 0\\
0 & 0 & 0 & 1
\end{pmatrix}
$\\
&  & \\\hline
\end{tabular}
\]

\begin{center}
Table \ref{S4}.2: The elements of $\left(  \Omega_{p}\right\vert .$
\end{center}

\subsection{\label{s4.3}The metric in the matrix formalism}

The metric norm of $Q,$ as given by (\ref{2w9}), is the determinant of the
matrix $N,$%
\begin{equation}
M\left(  Q\right)  =\det\left(  N\right)  \in\mathbb{C} \label{4b1}%
\end{equation}

The following relation follows from the fact that the determinant of a product
of matrices is the product of the determinants of the factors:%
\begin{equation}
M\left(  QR\right)  =M\left(  Q\right)  M\left(  R\right)  \label{4b2}%
\end{equation}

Other properties of determinants yield the following identities:%
\begin{align}
M\left(  Q^{\dag}\right)   &  =\left[  M\left(  Q\right)  \right]  ^{\ast
}\label{4b3}\\
M\left(  Q^{\#}\right)   &  =M\left(  Q\right)  \label{4b4}%
\end{align}

Since $\det\left(  N\right)  $ may vanish even if $N\neq0,$ we have to make a
distinction between \textbf{regular quantions,} for which $M\left(  Q\right)
\neq0,$ and \textbf{singular quantions,} for which $M\left(  Q\right)  =0$
while\textbf{\ }$Q\neq0.$

It follows from the relation $Q^{-1}=Q^{\#}/M\left(  Q\right)  $ that the
regular quantions are the only ones which have an inverse. It is therefore
owing to the existence of singular quantions that the algebras $\mathcal{Q}$
and $\mathcal{P}$ fail to be division algebras.

If the matrix $N$ is Hermitian, $N^{\dag}=N,$ then $\det\left(  N\right)  $ is
real. Hence, \emph{the metric norm of a Hermitian quantion is a real number.}

We get an alternative expression for the metric norm from relation
(\ref{2w11}) by taking the expectation value of both sides:%
\begin{equation}
M\left(  Q\right)  =\left\langle Q^{\#}Q\right\rangle \label{4b5}%
\end{equation}

The metric norm $M\left(  Q\right)  $ gives rise to the scalar product
$\left(  Q,R\right)  $ of two arbitrary quantions by the polarization
procedure, which amounts to computing the norm of the sum $Q+R,$%
\begin{align*}
M\left(  Q+R\right)  \ I  &  =\left(  Q+R\right)  ^{\#}\left(  Q+R\right) \\
&  =Q^{\#}Q+R^{\#}R+Q^{\#}R+R^{\#}Q
\end{align*}
and collecting the three metric norms into a single complex number on the
right-hand side,%
\[
\left(  Q^{\#}R+R^{\#}Q\right)  =\left[  M\left(  Q+R\right)  -M\left(
Q\right)  -M\left(  R\right)  \right]  \ I
\]
The left-hand side is therefore proportional to the unit matrix. Let us denote
the complex factor of proportionality by $2\left(  Q,R\right)  .$ Owing to the
anti-isomorphism (\ref{2w12}), one obtains%
\[
\left(  Q,R\right)  \ I=\frac{1}{2}\left(  Q^{\#}R+R^{\#}Q\right)  =\frac
{1}{2}\left[  Q^{\#}R+\left(  Q^{\#}R\right)  ^{\#}\right]
\]
Taking the expectation value of both sides eliminates the unit matrix, while
relation (\ref{4b5}) reduces the right-hand side to a single term. We may
therefore define the \textbf{scalar product of quantions} as the complex
symmetric function%
\begin{equation}
\left(  Q,R\right)  =\left\langle Q^{\#}R\right\rangle \in\mathbb{C}
\label{4b6}%
\end{equation}
As it should be, the scalar product of a quantion with itself is its metric
norm%
\begin{equation}
\left(  Q,Q\right)  =\left\langle Q^{\#}Q\right\rangle =M\left(  Q\right)
\in\mathbb{C} \label{4b7}%
\end{equation}

The expressions \textquotedblleft scalar product\textquotedblright\ and
\textquotedblleft inner product\textquotedblright\ are usually
interchangeable, but since we need them in two different quantionic
structures, we shall use the following conventions:

The \textbf{inner product} $\left(  p|q\right)  =p_{1}^{\ast}q_{1}+p_{2}%
^{\ast}q_{2}+p_{3}^{\ast}q_{3}+p_{4}^{\ast}q_{4}$ is defined in the
representation space $\mathcal{H}.$

The \textbf{scalar product} $\left(  Q,R\right)  =\left\langle Q^{\#}%
R\right\rangle $ is defined in the quantionic algebras $\mathcal{Q}$ and
$\mathcal{P}.$

\subsection{\label{s4.4}The metric in the reciprocal space}

Restricting the general decomposition (\ref{3t10}) to the algebras
$\mathcal{Q}$ and $\mathcal{P},$%
\begin{align}
X  &  =x^{0}I+x^{1}\Theta_{1}+x^{2}\Theta_{2}+x^{3}\Theta_{3}=x^{\mu}%
\Theta_{\mu}\in\mathcal{Q}\label{4c1}\\
P  &  =p_{0}I+p_{1}\Theta^{1}+p_{2}\Theta^{2}+p_{3}\Theta^{3}=p_{\nu}%
\Theta^{\nu}\in\mathcal{P} \label{4c2}%
\end{align}
we shall now consider the spaces of coefficients $x^{\alpha}$ and $p_{\alpha
}.$ They are analogous to the real Gaussian plane of coefficients $x,y$ in
$z=x+yi\in\mathbb{C},$ and to the real quaternionic Gaussian space of
coefficients $w,x,y,z$ in the decomposition (\ref{2kb}) if quaternions. We
shall therefore denote the complex four-dimensional spaces of coefficients
$x^{\alpha}$ and $p_{\alpha}$ by $M_{q}^{4}$ and $M_{p}^{4}$ respectively, and
refer to them as \textbf{quantionic Gaussian spaces.} These spaces are
obviously contained in the space $\mathcal{K}$ of reciprocal matrices
(\ref{4u27}): $M_{q}^{4}$ is the first row, $M_{p}^{4}$ the first column of
these matrices.

Since the basis matrices are Hermitian, the Hermitian conjugates are obtained
by taking the complex conjugates of the coefficients:%
\begin{align}
X^{\dag}  &  =\left(  x^{\mu}\right)  ^{\ast}\ \Theta_{\mu}\in\mathcal{Q}%
\label{4c3}\\
P^{\dag}  &  =\left(  p_{\nu}\right)  ^{\ast}\ \Theta^{\nu}\in\mathcal{P}
\label{4c4}%
\end{align}

We get the matrix representations of $X$ and $P$ from Table \ref{S3}.2 on page
\pageref{pTable2}:%
\begin{equation}
X=%
\begin{pmatrix}
x^{0}+x^{3} & x^{1}-ix^{2} & 0 & 0\\
x^{1}+ix^{2} & x^{0}-x^{3} & 0 & 0\\
0 & 0 & x^{0}+x^{3} & x^{1}-ix^{2}\\
0 & 0 & x^{1}+ix^{2} & x^{0}-x^{3}%
\end{pmatrix}
\label{4c5}%
\end{equation}%
\begin{equation}
P=%
\begin{pmatrix}
p_{0}+p_{3} & 0 & p_{1}+ip_{2} & 0\\
0 & p_{0}+p_{3} & 0 & p_{1}+ip_{2}\\
p_{1}-ip_{2} & 0 & p_{0}-p_{3} & 0\\
0 & p_{1}-ip_{2} & 0 & p_{0}-p_{3}%
\end{pmatrix}
\label{4c6}%
\end{equation}

The expressions for the metric duals follow from relations (\ref{2w7}) and
(\ref{2w8}):%
\begin{equation}
X^{\#}=%
\begin{pmatrix}
x^{0}-x^{3} & -x^{1}+ix^{2} & 0 & 0\\
-x^{1}-ix^{2} & x^{0}+x^{3} & 0 & 0\\
0 & 0 & x^{0}-x^{3} & -x^{1}+ix^{2}\\
0 & 0 & -x^{1}-ix^{2} & x^{0}+x^{3}%
\end{pmatrix}
\label{4c7}%
\end{equation}%
\begin{equation}
P^{\#}=%
\begin{pmatrix}
p_{0}-p_{3} & 0 & -p_{1}-ip_{2} & 0\\
0 & p_{0}-p_{3} & 0 & -p_{1}-ip_{2}\\
-p_{1}+ip_{2} & 0 & p_{0}+p_{3} & 0\\
0 & -p_{1}+ip_{2} & 0 & p_{0}+p_{3}%
\end{pmatrix}
\label{4c8}%
\end{equation}

Computing the metric duals of the basis matrices by the rules (\ref{2w7}) and
(\ref{2w8}), we get%
\begin{align*}
I^{\#}  &  =I\\
\left(  \Theta_{i}\right)  ^{\#}  &  =-\Theta_{i}\\
\left(  \Theta^{i}\right)  ^{\#}  &  =-\Theta^{i}%
\end{align*}
The expressions (\ref{4c7}) and (\ref{4c8}) may therefore be written in the
compact form%
\begin{align*}
X^{\#}  &  =x^{\mu}\left(  \Theta_{\mu}\right)  ^{\#}\\
P^{\#}  &  =p_{\nu}\left(  \Theta^{\nu}\right)  ^{\#}%
\end{align*}

Comparison of these expressions with the expressions (\ref{4c3}) and (\ref{4c4}) yields the
following self-evident rules:%
\begin{equation}
\left.
\begin{tabular}
[c]{lllll}%
\emph{Hermitian conjugation} affects the coefficients: &  & $Q=z^{\mu\ast
}\Theta_{\mu}$ &  & $P=p_{\nu}^{\ast}\Theta^{\nu}$\\
\emph{Algebraic duality} affects the basis matrices: &  & $Q^{\#}=z^{\mu
}\Theta_{\mu}^{\#}$ &  & $P^{\#}=p_{\nu}\Theta^{\nu\#}$\\
\emph{Complex conjugation} affects both: &  & $Q^{\ast}=z^{\mu\ast}\Theta
_{\mu}^{\ast}$ &  & $P^{\ast}=p_{\nu}^{\ast}\Theta^{\nu\ast}$%
\end{tabular}
\right\}  \label{4c15}%
\end{equation}

The duality operator \textquotedblleft$\#$\textquotedblright\ in
$\mathcal{Q}_{h}$ (or in $\mathcal{P}_{h}$) is therefore equivalent to the
parity operator $\mathbf{P}$ in $M_{q}^{4}$ (or in $M_{p}^{4}$):%
\begin{equation}
\left.
\begin{array}
[c]{c}%
x^{\mu}\Theta_{\mu}^{\#}\equiv\left(  \mathbf{P}x^{\mu}\right)  \Theta_{\mu}\\
p_{\nu}\Theta^{\nu\#}\equiv\left(  \mathbf{P}p_{\nu}\right)  \Theta^{\nu}%
\end{array}
\right\}  \label{4c16}%
\end{equation}

It follows from these results that both $M_{q}^{4}$ and $M_{p}^{4}$ \emph{are
linear Minkowski spaces}  $M_{0}^{4}$, and therefore%
\begin{equation}%
\begin{tabular}
[c]{|ll|l|}\hline
$M\left(  Q\right)  =\eta_{\mu\nu}z^{\mu}z^{\nu}$ &  & $M\left(  P\right)
=\eta^{\mu\nu}p_{\mu}p_{\nu}$\\
$\Theta_{k}^{\#}=-\Theta_{k}$ &  & $\Theta^{k\#}=-\Theta^{k}$\\
$Q^{\#}=p_{\mu}\Theta_{\mu}^{\#}$ &  & $P^{\#}=p_{\mu}\Theta^{\mu\#}$\\
$Q^{\dag}=z^{\mu\ast}\Theta_{\mu}$ &  & $P^{\dag}=p_{\mu}^{\ast}\Theta^{\mu}%
$\\\hline
\end{tabular}
\label{4c17}%
\end{equation}
where $\eta^{\mu\nu}$ is the Minkowskian metric tensor in the $\left(
1,3\right)  $-form:%
\begin{equation}
\left(  \eta^{\mu\nu}\right)  =\left(  \eta_{\mu\nu}\right)  =%
\begin{pmatrix}
1 & 0 & 0 & 0\\
0 & -1 & 0 & 0\\
0 & 0 & -1 & 0\\
0 & 0 & 0 & -1
\end{pmatrix}
\label{4c18}%
\end{equation}

The first equations in (\ref{4c17}) yield an important conclusion: \emph{The
quantionic metric norm and the Minkowskian metric norm are equivalent.} It
follows that \emph{the singular quantions are in one-to-one correspondence
with the null vectors.}

Since complex conjugation will appear frequently in the quantionic field
equations, it is worth not forgetting that it changes the signs of the
matrices $\Theta_{2}$ and $\Theta^{2}:$%
\begin{equation}
\left.
\begin{array}
[c]{l}%
Q^{\ast}=z^{0\ast}I+z^{1\ast}\Theta_{1}-z^{2\ast}\Theta_{2}+z^{3\ast}%
\Theta_{3}\\
P^{\ast}=p_{0}^{\ast}I+p_{1}^{\ast}\Theta^{1}-p_{2}^{\ast}\Theta^{2}%
+p_{3}^{\ast}\Theta^{3}%
\end{array}
\right\}  \label{4c20}%
\end{equation}
We may therefore expect these basis matrices to play distinguished roles in
quantionic physics --- though this is still a hypothetical concept.

Finally, let us point out that the nondegeneracy of quantions, $Q^{\dag}\neq
Q^{\#},$ manifests itself most clearly in the quantionic basis:%
\[
Q^{\dag}=z^{\mu\ast}\Theta_{\mu}\neq z^{\mu}\Theta_{\mu}^{\#}=Q^{\#}%
\]

\subsection{\label{s4.5}The quantionic derivation operator}

Let us consider a p-quantion given by relation (\ref{4c2}), that is%
\[
P=\Theta^{\nu}p_{\nu}\in\mathcal{P}%
\]
and let the covariant vector $p_{\nu}$ be the gradient of some smooth scalar
function $p\left(  x\right)  $ in a Riemannian space. Then, by associativity,%
\[
P\left(  x\right)  =\Theta^{\nu}\left(  \partial_{\nu}p\left(  x\right)
\right)  =\left(  \Theta^{\nu}\partial_{\nu}\right)  p\left(  x\right)
\]
This suggests that a differential operator $\mathcal{D}$  should be defined as%
\begin{equation}
\mathcal{D}=\Theta^{\nu}\partial_{\nu} \label{4d1}%
\end{equation}
By Table \ref{S3}.1, it is of the form (\ref{2v20}),
\begin{align}
\mathcal{D}  &  =%
\begin{pmatrix}
I & 0\\
0 & I
\end{pmatrix}
\partial_{0}+%
\begin{pmatrix}
0 & I\\
I & 0
\end{pmatrix}
\partial_{1}+%
\begin{pmatrix}
0 & iI\\
-iI & 0
\end{pmatrix}
\partial_{2}+%
\begin{pmatrix}
I & 0\\
0 & -I
\end{pmatrix}
\partial_{3}\nonumber\\
&  =%
\begin{pmatrix}
\left(  \partial_{0}+\partial_{3}\right)  I & \left(  \partial_{1}%
+i\partial_{2}\right)  I\\
\left(  \partial_{1}-i\partial_{2}\right)  I & \left(  \partial_{0}%
-\partial_{3}\right)  I
\end{pmatrix}
=%
\begin{pmatrix}
ID & I\delta\\
I\delta^{\ast} & I\Delta
\end{pmatrix}
\label{2d1a}%
\end{align}
where $D,$ $\Delta,$ and $\delta$ are the standard Newman-Penrose symbols
introduced in \cite{NP}.

The p-quantionic field $P\left(  x\right)  $ now assumes the simple form%
\[
P\left(  x\right)  =\mathcal{D}p\left(  x\right)
\]

The operator $\mathcal{D}$ will be referred to as the \textbf{quantionic
derivation operator.}

For its metric norm, computed as the determinant of the sub-matrix%
\begin{equation}
\mathcal{D}_{0}\overset{def}{=}%
\begin{pmatrix}
\left(  \partial_{0}+\partial_{3}\right)  & \left(  \partial_{1}+i\partial
_{2}\right) \\
\left(  \partial_{1}-i\partial_{2}\right)  & \left(  \partial_{0}-\partial
_{3}\right)
\end{pmatrix}
=%
\begin{pmatrix}
D & \delta\\
\delta^{\ast} & \Delta
\end{pmatrix}
\label{4d.2}%
\end{equation}
we get%
\begin{equation}
M\left(  \mathcal{D}\right)  =\det\left(  \mathcal{D}_{0}\right)  =\det%
\begin{pmatrix}
\left(  \partial_{0}+\partial_{3}\right)  & \left(  \partial_{1}+i\partial
_{2}\right) \\
\left(  \partial_{1}-i\partial_{2}\right)  & \left(  \partial_{0}-\partial
_{3}\right)
\end{pmatrix}
=\partial_{0}^{2}-\partial_{1}^{2}-\partial_{2}^{2}-\partial_{3}^{2}%
\end{equation}
The metric norm of the derivation operator $\mathcal{D}$ is therefore the
D'Alambertian,%
\begin{equation}
M\left(  \mathcal{D}\right)  =\eta^{\mu\nu}\partial_{\mu}\partial_{\nu
}=\square\label{4d3}%
\end{equation}

It is expected by many physicists --- as clearly stated, for example, by
Vlatko Vedral in a recent Scientific American article, \cite{VV} --- that
space and time should both somehow emerge from a spaceless and timeless
physics. In the quantionic approach, this desideratum is satisfied better than
expected: Space and time do not emerge from fundamentally spaceless and
timeless physics, but from a number system which is distinguished as the
structurally nearest generalization of the complex numbers outside the family
of division algebras. Physical arguments play no role in its derivation. The
author regards this result as a probably necessary but far from sufficient condition
for the reconciliation of quantum physics with general relativity.

The following observations suggest the approach to be pursued.

(1) A real linear Minkowski space, $M_{0}^{4},$ supports tensors and
differential operators.

(2) A real locally Minkowskian Riemannian space, $\mathcal{R}^{4},$ supports
general relativity.

(3) A real affine Minkowski space, $M^{4},$ supports quantum field theory.

We are to obtain these spaces without introducing new postulates.

(1) The real linear space $M_{0}^{4}$ has just been shown to be of quantionic origin.

(2) The locally Minkowskian Riemannian space $\mathcal{R}^{4}$ is a Riemannian
manifold whose cotangent space at every point is a linear Minkowski space
$M_{0}^{4}.$ We may therefore define $\mathcal{R}^{4}$ in a very natural way
as the base space of a fiber bundle in which $M_{q}^{4}$ is the fiber.

(3) The real affine space $M^{4}$ is to be regarded as a special case (zero
curvature) of the Riemannian space $\mathcal{R}^{4},$ not as the affine
generalization of the linear space $M_{0}^{4}.$ This avoids introducing the
translation group, whose disadvantage is that it locks-in the flatness of
spacetime as a structural property which axiomatically precludes gravitation.

Let us conclude with a comparative summary of some insights obtained so far.
As brought out in the following table, the relativistic structure of spacetime
(complex in general, real for Hermitian quantions) is nothing more than the
metric structure of the quantionic Gaussian space.
\[%
\begin{tabular}
[c]{|c|c|c|c|}\hline
Number systems & Associated Gaussian spaces & Metric & Hurwitz\\\hline\hline
$\mathbb{R}$ & $\mathbb{R}^{1}$ & Pythagorean & yes\\
$\mathbb{C}$ & $\mathbb{R}^{2}$ & Pythagorean & yes\\
$\mathbb{H}$ & $\mathbb{R}^{4}$ & Pythagorean & yes\\\hline
Duality:$\ \left\{
\begin{tabular}
[c]{l}%
$\mathcal{Q}$\\
$\mathcal{P}$%
\end{tabular}
\ \right.  $ & $\left.
\begin{tabular}
[c]{l}%
$M_{q}^{4}$\\
$M_{p}^{4}$%
\end{tabular}
\ \right\}  =M_{0}^{4}\ \ \left(  \mathbb{C}^{4}\ \text{or\ }\mathbb{R}%
^{4}\right)  $ & Minkowskian & no\\\hline
\end{tabular}
\
\]

Since the algebra of quantions has not been adjusted to yield the Minkowski
metric (which would even be impossible owing to the fact that mathematical
structures, especially the rich ones, cannot be invented at will to suit our
wishes) but came to light as the unique Leibnizian generalization of the field
of complex numbers, this algebra is more than a \textquotedblleft unifying
number system\textquotedblright. It is a structurally relativistic number
system which guarantees the relativistic nature of a quantum theory built over
it --- though it remains to be shown that such a theory can be built and that it agrees with experiments.

It is therefore clear that a structural unification of relativity and quantum
mechanics cannot be built over the division algebras. These higher algebras
(quaternions and octonions) may well be useful in some areas of physics, as
pointed out in \cite{GC}, but expecting them to be of fundamental physical
importance, as argued by Dixon \cite{Dixon}, does not seem to be justified.

\subsection{\label{s4.6}The basis tetrads in $\mathcal{P}$ and $\mathcal{Q}$}

Let us regard the basis matrices $\Theta_{\mu}$ and
$\Theta^{\mu}$ as basis tetrad vectors in $M_{q}^{4}$ and $M_{p}^{4}$
respectively, and derive their corresponding orthogonality relations with
respect to the scalar product defined by (\ref{4b6}), that is%
\begin{equation}
\left(  X,Y\right)  =\left\langle X^{\#}Y\right\rangle \label{4f1}%
\end{equation}
where $X$ and $Y$ are either in $\mathcal{Q}$ or in $\mathcal{P},$ but it
suffices to work in the former.

Substitution of all pairs of tetrad vectors in turn into (\ref{4f1}) yields%
\begin{align*}
\left(  I,I\right)   &  =\left\langle I^{\#}I\right\rangle =\left\langle
I\right\rangle =1\\
\left(  \Theta_{i},\Theta_{i}\right)   &  =\left\langle \Theta_{i}^{\#}%
\Theta_{i}\right\rangle =-\left\langle \Theta_{i}\Theta_{i}\right\rangle
=-\left\langle I\right\rangle =-1\\
\left(  I,\Theta_{i}\right)   &  =\left\langle I^{\#}\Theta_{i}\right\rangle
=\left\langle \Theta_{i}\right\rangle =0\\
\left(  \Theta_{i},\Theta_{j}\right)   &  =\left\langle \Theta_{i}^{\#}%
\Theta_{j}\right\rangle =-\left\langle \Theta_{i}\Theta_{j}\right\rangle
=-i\left\langle \Theta_{k}\right\rangle =0
\end{align*}
We see that $I$ is a timelike unit vector, that the $\Theta_{i}$ are spacelike
unit vectors, and that all vectors are mutually orthogonal. Thus, in compact
form,%
\begin{equation}
\left(  \Theta_{\mu},\Theta_{\nu}\right)  =\eta_{\mu\nu} \label{4f2}%
\end{equation}
For $X=x^{\mu}\Theta_{\mu}$ and $Y=y^{\mu}\Theta_{\mu},$ one therefore obtains
the Minkowskian scalar product:%
\[
\left(  X,Y\right)  =\left\langle x^{\mu}\Theta_{\mu}^{\#}y^{\nu}\Theta_{\nu
}\right\rangle =x^{\mu}y^{\nu}\left\langle \Theta_{\mu}^{\#}\Theta_{\nu
}\right\rangle =\left(  \Theta_{\mu},\Theta_{\nu}\right)  x^{\mu}y^{\nu}%
=\eta_{\mu\nu}x^{\mu}y^{\nu}%
\]
The same conclusions hold for the tetrad vectors $\Theta^{\mu}$ in $M_{p}%
^{4}.$

Turning to the tables \ref{S4}.1 and \ref{S4}.2 in subsection \ref{s4.2}, we
note that the two sets of matrices $\lambda_{1}$ to $\lambda_{4}$ and $\pi
_{1}$ to $\pi_{4}$ may also be regarded as basis tetrads in a linear Minkowski
space. Let us compute their orthogonality relations.

Writing the definitions of the vectors $\lambda_{1}$ to $\lambda_{4}$ in the
first columns, their duals are listed in the second columns:%
\begin{equation}
\left.
\begin{tabular}
[c]{lll}%
$\lambda_{1}=\frac{1}{\sqrt{2}}\left(  I+\Theta_{3}\right)  $ &  &
$\lambda_{1}^{\#}=\frac{1}{\sqrt{2}}\left(  I-\Theta_{3}\right)  $\\
$\lambda_{2}=\frac{1}{\sqrt{2}}\left(  \Theta_{1}-i\Theta_{2}\right)  $ &  &
$\lambda_{2}^{\#}=\frac{1}{\sqrt{2}}\left(  -\Theta_{1}+i\Theta_{2}\right)
$\\
$\lambda_{3}=\frac{1}{\sqrt{2}}\left(  \Theta_{1}+i\Theta_{2}\right)  $ &  &
$\lambda_{3}^{\#}=\frac{1}{\sqrt{2}}\left(  -\Theta_{1}-i\Theta_{2}\right)
$\\
$\lambda_{4}=\frac{1}{\sqrt{2}}\left(  I-\Theta_{3}\right)  $ &  &
$\lambda_{4}^{\#}=\frac{1}{\sqrt{2}}\left(  I+\Theta_{3}\right)  $%
\end{tabular}
\ \right\}  \label{4f4}%
\end{equation}
The ten products $\lambda_{r}^{\#}\lambda_{s}$ for $r\leq s$ are therefore:%
\[%
\begin{tabular}
[c]{ll}%
$\lambda_{1}^{\#}\lambda_{1}=\frac{1}{2}\left(  I-I\right)  =0$ & $\lambda
_{1}^{\#}\lambda_{3}=\frac{1}{2}\left(  \Theta_{1}+i\Theta_{2}-i\Theta
_{2}-\Theta_{1}\right)  =0$\\
$\lambda_{2}^{\#}\lambda_{2}=\frac{1}{2}\left(  -I+I\right)  =0$ &
$\lambda_{2}^{\#}\lambda_{4}=\frac{1}{2}\left(  -\Theta_{1}+i\Theta_{2}%
+\Theta_{1}-i\Theta_{2}\right)  =0$\\
$\lambda_{3}^{\#}\lambda_{3}=\frac{1}{2}\left(  -I+I\right)  =0$ &
$\lambda_{1}^{\#}\lambda_{2}=\frac{1}{2}\left(  \Theta_{1}-i\Theta_{2}%
-i\Theta_{2}+\Theta_{1}\right)  =\Theta_{1}-i\Theta_{2}$\\
$\lambda_{4}^{\#}\lambda_{4}=\frac{1}{2}\left(  I-I\right)  =0$ & $\lambda
_{1}^{\#}\lambda_{4}=\frac{1}{2}\left(  I-\Theta_{3}-\Theta_{3}+I\right)
=I-\Theta_{3}$\\
& $\lambda_{2}^{\#}\lambda_{3}=\frac{1}{2}\left(  -I+\Theta_{3}+\Theta
_{3}-I\right)  =-I+\Theta_{3}$\\
& $\lambda_{3}^{\#}\lambda_{4}=\frac{1}{2}\left(  -\Theta_{1}+i\Theta
_{2}+\Theta_{1}-i\Theta_{2}\right)  =-\Theta_{1}-i\Theta_{2}$%
\end{tabular}
\
\]
Substitution of these expressions into (\ref{4f1}) yields%
\begin{equation}
\left.
\begin{array}
[c]{l}%
\left(  \lambda_{1},\lambda_{1}\right)  =\left(  \lambda_{2},\lambda
_{2}\right)  =\left(  \lambda_{3},\lambda_{3}\right)  =\left(  \lambda
_{4},\lambda_{4}\right)  =0\ \ \ \ \ \\
\left(  \lambda_{1},\lambda_{2}\right)  =\left(  \lambda_{1},\lambda
_{3}\right)  =\left(  \lambda_{2},\lambda_{4}\right)  =\left(  \lambda
_{3},\lambda_{4}\right)  =0\\
\left(  \lambda_{1},\lambda_{4}\right)  =1\\
\left(  \lambda_{2},\lambda_{3}\right)  =-1
\end{array}
\right\}  \label{4f5}%
\end{equation}

For the $\pi$ matrices, the corresponding relations are%
\begin{equation}
\left.
\begin{tabular}
[c]{lll}%
$\pi_{1}=\frac{1}{\sqrt{2}}\left(  I+\Theta^{3}\right)  $ &  & $\pi_{1}%
^{\#}=\frac{1}{\sqrt{2}}\left(  I-\Theta^{3}\right)  $\\
$\pi_{2}=\frac{1}{\sqrt{2}}\left(  \Theta^{1}-i\Theta^{2}\right)  $ &  &
$\pi_{2}^{\#}=\frac{1}{\sqrt{2}}\left(  -\Theta^{1}+i\Theta^{2}\right)  $\\
$\pi_{3}=\frac{1}{\sqrt{2}}\left(  \Theta^{1}+i\Theta^{2}\right)  $ &  &
$\pi_{3}^{\#}=\frac{1}{\sqrt{2}}\left(  -\Theta^{1}-i\Theta^{2}\right)  $\\
$\pi_{4}=\frac{1}{\sqrt{2}}\left(  I-\Theta^{3}\right)  $ &  & $\pi_{4}%
^{\#}=\frac{1}{\sqrt{2}}\left(  I+\Theta^{3}\right)  $%
\end{tabular}
\right\}  \label{4f6}%
\end{equation}
and the only nonvanishing scalar products are the same as in (\ref{4f5}):%
\begin{equation}
\left.
\begin{array}
[c]{l}%
\left(  \pi_{1},\pi_{4}\right)  =1\\
\left(  \pi_{2},\pi_{3}\right)  =-1
\end{array}
\right\}  \label{4f6a}%
\end{equation}

The algebraic products of the $\lambda$ and $\pi$ matrices are displayed in
the following tables:%

\vspace{5pt}%
\begin{equation}%
\begin{tabular}
[c]{|c|cccc|}\hline
$\lambda_{a}\lambda_{b}$ & $\lambda_{1}$ & $\lambda_{2}$ & $\lambda_{3}$ &
$\lambda_{4}$\\\hline
$\lambda_{1}$ & $\sqrt{2}\lambda_{1}$ & $0$ & $\sqrt{2}\lambda_{3}$ & $0$\\
$\lambda_{2}$ & $\sqrt{2}\lambda_{2}$ & $0$ & $\sqrt{2}\lambda_{4}$ & $0$\\
$\lambda_{3}$ & $0$ & $\sqrt{2}\lambda_{1}$ & $0$ & $\sqrt{2}\lambda_{3}$\\
$\lambda_{4}$ & $0$ & $\sqrt{2}\lambda_{2}$ & $0$ & $\sqrt{2}\lambda_{4}%
$\\\hline
\end{tabular}
\label{4f7}%
\end{equation}

\vspace{5pt}%
\begin{equation}%
\begin{tabular}
[c]{|c|cccc|}\hline
$\pi_{a}\pi_{b}$ & $\pi_{1}$ & $\pi_{2}$ & $\pi_{3}$ & $\pi_{4}$\\\hline
$\pi_{1}$ & $\sqrt{2}\pi_{1}$ & $\sqrt{2}\pi_{2}$ & $0$ & $0$\\
$\pi_{2}$ & $0$ & $0$ & $\sqrt{2}\pi_{1}$ & $\sqrt{2}\pi_{2}$\\
$\pi_{3}$ & $\sqrt{2}\pi_{3}$ & $\sqrt{2}\pi_{4}$ & $0$ & $0$\\
$\pi_{4}$ & $0$ & $0$ & $\sqrt{2}\pi_{3}$ & $\sqrt{2}\pi_{4}$\\\hline
\end{tabular}
\ \label{4f8}%
\end{equation}
\vspace{5pt}%

We observe that the relations (\ref{4f5}) and (\ref{4f6a}) coincide with
the `orthogonality' relations for the Newman-Penrose null-tetrad of
vectors $l,$ $n,$ $m,$ and $\bar{m}$ (see \cite{NP}, \cite{PR}, or
\cite{Chan} for an application), for which the only non-vanishing scalar products are $\left(
l,n\right)  =1$ and $\left(  m,\bar{m}\right)  =-1.$ The $\lambda$ and $\pi$
tetrads are thus tetrads of null-vectors whose components are not numbers but matrices.

This is unexpected because the matrices $\lambda_{r}$ and $\pi_{r}$ came to
light in section (\ref{s4.2}) as the elements the \textquotedblleft inverse
linking vectors\textquotedblright\ $\omega^{-1}$ defined in the algebra
$\mathcal{M}$ of all $4\times4$ matrices. There was no indication that they
could be related to a geometric null-tetrad. Inverting this observation, it is
also true that the general-relativistic investigations by Ted Newman and Roger
Penrose, which gave rise in 1962 to the null-tetrad vectors, contained no
indication that these vectors had an algebraic structure. Yet, the
identification%

\vspace{5pt}%

\begin{equation}
\left.
\begin{array}
[c]{c}%
l=\lambda_{1}\\
m=\lambda_{2}\\
\bar{m}=\lambda_{3}\\
n=\lambda_{4}%
\end{array}
\right\}  \label{5f9}%
\end{equation}
implies the following multiplication rules:
\begin{equation}%
\begin{tabular}
[c]{|c|cccc|}\hline
& $l$ & $m$ & $\bar{m}$ & $n$\\\hline
$l$ & $\sqrt{2}l$ & $0$ & $\sqrt{2}\bar{m}$ & $0$\\
$m$ & $\sqrt{2}m$ & $0$ & $\sqrt{2}n$ & $0$\\
$\bar{m}$ & $0$ & $\sqrt{2}l$ & $0$ & $\sqrt{2}\bar{m}$\\
$n$ & $0$ & $\sqrt{2}m$ & $0$ & $\sqrt{2}n$\\\hline
\end{tabular}
\ \ \ \label{5f10}%
\end{equation}
\vspace{5pt}%

This associative algebra of the basis tetrad vectors plays no role in standard general relativity, but it
might play an essential one in a hypothetical quantum theory of gravitation. It is indeed strongly reminiscent of the basic idea of non-commutative geometry in one of the approaches to quantum gravity (where the coordinates are assumeded to be non-commutative ).

\subsection{\label{s4.7}The algebraic norm}

Quantions being matrices, their Hermitian conjugation is an anti-automorphism:%
\[
\left(  PQ\right)  ^{\dag}=Q^{\dag}P^{\dag}%
\]
The algebraic norm of an arbitrary quantion $Q$ is therefore a Hermitian
quantion:%
\[
\left[  A\left(  Q\right)  \right]  ^{\dag}=\left(  Q^{\dag}Q\right)  ^{\dag
}=Q^{\dag}Q^{\dag\dag}=Q^{\dag}Q=A\left(  Q\right)
\]

The metric and algebraic norms of a quantion $Q$ being products of two
quantions, namely $Q^{\#}Q$ and $Q^{\dag}Q$ respectively, they are quantions
themselves. The norm of a norm is therefore a quantion as well. Let us compute
the four cases in turn.

\textbf{The metric norm of the metric norm}

Since the metric norm of a quantion is a complex number, it is self-dual,%
\[
\left[  M\left(  Q\right)  \right]  ^{\#}=M\left(  Q\right)
\]
Hence%
\begin{equation}
M\left(  M\left(  Q\right)  \right)  =\left[  M\left(  Q\right)  \right]  ^{2}
\label{4g1}%
\end{equation}

\textbf{The algebraic norm of the algebraic norm}

Since the algebraic norm of a quantion is a Hermitian quantion, it is
self-conjugate,%
\[
\left[  A\left(  Q\right)  \right]  ^{\dag}=A\left(  Q\right)
\]
Hence,%
\begin{equation}
A\left(  A\left(  Q\right)  \right)  =\left[  A\left(  Q\right)  \right]  ^{2}
\label{4g2}%
\end{equation}

\textbf{The mixed norms}

Let us expand the metric norm of the algebraic norm:%
\[%
\begin{tabular}
[c]{llll}%
$M\left(  A\left(  Q\right)  \right)  $ & $=M\left(  Q^{\dag}Q\right)  $ &
$\cdots$ & by the definition (\ref{2w5}) of $A\left(  Q\right)  $\\
& $=M\left(  Q^{\dag}\right)  M\left(  Q\right)  $ & $\cdots$ & by relation
(\ref{4b2})\\
& $=\left[  M\left(  Q\right)  \right]  ^{\ast}M\left(  Q\right)  $ & $\cdots$
& because $M\left(  Q\right)  $ is a complex number\\
& $=A\left(  M\left(  Q\right)  \right)  $ & $\cdots$ & because $z^{\ast
}\equiv z^{\dag}$ for complex numbers
\end{tabular}
\
\]
Thus, the two norm functions commute,%
\begin{equation}
M\left(  A\left(  Q\right)  \right)  =A\left(  M\left(  Q\right)  \right)
\label{4g3}%
\end{equation}
Symbolically,%
\begin{equation}
\left[  A,M\right]  =0 \label{4g4}%
\end{equation}

It follows from (\ref{4g3}) that the metric norm of the algebraic norm is a
non-negative real number,%
\begin{equation}
MA\left(  Q\right)  =AM\left(  Q\right)  =\left\Vert M\left(  Q\right)
\right\Vert ^{2}\geqslant0 \label{4g5}%
\end{equation}
which vanishes if and only if $Q$ is singular, that is, if $\det\left(
Q\right)  =0.$

\subsubsection*{The geometric properties of the algebraic norm}

The algebraic norm $A\left(  Q\right)  $ of a q-quantion may be expanded in
the $\Theta_{\mu}$ basis:%
\begin{equation}
A\left(  Q\right)  =j^{\mu}\Theta_{\mu} \label{4g6}%
\end{equation}
To compute the coefficients $j^{\mu},$ we multiply both sides of this relation
by $\Theta^{\nu},$%
\[
j^{\mu}\Theta_{\mu}\Theta^{\nu}=Q^{\dag}Q\Theta^{\nu}%
\]
and take the expectation values (reminder: the p-quantions $\Theta^{\nu}$
commute with every q-quantion $Q$):%
\[
j^{\mu}\left\langle \Theta_{\mu}^{\nu}\right\rangle =\left\langle Q^{\dag
}Q\Theta^{\nu}\right\rangle \equiv\left\langle Q^{\dag}\Theta^{\nu
}Q\right\rangle
\]
Using the relation (\ref{4a19}), we get the following equivalent expressions%
\begin{align}
j^{\nu}  &  =\left\langle Q^{\dag}\Theta^{\nu}Q\right\rangle \label{4g8}\\
j^{\nu}  &  =\left(  Q\left\vert \Theta^{\nu}\right\vert Q\right)  \label{4g9}%
\end{align}

It follows from (\ref{4g9}) that the 0-component (the `time component') of
$j^{\nu}$ is positive definite,%
\begin{equation}
j^{0}=\left\langle I\right\rangle =\left(  Q|Q\right)  =\sum_{i=1}^{4}%
x_{i}^{\ast}x_{i}\geqslant0 \label{4g10}%
\end{equation}
meaning that $j^{\mu}$ is future-oriented.

Let us compute the metric norm of $A\left(  Q\right)  :$
\[
MA\left(  Q\right)  =\left(  j^{\mu}\Theta_{\mu}\right)  ^{\#}\left(  j^{\nu
}\Theta_{\nu}\right)  =j^{\mu}j^{\nu}\Theta_{\mu}^{\#}\Theta_{\nu}%
\]
By (\ref{4g5}), the left-hand side is a positive real number. Taking the
expectation value of both sides, we get%
\[
MA\left(  Q\right)  =j^{\mu}j^{\nu}\left\langle \Theta_{\mu}^{\#}\Theta_{\nu
}\right\rangle =\ j^{\mu}j^{\nu}\eta_{\mu\nu}\geqslant0
\]

Defining a \textbf{causal vector} as a \emph{future-oriented timelike or null
vector:}%
\begin{equation}
\left.
\begin{array}
[c]{c}%
j^{0}>0\\
\eta_{\mu\nu}\ j^{\mu}j^{\nu}\geqslant0
\end{array}
\right\}  \label{4g11}%
\end{equation}
we arrive at the following essential conclusion:%
\begin{equation}%
\begin{tabular}
[c]{|l|}\hline
\emph{The algebraic norm of an arbitrary quantion is a causal Minkowski
vector}\\\hline
\end{tabular}
\label{4g12}%
\end{equation}

The future orientation, which is \emph{arbitrary in classical relativity,} is
therefore \emph{distinguished in quantionic relativity.}

Since this conclusion clashes with our long-standing physical intuitions, let
us parse it conceptually in terms of simple observations:

(1) `Pure' geometric spaces are as isotropic as the metric tensor will allow.

(2) The non-quantum spacetime metric, which is of macroscopic origin,
distinguishes three classes of directions: present, past, and future.

(3) While past and future are not interchangeable, neither is metrically
distinguished. What we call `future' is a matter of convention (maybe not in
weak interactions, but these are outside the macroscopic domain).

(4) Unlike geometry, which has no distinguished direction, algebra does have a
distinguished element: the unit $I.$

(5) In quantionic spacetime, the distinguished algebraic element admits a
geometric expansion%
\begin{equation}
I=\omega^{\mu}\Theta_{\mu} \label{4g12a}%
\end{equation}
which gives rise to the causal unit vector $\omega^{\mu}:$%
\begin{equation}
\left(  \omega^{\mu}\right)  =\left(
\begin{array}
[c]{c}%
1\\
0\\
0\\
0
\end{array}
\right)  \in\emph{\ }M_{q}^{4} \label{4g13}%
\end{equation}

We shall refer to the distinguished vector $\omega^{\mu}$ as the
\textbf{structure vector.} With its help, causal vectors may be defined by the
covariant inequalities%
\begin{equation}
\left.
\begin{array}
[c]{c}%
j_{\mu}j^{\mu}\geqslant0\\
\omega_{\mu}j^{\mu}>0
\end{array}
\right\}  \label{4g14}%
\end{equation}
\textbf{\ }

Let us conclude with a speculative observation:

Judging by its coordinates-related form (\ref{4g13}), the structure vector may seem to be
a concept related to coordinates. But let's look at it from a different point of
view by stepping into a Riemannian space of general relativity and considering
the world line of a macroscopic object --- let's say of a bubble chamber with its proper time. At
every instant, the tangent vector specifies the structure vector of the
quantionic Minkowski space in which elementary interactions take place. While
the vector $\omega^{\mu}$ is frozen in this space, it is variable in the
Riemannian space. The reconciliation of the two viewpoints could therefore be
a source of relations between gravitational and quantum effects. It might be particularly instructive to see how this idea is affected by a black hole.

\subsection{\label{s4.8}The polar representation of quantions}

The following self-explanatory diagram comparatively displays the Cartesian
decompositions of the field of complex numbers and of the algebra of quantions
(discussed in $\mathcal{Q},$ but also valid in $\mathcal{P}$).%
\begin{equation}
\left.
\begin{tabular}
[c]{ccccc}%
$\mathbb{C}$ & $=$ & $\mathbb{R}$ & $\oplus$ & $i~\mathbb{R}$\\
$\updownarrow$ &  & $\updownarrow$ &  & $\updownarrow$\\
$\mathcal{Q}$ & $=$ & $\mathcal{Q}_{h}$ & $\oplus$ & $i\mathcal{Q}_{h}$%
\end{tabular}
\right\}  \label{4h1}%
\end{equation}

We shall extend the polar factorization of complex numbers to quantions
according to the following analogies:%
\begin{equation}
\left.
\begin{tabular}
[c]{cccc}%
$z$ & $=$ & $e^{i\chi}$ & $r$\\
$\updownarrow$ &  & $\updownarrow$ & $\updownarrow$\\
$Q$ & $=$ & $E$ & $R$%
\end{tabular}
\right\}  \label{4h2}%
\end{equation}
where $E\in\mathcal{Q}$ will be referred to as the \textbf{quantionic phase
factor} and $R\in\mathcal{Q}_{h}$ as the \textbf{quantionic modulus.}
Substitution of these relations into the algebraic norms yields the following
correspondences:%
\begin{equation}
\left.
\begin{tabular}
[c]{cccccccccc}%
$A\left(  z\right)  $ & $=$ & $z^{\ast}z$ & $=$ & $r$ & $e^{-i\chi}$ &
$e^{i\chi}$ & $r$ & $=$ & $r^{2}$\\
&  &  &  &  &  &  &  &  & \\
$\updownarrow$ &  & $\updownarrow$ &  & $\updownarrow$ & $\updownarrow$ &
$\updownarrow$ & $\updownarrow$ &  & $\updownarrow$\\
&  &  &  &  &  &  &  &  & \\
$A\left(  Q\right)  $ & $=$ & $Q^{\dag}Q$ & $=$ & $R$ & $E^{\dag}$ & $E$ & $R
$ & $=$ & $R^{2}$%
\end{tabular}
\right\}  \label{4h3}%
\end{equation}
We therefore conclude that the phase factor $E$ belongs to a unitary group
which is a substructure of the algebra $\mathcal{Q}.$ Since it generalizes the
complex gauge group $U\left(  1\right)  ,$ we shall denote it by $U_{q}\left(
1\right)  $ and refer to it as the \textbf{quantionic gauge group. }

The factorization%
\begin{equation}
Q=ER \label{4h0}%
\end{equation}
is unique up to signs, for if it were not, there would exist another phase
factor, $F,$ and another modulus, $S,$ such that
\[
FS=ER
\]
implying%
\[
S=\left(  F^{\dag}E\right)  R
\]
Since $S$ and $R$ are both Hermitian while $\left(  F^{\dag}E\right)  $ is
unitary, the latter cannot be more general than $\varepsilon I,$ where
$\varepsilon=\pm1.$ It follows that $F=\varepsilon E$ and $S=\varepsilon R.$
In the factorization $re^{\phi}$ of complex numbers, we take $r>0,$ but since
$R$ is not a real number, we cannot naively take $R>0$ without knowing whether
this condition has a meaning for Hermitian quantions.

\subsubsection*{The quantionic phase factor}

A general quantion $Q\in\mathcal{Q}$ is defined by eight real variables, while
a Hermitian quantion $R\in\mathcal{Q}_{h}$ is defined by four real variables.
It thus follows that $U_{q}\left(  1\right)  $ is a four-parametric group.
Furthermore, since%
\[
A\left(  e^{i\phi}E\right)  =\left(  e^{i\phi}E\right)  ^{\dag}\left(
e^{i\phi}E\right)  =A\left(  E\right)  =I
\]
\ for any factor $e^{i\phi},$ an arbitrary phase factor $E$ admits the unique
factorization%
\begin{equation}
E=e^{i\chi}E_{0} \label{4h8}%
\end{equation}
where $E_{0}$ contains no complex phase factor. To extract $e^{i\chi}$ from a
given quantionic phase factor $E,$ we take the metric norm of both sides of
(\ref{4h8}):%
\[
M\left(  E\right)  =e^{i2\chi}M\left(  E_{0}\right)
\]
We may therefore characterize $E_{0}$ (up to sign) by the condition%
\begin{equation}
M\left(  E_{0}\right)  =I \label{4h9}%
\end{equation}
We refer to $E_{0}$ as a \textbf{pure quantionic phase factor.} Since%
\[
M\left(  E_{0}F_{0}\right)  =M\left(  E_{0}\right)  M\left(  F_{0}\right)  =I
\]
the pure quantionic phase factors form a subgroup of $U_{q}\left(  1\right)
.$ Thus, given an arbitrary phase factor $E\in U_{q}\left(  1\right)  ,$ its
\textbf{complex phase factor} $e^{i\chi}$ and its pure quantionic phase factor
$E_{0}$ are defined, up to signs, by the relations%
\begin{equation}
\left.
\begin{array}
[c]{c}%
e^{i2\chi}=\left\langle M\left(  E\right)  \right\rangle \\
E_{0}=e^{-i\chi}E
\end{array}
\right\}  \label{4h10}%
\end{equation}

In block form,%
\begin{equation}
E_{0}=%
\begin{pmatrix}
U_{0} & 0\\
0 & U_{0}%
\end{pmatrix}
\label{4h11}%
\end{equation}
and thus%
\[
\det U_{0}=\left\langle M\left(  E_{0}\right)  \right\rangle =1
\]
It follows that $U_{0}$ is an arbitrary unitary $2\times2$ unitary matrix.
Therefore%
\begin{equation}
U_{q}\left(  1\right)  =U\left(  1\right)  \times SU\left(  2\right)
\label{4h12}%
\end{equation}

In the $\Theta_{\mu}$ basis, the most general unitary matrix $U_{0}$ may be
written in the form%
\begin{equation}
E_{0}=e^{i\phi\left(  \vec{m}\cdot\vec{\Theta}\right)  }=\cos\phi\ I+i\sin
\phi\ \vec{m}\cdot\vec{\Theta} \label{4h13}%
\end{equation}
where $\vec{m}$ is an arbitrary unit vector, $\vec{m}\cdot\vec{m}=1.$

Let us point out that the factor ordering $ER$ in the factorization
(\ref{4h0}) has been selected for convenience. It can be reversed by adjusting
the modulus. This follows from the identities%
\[
ER\equiv ER\left(  E^{\dag}E\right)  \equiv\left(  ERE^{\dag}\right)  E\equiv
SE
\]
Since the factorization of $Q$ is unique up to sign once the ordering has been
selected, we have the two extreme options%
\[
ER=Q=SE
\]
where%
\[
S=ERE^{\dag}%
\]
and an infinity of mixed factorizations $Q=E^{\prime}R^{\prime}F^{\prime}, $
provided $E^{\prime}F^{\prime}=E.$

\subsubsection*{The quantionic modulus}

The modulus $R$ of an arbitrary quantion $Q$ is the Hermitian quantion%
\begin{equation}
R=\pm\sqrt{A\left(  Q\right)  } \label{4h4}%
\end{equation}

If $Q$ is regular, the Minkowski vector that corresponds to $A\left(
Q\right)  =R^{2}$ is a future-oriented timelike vector in $M_{q}^{4}.$ It may
therefore be written in the parametric form%
\begin{equation}
R^{2}=r^{\mu}\Theta_{\mu}=a^{2}\left(  \cosh2\sigma\ I+\sinh2\sigma\ \vec
{n}\cdot\vec{\Theta}\right)  \label{4h14}%
\end{equation}
where $\vec{n}$ is a unit vector. This expression has three different types of
square roots, $R_{t},$ $R_{n},$ and $R_{s},$ with the following
interpretations:%
\begin{equation}
\left.
\begin{tabular}
[c]{lll}%
$R_{t}=\varepsilon\left\vert a\right\vert \left(  \cosh\sigma\ I+\sinh
\sigma\ \vec{n}\cdot\vec{\Theta}\right)  $ & $\cdots$ & timelike\\
$R_{n}=\varepsilon\left\vert a\right\vert \left(  \ I+\ \vec{n}\cdot
\vec{\Theta}\right)  $ & $\cdots$ & null\\
$R_{s}=\left\vert a\right\vert \left(  \sinh\sigma\ I+\cosh\sigma\ \vec
{n}\cdot\vec{\Theta}\right)  $ & $\cdots$ & spacelike
\end{tabular}
\ \ \right\}  \label{4h15}%
\end{equation}
where $\left\vert a\right\vert $ is a positive real coefficient, and
$\varepsilon=\pm1$ is the sign indicator.

In the spacelike case, $\varepsilon$ is undefined because it can be absorbed
by $\sigma$ and $\vec{n}:$%
\[
\sinh\left(  \varepsilon\sigma\right)  \ I+\cosh\sigma\ \left(  \varepsilon
\vec{n}\right)  \cdot\vec{\Theta}\equiv\varepsilon\left(  \sinh\sigma
\ I+\cosh\sigma\ \vec{n}\cdot\vec{\Theta}\right)
\]
In the timelike and null cases, it is the sign of the coefficient of the unit
matrix. Thus,
\begin{equation}
\varepsilon=sign\left\langle R_{t}\right\rangle \ \ \ \text{or\ \ \ }%
\varepsilon=sign\left\langle R_{n}\right\rangle \label{4h16}%
\end{equation}

In the latter two cases, we may therefore speak of \textbf{positive quantions}
and \textbf{negative quantions.} Clearly, the algebraic norm of every quantion
is a positive quantion, and the vector representing a positive quantion is a
causal vector.

This completes the polar formulation of quantions, but let us conclude with
some tangential observations.%

\vspace{10pt}%

While the polar representation of complex numbers can be extended
to quaternions,%
\[
Q=re^{i\alpha}e^{j\beta}e^{k\gamma}\in\mathbb{H}%
\]
and even very much beyond quaternions (see \cite{Sudar}), if we stay within the number systems, it is
only for complex numbers and quantions that the modulus and the phase factor
have the same number of degrees of freedom, namely one and four respectively.
This is related to the symmetric splitting of these number systems into
Hermitian and antihermitian parts:%
\begin{align*}
\mathbb{C}  &  =\mathbb{R}\oplus i\mathbb{R}\\
\mathcal{Q}  &  =\mathcal{Q}_{h}\oplus i\mathcal{Q}_{h}%
\end{align*}

In quantum mechanics, the modulus squared of the wave function is assigned a
physical meaning by Born's interpretation. The phase factor is not directly
observable, but the gauge group $U\left(  1\right)  $ gives rise to a
differential connection interpreted as the electromagnetic potential.

We shall show in Part 2 of the present work that these are special cases of
their counterparts in the quantionic domain.

\subsection{\label{s4.9}The quantionic field equation}

To arrive at a quantionic field equation, it would be desirable to have a
general constructive idea (like a principle) that generates differential
equations from the properties of quantions developed so far. Such an idea was
suggested by Nikola Zovko.

At this point in the present paper, the mathematics of quantions consists of
the algebra $\mathcal{Q},$ of the Minkowski spaces $M_{q}^{4}$ and $M_{q}%
^{4},$ of an inner product space $\mathcal{H},$ and of the quantionic
derivation operator $\mathcal{D}.$ This derivation operator would enable us to
write structurally sound differential equations that might have physical
meaning if we had a physically meaningful mechanism for generating such equations. Such a mechanism is the observation
(\ref{4g12}).\ Let us emphasize it once again, for it supports all
subsequent physical applications of quantions:%
\[%
\begin{tabular}
[c]{|c|}\hline
\textbf{A strictly mathematical object}\\
(the algebraic norm of an arbitrary quantion)\\
\textbf{is equivalent to an apparently physical object}\\
(a causal Minkowski vector)\\\hline
\end{tabular}
\]
This causal vector calls for a physical interpretation. If correctly guessed,
this interpretation could have far-reaching consequences --- as was the case
with Born's interpretation.

Having noticed the above relationship in the very early days of the quantionic
approach, the author took it for granted\ that the physical interpretation of
the causal vector `must be' four-momentum. After much fruitless effort at
trying to verify this hypothesis, Nikola Zovko, assuming a field of quantions,
asked the right question: \emph{Could the algebraic norm satisfy the equation
of continuity? }This was equivalent to suggesting that the causal vector is to
be regarded as a current (of whatever charge). We refer to this idea as
\textbf{Zovko's interpretation.}

This interpretation implies that the vector\footnote{From now on, we shall use
lower case letters for vectors in $\mathcal{H}.$ Writing $\left\vert q\right)
$ instead of $\left\vert Q\right)  $ avoids confusing the vector components
$q_{i}$ of $\left\vert q\right)  $ with indexed quantions $Q_{i}.$}%
\begin{equation}
j^{\mu}=\left(  q\left\vert \Theta^{\mu}\right\vert q\right)  \label{4k1}%
\end{equation}
is to satisfy the equation of continuity%
\begin{equation}
\partial_{\mu}j^{\mu}=\partial_{\mu}\left(  q\left\vert \Theta^{\mu
}\right\vert q\right)  =0 \label{4k2}%
\end{equation}
We shall assume that the four complex components of $Q\left(  x\right)  ,$%
\begin{equation}
\left\vert q\right)  =\left(
\begin{array}
[c]{c}%
q_{1}\left(  x\right) \\
q_{2}\left(  x\right) \\
q_{3}\left(  x\right) \\
q_{4}\left(  x\right)
\end{array}
\right)  \label{4k3}%
\end{equation}
allow unrestricted differentiation.

Since the expression $\left(  q\left\vert \Theta^{\mu}\right\vert q\right)  $
is a linear combination of 16 terms of type $q_{i}^{\ast}q_{j}$ for each value
of $\mu,$ the Leibniz rule for the derivation of a product is applicable to
each term separately,%
\[
\partial_{\mu}j^{\mu}=\left(  \partial_{\mu}q\left\vert \Theta^{\mu
}\right\vert q\right)  +\left(  q\left\vert \Theta^{\mu}\right\vert
\partial_{\mu}q\right)  =0
\]
The basis matrices $\Theta^{\mu}$ being Hermitian, this equation may be
rewritten in the form%
\begin{equation}
\partial_{\mu}j^{\mu}=\left(  \Theta^{\mu}\partial_{\mu}q|q\right)  +\left(
q|\Theta^{\mu}\partial_{\mu}q\right)  =0 \label{4k4}%
\end{equation}
We recognize in these expressions the quantionic derivation operator
$\mathcal{D}$ defined by relation (\ref{4d1}),%
\begin{equation}
\partial_{\mu}j^{\mu}=\left(  \mathcal{D}q|q\right)  +\left(  q|\mathcal{D}%
q\right)  =0 \label{4k5}%
\end{equation}
where\ $\left\vert \mathcal{D}q\right)  =\mathcal{D}\left\vert q\right)  ,$
and $\left(  \mathcal{D}q\right\vert =\left[  \mathcal{D}\left\vert q\right)
\right]  ^{\dag}.$

From the point of view of functional analysis, which is not relevant here, the
operator $\mathcal{D}$ would be antihermitian, but it is Hermitian from the
point of view of matrix algebra. Thus, $\mathcal{D}^{\dag}\ $is not a new
operator. On the other hand, the complex conjugate $\mathcal{D}^{\ast},$ the
metric dual $\mathcal{D}^{\#},$ and the conjugate of the dual $\mathcal{D}%
^{\#\ast}=\mathcal{D}^{\ast\#},$ are all different. Figure \ref{S4}.2
shows their mutual relationships.
Since $\Theta^{2}$ anticommutes with $\Theta^{1}$ and $\Theta^{3},$ the
diagonally opposite operators are related by the commutation relations%
\begin{align}
\mathcal{D}^{\#}\Theta^{2}  &  =\Theta^{2}\mathcal{D}^{\ast}\label{4k8}\\
\mathcal{D}^{\#\ast}\Theta^{2}  &  =\Theta^{2}\mathcal{D}\ \label{4k9}%
\end{align}

\begin{center}%
\setlength{\unitlength}{0.01in}
\begin{picture}(500,180)
\put(250,150){\makebox(0,0){$\mathcal{D}= I \partial_0 +\Theta^1 \partial
_1+\Theta^2 \partial_2+\Theta^3 \partial_3$}}
\put(100,80){\makebox(0,0){$\mathcal{D}^\#= I \partial_0 -\Theta
^1 \partial_1-\Theta^2 \partial_2-\Theta^3 \partial_3$}}
\put(400,80){\makebox(0,0){$\mathcal{D}^*= I \partial_0 +\Theta^1 \partial
_1-\Theta^2 \partial_2+\Theta^3 \partial_3$}}
\put(250,10){\makebox(0,0){$\mathcal{D}^{*\#}= I \partial_0 -\Theta
^1 \partial_1+\Theta^2 \partial_2-\Theta^3 \partial_3$}}
\put(250,150){\oval(250,25)}
\put(100,80){\oval(250,25)}
\put(400,80){\oval(250,25)}
\put(250,10){\oval(250,25)}
\put(200,137){\vector(-1,-1){44}}
\put(300,137){\vector(1,-1){44}}
\put(156,67){\vector(1,-1){44}}
\put(344,67){\vector(-1,-1){44}}
\put(170,115){\makebox(0,0){\#}}
\put(330,115){\makebox(0,0){*}}
\put(170,45){\makebox(0,0){*}}
\put(330,45){\makebox(0,0){\#}}
\end{picture}%

\vspace{0.1in}%

Figure \ref{S4}.2: The four quantionic derivation operators%

\vspace{0.1in}%

\end{center}

The relation%
\begin{equation}
\mathcal{D}^{\#}\mathcal{D}=\square\label{4k10}%
\end{equation}
follows from the symmetry $\partial_{\mu}\partial_{\nu}=\partial_{\nu}%
\partial_{\mu}=\partial_{\mu\nu}$ of second partial derivatives:
\[
\mathcal{D}^{\#}\mathcal{D}=\Theta^{\mu\#}\ \Theta^{\nu}\partial_{\mu\nu
}=\frac{1}{2}\left(  \Theta^{\nu\#}\ \Theta^{\mu}+\Theta^{\mu\#}\ \Theta^{\mu
}\right)  \partial_{\mu\nu}=I\ \eta^{\mu\nu}\partial_{\mu\nu}=I\ \square
\]

Since $\left(  \mathcal{D}q|q\right)  $ is a complex number, the divergence
(\ref{4k5}) may be rewritten as%
\[
\partial_{\mu}j^{\mu}=\left(  q|\mathcal{D}q\right)  ^{\ast}+\left(
q|\mathcal{D}q\right)
\]
The equation of continuity thus assumes the simple form%
\begin{equation}
\Re\left(  q\left\vert \mathcal{D}\right\vert q\right)  =0 \label{4k12}%
\end{equation}

This is still the classical equation of continuity because $j^{\mu}$ is a
classical vector. A quantionic field equation would have to be of the type%
\begin{equation}
\mathcal{D}\left\vert q\right)  =\left\vert F\right)  \label{4k13}%
\end{equation}
where the most general components $F_{i}$ would be some complex function
$F_{i}\left(  x,Q,Q^{\dag}\right)  $ such that%
\begin{equation}
\Re\left(  q|F\right)  =0 \label{4k14}%
\end{equation}
This condition being much too general, we shall tighten it with two additional requirements:

\vspace{7pt}%

\textbf{Requirement 1:} \emph{\ Equation (\ref{4k14}) is to be interpreted as
an identity} --- meaning that it must be satisfied for an arbitrary vector
$\left\vert q\right)  .$ The most general solution is the linear combination%
\begin{equation}
\left\vert F\right)  =iH\left\vert q\right)  +M\left\vert q^{\ast}\right)
\label{4k16}%
\end{equation}
where $H$ is an arbitrary Hermitian matrix, and $M$ a matrix whose properties
remain to be investigated (an arbitrary antisymmetric matrix $M$ is a
solution, but not the only one). \textit{A priori,} these matrices could be
functions of $x,$ of $Q,$ and of $Q^{\dag},$ which is still too general.

\vspace{7pt}%

\textbf{Requirement 2:} \emph{The fields }$\left\vert q\right)  $\emph{\ must
not be directly self-interacting} --- implying that $H$ and $M$ are not
functions of $Q$ or $Q^{\dag}.$

\vspace{7pt}%

The requirements 1 and 2 taken together will be referred to as
\textbf{structural quantization.} We take this as a postulate. It seems
possibe to relax these requirements without losing the underlying algebraic
structure developed up to this point.\footnote{See the reference to \v{Z}ubrini\'{c}'s observation on the
second page of the introduction.}

Since the field equation (\ref{4k13}) may be written in either of the
following two forms,%
\begin{equation}
\mathcal{D}\left\vert q\right)  =iH\left\vert q\right)  +M\left\vert q^{\ast
}\right)  \label{4k19}%
\end{equation}%
\begin{equation}
\left(  \mathcal{D}-iH\right)  \left\vert q\right)  =M\left\vert q^{\ast
}\right)  \label{4k20}%
\end{equation}
the matrix $H$ may be interpreted as an external potential (the case of
(\ref{4k19})), or as a differential connection (the case of (\ref{4k20})). 

We may therefore refer to the matrix  $H$  as the \textbf{matrix potential} or the
\textbf{quantionic connection.} We shall refer to $M$ as the \textbf{mass
matrix.}

\subsubsection*{Structural versus canonical quantization}

These two quantizations are complementary in some sense.

\vspace{7pt}%

\emph{Canonical quantization} refers to the mapping
\[
\left(
\begin{array}
[c]{c}%
E\\
\vec{p}%
\end{array}
\right)  \rightarrow\left(
\begin{array}
[c]{c}%
i\hbar\partial_{t}\\
-i\hbar\mathcal{\nabla}%
\end{array}
\right)
\]
which is only phenomenologically justified. It is therefore a postulate from
the viewpoint of mathematics. In the opposite direction, the unobservable
function $\psi$ gives rise to the observable object $\left\langle \psi
|\psi\right\rangle ,$ which, by Born's interpretation, represents a classical
probability density. Schr\"{o}dinger equation of motion came first; the need
for an interpretation followed.

\vspace{7pt}%

\emph{Structural quantization} refers to the linearization of the equation of
continuity described above. It has no phenomenological support, but suggests
itself on mathematical grounds. Zovko's interpretation comes first, the
equations of motion follow from it.

\vspace{7pt}%

Structural quantization and the derivation of Dirac's equation are analogous
procedures: 

\vspace{5pt}%

Structural quantization reduces an \emph{algebraically second order equation}
to a first order one by taking the `square root' of the current $\left(
q\left\vert \Theta^{\mu}\right\vert q\right)  ;$ 

\vspace{5pt}%

Dirac's procedure reduces a \emph{differentially second order equation} to a first order one by taking the
`square root' of the Klein-Gordon operator $\left(  \square+m^{2}\right)  .$

\vspace{5pt}%

We shall see in Part 2 that Dirac's derivation operator $\mathcal{D}%
_{d}=\gamma^{\mu}\partial_{\mu}$ and the quantionic derivation operator
$\mathcal{D}=\Theta^{\mu}\partial_{\mu}$ emphasize two different aspects of
derivation:\ Dirac's operator emphasizes the geometric (or relativistic)
aspect; the quantionic operator emphasizes the algebraic (or quantum
mechanical) aspect.%

\section{\label{S8}Nonrelativistic fields (Schr\"{o}dinger's equation)}

Leaving the general quantionic field equation (\ref{4k19}) to Part 2 of the present work on account of length, we shall conclude with its non-relativistic limit: the Schr\"{o}dinger equation.

\vspace{7pt}%

\subsection{\label{s8.1}The nonrelativistic formalism}

The transition from the manifestly covariant 4-vector formalism to the
nonrelativistic formalism of classical physics involves the insertion of the
dimensional factor $c$ into the timelike direction,
\begin{equation}
\partial_{0}=\frac{\partial}{\partial x^{0}}=\frac{1}{c}\frac{\partial
}{\partial t} \label{8.00}%
\end{equation}
and the separation of the timelike and spacelike vector components. We shall
do the latter by writing Minkowski vectors as $1\times2$ or $2\times1$
matrices whose first component is a scalar and the second a 3-vector. The
covariant and contravariant forms of a vector are thus
\begin{equation}
v_{\mu}=%
\begin{pmatrix}
s & \vec{V}%
\end{pmatrix}
\rightleftarrows v^{\mu}=\left(
\begin{array}
[c]{c}%
s\\
-\vec{V}%
\end{array}
\right)  \label{8.0}%
\end{equation}
For the gradient, the most natural conventions are%
\begin{equation}
\left.
\begin{tabular}
[c]{l}%
$\partial_{\mu}=%
\begin{pmatrix}
\frac{1}{c}\partial_{t} & \mathcal{\nabla}%
\end{pmatrix}
$\\
\\
$\partial^{\mu}=\left(
\begin{array}
[c]{c}%
\frac{1}{c}\partial_{t}\\
-\mathcal{\nabla}%
\end{array}
\right)  $%
\end{tabular}
\right\}  \label{8.1}%
\end{equation}

For the basis matrices in the algebra of quantions, we shall write%
\begin{equation}
\Theta_{\mu}=%
\begin{pmatrix}
I & \vec{\Theta}%
\end{pmatrix}
\label{8.3}%
\end{equation}
The matrices $\Theta^{\mu}$ will not be needed.

\subsection{\label{s8.2}The algebraic nonrelativistic limit}

In the above notations, a q-quantion is of the form%
\begin{equation}
Q=z^{\mu}\Theta_{\mu}\equiv\Theta_{\mu}z^{\mu}=%
\begin{pmatrix}
I & \vec{\Theta}%
\end{pmatrix}
\left(
\begin{array}
[c]{c}%
\psi\\
\vec{w}%
\end{array}
\right)  =\psi I+\vec{w}\cdot\vec{\Theta} \label{8.2}%
\end{equation}
where $\psi$ is a complex number and $\vec{w}$ a complex 3-vector. While this
expression is not manifestly covariant, it is nevertheless relativistic if the
coefficients $\psi$ and $\vec{w}$ obey the Lorentz transformations.

Since we are interested in the nonrelativistic limit, the quantionic function
$Q\left(  x\right)  $ must remain in the infinitesimal neighborhood of the
field $\mathbb{C}$ of complex numbers. This neighborhood is specified by the
requirement
\begin{equation}
\vec{w}^{\ast}\cdot\vec{w}<<\psi^{\ast}\psi\label{8.6}%
\end{equation}

Taking the condition (\ref{8.6}) into account and dropping the infinitesimal
part, the algebraic norm of $Q$ becomes%
\begin{align}
A\left(  Q\right)   &  =Q^{\dag}Q=\left(  \psi^{\ast}I+\vec{w}^{\ast}\cdot
\vec{\Theta}\right)  \left(  \psi I+\vec{w}\cdot\vec{\Theta}\right)
\nonumber\\
&  =\left(  \psi^{\ast}\psi\right)  I+\left(  \psi\vec{w}^{\ast}+\psi^{\ast
}\vec{w}\right)  \cdot\vec{\Theta} \label{8.7}%
\end{align}

By Zovko's interpretation, the vector $j^{\mu}$ defined by (\ref{4g6}), that
is,%
\[
A\left(  Q\right)  =j^{\mu}\Theta_{\mu}=\rho I+\vec{j}\cdot\vec{\Theta}%
\]
is the current associated to the quantion (\ref{8.2}). Thus,%
\begin{equation}
\left(  j^{\mu}\right)  =\left(
\begin{array}
[c]{c}%
\rho\\
\\
\frac{1}{c}\vec{j}%
\end{array}
\right)  =\left(
\begin{array}
[c]{c}%
\psi^{\ast}\psi\\
\\
\psi\vec{w}^{\ast}+\psi^{\ast}\vec{w}%
\end{array}
\right)  \label{8.9}%
\end{equation}
where, by Born's interpretation, $\rho$ is a charge density (or probability
density),%
\begin{equation}
\rho=\psi^{\ast}\psi\label{8.10}%
\end{equation}
Hence,%
\begin{equation}
\vec{j}=c\left(  \psi\vec{w}^{\ast}+\psi^{\ast}\vec{w}\right)  \label{8.12}%
\end{equation}
is a 3-current.

\subsection{\label{s8.3}Structural quantization}

Substitution of (\ref{8.1}) and (\ref{8.9}) into the equation of continuity
$\partial_{\mu}j^{\mu}=0$ verifies the interpretation of $\vec{j}$ as a
current,%
\[%
\begin{pmatrix}
\frac{1}{c}\partial_{t} & \mathcal{\nabla}%
\end{pmatrix}
\left(
\begin{array}
[c]{c}%
\psi^{\ast}\psi\\
\psi\vec{w}^{\ast}+\psi^{\ast}\vec{w}%
\end{array}
\right)  =0
\]
that is,%
\begin{equation}
\partial_{t}\rho+\mathcal{\nabla}\cdot\vec{j}=0 \label{8.11}%
\end{equation}

The non-linear equation (\ref{8.11}) is to be reduced to a linear one.
Applying the Leibniz rule to%
\[
\partial_{t}\left(  \psi^{\ast}\psi\right)  +\mathcal{\nabla}\cdot\left(
\psi\vec{w}^{\ast}+\psi^{\ast}\vec{w}\right)  =0
\]
yields
\[
\psi^{\ast}\left(  \partial_{t}\psi\right)  +\psi\left(  \partial_{t}%
\psi\right)  ^{\ast}+c\left[  \left(  \mathcal{\nabla}\psi\right)  \cdot
\vec{w}^{\ast}+\psi\left(  \mathcal{\nabla}\cdot\vec{w}\right)  ^{\ast
}+\left(  \mathcal{\nabla}\psi\right)  ^{\ast}\cdot\vec{w}+\psi^{\ast}\left(
\mathcal{\nabla}\cdot\vec{w}\right)  \right]  =0
\]
In compact form, we may write%
\[
\Re\left[  \psi^{\ast}\left(  \partial_{t}\psi+c\mathcal{\nabla}\cdot\vec
{w}\right)  +c\vec{w}^{\ast}\cdot\left(  \mathcal{\nabla}\psi\right)  \right]
=0
\]
or, equivalently,%
\begin{equation}
\psi^{\ast}\left(  \partial_{t}\psi+c\mathcal{\nabla}\cdot\vec{w}\right)
+c\vec{w}^{\ast}\cdot\left(  \mathcal{\nabla}\psi\right)  =iX \label{8.12a}%
\end{equation}
where the most general term $X$ is an arbitrary real function of $\vec
{r},t,\psi,$ and $\vec{w}.$

Let us now interpret the condition (\ref{8.12a}) as an identity, in the sense
that the two terms on the left-hand side do not depend on each other for the
cancellation of their real parts. Each term is therefore identically
imaginary, which is the case if and only if
\begin{align}
\partial_{t}\psi+c\mathcal{\nabla}\cdot\vec{w}  &  =iF\psi\label{8.13}\\
\mathcal{\nabla}\psi &  =iN\vec{w} \label{8.14}%
\end{align}
for \textit{a priori} arbitrary real functions $F=F\left(  \vec{r},t,\psi
,\vec{w}\right)  $ and $N=N\left(  \vec{r},t,\psi,\vec{w}\right)  .$ More
generally, $N$ could be a $3\times3$ Hermitian tensor.

If we also require linearity (no self-interaction), the functions $F$ and $N$
do not depend on the fields $\psi$ and $\vec{w}.$ We assume this in the sequel.

This completes the structural quantization, that is, the linearization of the
equation of continuity.

\subsection{\label{s8.4}The Schr\"{o}dinger equation}

We are now to obtain a single equations for $\psi.$

Clearly, $N\vec{w}\neq0,$ for else (\ref{8.14}) would imply that $\psi$ is
constant. We may therefore write
\begin{equation}
\vec{w}=-iN^{-1}\mathcal{\nabla}\psi\label{8.15}%
\end{equation}
Substitution of this expression into equation (\ref{8.13}) yields%
\[
\frac{\partial}{\partial t}\psi-ic\mathcal{\nabla}\cdot\left(  N^{-1}%
\mathcal{\nabla}\psi\right)  =iF\psi
\]
and, after expansion of the divergence,%
\[
\frac{\partial}{\partial t}\psi-icN^{-1}\Delta\psi-ic\left(  \mathcal{\nabla
}N^{-1}\right)  \cdot\left(  \mathcal{\nabla}\psi\right)  =iF\psi
\]
Let us multiply all terms of this equation by $i\hbar,$ and rearrange them to%
\begin{equation}
i\hbar\frac{\partial}{\partial t}\psi=\left(  -\hbar cN^{-1}\Delta-\hbar
F\right)  \psi-\hbar c\left(  \mathcal{\nabla}N^{-1}\right)  \cdot\left(
\mathcal{\nabla}\psi\right)  \label{8.16}%
\end{equation}
Temporarily dropping the term with $\left(  \mathcal{\nabla}N^{-1}\right)  ,$
equation (\ref{8.16}) can be directly compared with the Schr\"{o}dinger
equation%
\begin{equation}
i\hbar\frac{\partial}{\partial t}\psi=\left(  -\frac{\hbar^{2}}{2m}%
\Delta+V\right)  \psi\label{8.17}%
\end{equation}
This yields the relations%
\begin{align*}
V  &  =-\hbar F\\
\frac{\hbar^{2}}{2m}  &  =\hbar cN^{-1}%
\end{align*}
Thus, $F$ is essentially the potential while $N$ is a scalar,%
\begin{equation}
N=\frac{2mc}{\hbar} \label{8.18}%
\end{equation}
If the mass parameter $m$ is constant, the extra term in (\ref{8.16}) vanishes.

If $m$ is a function of space, substitution of (\ref{8.18}) into equation
(\ref{8.16}) yields, after some rearrangement, the generalized Schr\"{o}dinger
equation%
\begin{equation}
\left(  -\frac{\hbar^{2}}{2}\mathcal{\nabla}\frac{1}{m}\mathcal{\nabla
}+V\right)  \psi=i\hbar\frac{\partial}{\partial t}\psi\label{8.20}%
\end{equation}

The idea of a space-dependent mass has been considered in the physics of
condensed matter, but there was apparently no good argument to select one of
the following two Hermitian generalizations of the Laplaceian%
\[%
\begin{tabular}
[c]{rccl}%
First choice: & $\frac{1}{m}\Delta$ & $\rightarrow$ & $\mathcal{\nabla}%
\frac{1}{m}\mathcal{\nabla}$\\
&  &  & \\
Second choice: & $\frac{1}{m}\Delta$ & $\rightarrow$ & $\frac{1}{m}%
\Delta+\Delta\frac{1}{m}$%
\end{tabular}
\]
We see that the choice consistent with the quantionic derivation of the
Schr\"{o}dinger equation is the first one .

\subsubsection*{The Schr\"{o}dinger current}

Substitution of (\ref{8.18}) into (\ref{8.15}) yields%
\begin{align*}
\vec{w}  &  =-i\frac{\hbar}{2mc}\mathcal{\nabla}\psi\\
\vec{w}^{\ast}  &  =i\frac{\hbar}{2mc}\mathcal{\nabla}\psi^{\ast}%
\end{align*}
A subsequent substitution of these expressions into (\ref{8.12}) yields the
well-known expression for the Schr\"{o}dinger current:%
\begin{equation}
\vec{j}=i\frac{\hbar}{2m}\left(  \psi\mathcal{\nabla}\psi^{\ast}-\psi^{\ast
}\mathcal{\nabla}\psi\right)  \label{8.22}%
\end{equation}

In quantum mechanics, this expression for the current has been constructed so
as to satisfy the continuity equation%
\[
\frac{\partial}{\partial t}\rho+\mathcal{\nabla}\cdot\vec{j}=0
\]
In the quantionic approach, (\ref{8.22}) is postulated as Zovko's
interpretation in the nonrelativistic limit, while the generalized
Schr\"{o}dinger equation follows from it.%

\section{\label{S21}A survey of quantionic properties}

A conceptual overview of the mathematics of quantions\ is displayed in Figure
\ref{S21}.1. It brings to light a natural organization of concepts which was
neither intended nor evident in the previous sections dedicated to the
step-by-step development of quantionic properties.
The quantum relativistic objects are clustered in the upper half of the
diagram, the classically relativistic ones in the lower half. The origin of
the separation is the reciprocity $\mathcal{M}\rightleftarrows\mathcal{K}$ of
the two interpretations of the totality of $4\times4$ complex matrices: While
the structure of the former is primarily algebraic (and thus quantum
mechanical), the latter is linear with a Minkowski metric (and thus
relativistic). The left-right separation is with respect to differentiation:
The derivation operators are\ on one side, their operands on the other.

Both separations ultimately stem from the structure of $\mathcal{M}$ as the
tensor product,%
\[
\mathcal{M}=\mathcal{Q}\otimes\mathcal{P}%
\]
of two subalgebras which are mutually dual\ and mutual commutants. This
structuring is encoded in a system of basis matrices, $\Theta_{\nu}^{\mu},$
naturally adapted to the subalgebras $\mathcal{Q}$ and $\mathcal{P}$ and to
the reciprocal space $\mathcal{K}$ of matrices of complex coefficients.

The two top arrows (quantionic and Dirac's derivation) complete the diagram,
but the details will be developed in Part 2 for field $\left\vert q\left(
x\right)  \right)  ,$ and probably in Part 3 for fields $Q\left(  x\right)  .$
Even though these fields are \textquotedblleft linked\textquotedblright\ by the
equation
\[
\left\vert q\left(  x\right)  \right)  =Q\left(  x\right)  \left\vert
\omega\right)
\]
their physical interpretations are different.

While Figure \ref{S21}.1 illustrates the relational structure of the
mathematics of quantions, Table \ref{S21}.1 comparatively displays the
properties of the algebra of quantions.

\begin{center}%
\setlength{\unitlength}{0.01in}
\begin{picture}(580,555)
\put(190,25){\oval(40,40)}
\put(390,25){\oval(40,40)}
\put(50,25){\oval(40,40)}
\put(290,125){\oval(40,40)}
\put(290,25){\vector(1,0){78}}
\put(290,25){\vector(-1,0){78}}
\put(276,111){\vector(-1,-1){70}}
\put(304,111){\vector(1,-1){70}}
\put(290,200){\oval(40,40)}
\put(290,275){\oval(40,40)}
\put(290,275){\oval(35,35)}
\put(390,375){\oval(40,40)}
\put(190,375){\oval(40,40)}
\put(530,25){\oval(40,40)}
\put(530,375){\oval(40,40)}
\put(530,465){\oval(40,40)}
\put(50,375){\oval(40,40)}
\put(50,465){\oval(80,40)}
\put(290,375){\vector(1,0){78}}
\put(290,375){\vector(-1,0){78}}
\put(276,289){\vector(-1,1){70}}
\put(304,289){\vector(1,1){70}}
\put(286,145){\vector(0,1){35}}
\put(294,180){\vector(0,-1){35}}
\put(290,237.5){\vector(0,1){15}}
\put(290,237.5){\vector(0,-1){15}}
\put(120,375){\vector(-1,0){48}}
\put(120,375){\vector(1,0){48}}
\put(170,25){\vector(-1,0){98}}
\put(410,375){\line(1,0){100}}
\put(410,25){\line(1,0){100}}
\put(50,395){\vector(0,1){48}}
\put(530,395){\vector(0,1){48}}
\put(190,355){\vector(0,-1){308}}
\put(390,355){\vector(0,-1){308}}
\put(515,390){\line(-1,1){35}}
\put(100,425){\line(1,0){380}}
\put(100,425){\vector(-1,-1){34}}
\put(240,425){\vector(-1,-1){34}}
\put(510,465){\vector(-1,0){418}}
\put(515,10){\line(-1,-1){35}}
\put(100,-25){\vector(-1,1){34}}
\put(100,-25){\line(1,0){380}}
\put(530,200){\vector(0,1){152}}
\put(530,200){\vector(0,-1){152}}
\put(255,410){\vector(-1,0){25}}
\put(115,410){\vector(-1,0){25}}
\thicklines\put(20,275){\vector(0,1){70}}
\put(20,275){\vector(0,-1){70}}
\put(10,275){\line(1,0){20}}
\put(290,500){\line(0,1){40}}
\put(290,520){\vector(1,0){180}}
\put(290,520){\vector(-1,0){180}}
\thinlines\multiput(270,275)(-30,0){8}{\line(-1,0){14}}
\multiput(310,275)(30,0){7}{\line(1,0){14}}
\put(50,472){\makebox(0,0){Dirac}}
\put(50,458){\makebox(0,0){spinors}}
\put(190,25){\makebox(0,0){$M^4_q$}}
\put(390,25){\makebox(0,0){$M^4_p$}}
\put(50,25){\makebox(0,0){$\mathcal{R}^4$}}
\put(290,125){\makebox(0,0){$\mathcal{K}_r$}}
\put(290,200){\makebox(0,0){$\mathcal{K}$}}
\put(290,275){\makebox(0,0){$\mathcal{M}$}}
\put(390,375){\makebox(0,0){$\mathcal{P}$}}
\put(190,375){\makebox(0,0){$\mathcal{Q}$}}
\put(530,25){\makebox(0,0){$\partial_\mu$}}
\put(530,375){\makebox(0,0){$\mathcal{D}$}}
\put(530,465){\makebox(0,0){$\mathcal{D}_d$}}
\put(50,375){\makebox(0,0){$\mathcal{H}$}}
\put(290,475){\makebox(0,0){Dirac's derivation}}
\put(290,435){\makebox(0,0){Quantionic derivation}}
\put(290,-15){\makebox(0,0){Covariant derivation}}
\put(120,385){\makebox(0,0){Linking}}
\put(290,365){\makebox(0,0){Commutants}}
\put(290,385){\makebox(0,0){Algebraic duality}}
\put(460,385){\makebox(0,0){Leibniz}}
\put(183,186){\makebox(0,0)[r]{Associated}}
\put(183,172){\makebox(0,0)[r]{Gaussian}}
\put(183,158){\makebox(0,0)[r]{space}}
\put(397,186){\makebox(0,0)[l]{Associated}}
\put(397,172){\makebox(0,0)[l]{Gaussian}}
\put(397,158){\makebox(0,0)[l]{space}}
\put(297,238){\makebox(0,0)[l]{Reciprocity}}
\put(300,164){\makebox(0,0)[l]{$\mathbb{C}\rightarrow\mathbb{R}$}}
\put(280,164){\makebox(0,0)[r]{$\mathbb{R}\rightarrow\mathbb{C}$}}
\put(95,70){\makebox(0,0){Base space of a fiber bundle}}
\put(59,55){\makebox(0,0){Fiber = $\mathcal{H}$ or $\mathcal{Q}$}}
\put(290,35){\makebox(0,0){Metric duality}}
\put(290,86){\makebox(0,0){Minkowski}}
\put(290,71){\makebox(0,0){subspaces}}
\put(290,320){\makebox(0,0){Subalgebras}}
\scriptsize\put(260,410){\makebox(0,0)[l]{Part 2 or Part 3}}
\put(120,410){\makebox(0,0)[l]{Part 2}}
\put(430,472){\makebox(0,0){Part 2}}
\normalsize\bf\put(25,312){\makebox(0,0)[l]{Quantum relativistic}}
\put(70,298){\makebox(0,0)[l]{domain}}
\put(25,252){\makebox(0,0)[l]{Classical relativistic}}
\put(70,238){\makebox(0,0)[l]{domain}}
\put(200,530){\makebox(0,0){Operands}}
\put(380,530){\makebox(0,0){Operators}}
\end{picture}%

\vspace{0.4in}%

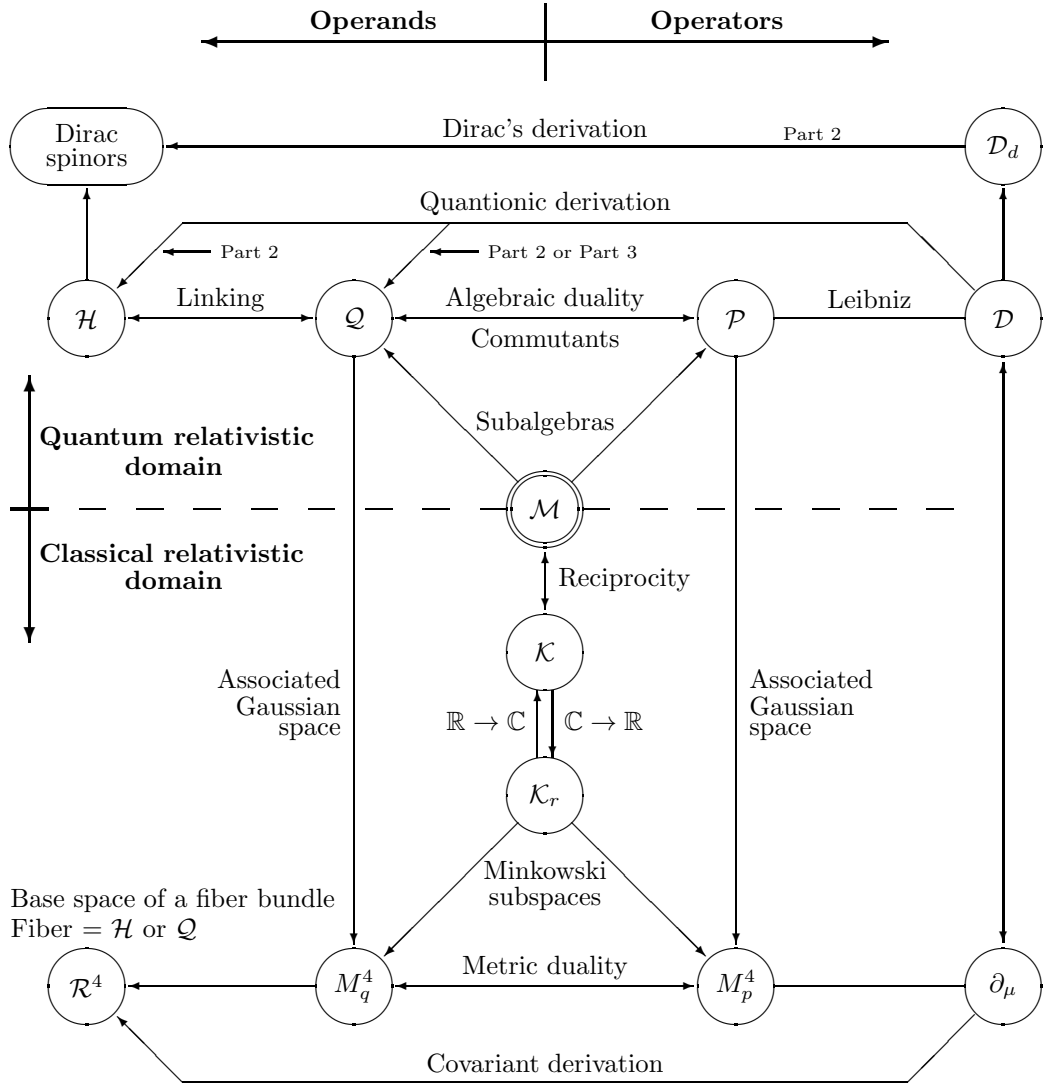
Figure \ref{S21}.1: A panoramic view of the quantionic structures.
\end{center}

\vspace{7pt}%

These properties are of three types, denoted by A, B and C.

\vspace{5pt}%

\emph{(A) Dimensionality.} The number of real or complex degrees of freedom is
obvious for each algebra. The number of quantionic degrees of freedom is not
applicable to quaternions and octonions.

A field with one degree of freedom is referred to as a \textbf{scalar field.}
Thus, the quantionic fields $\left\vert q\left(  x\right)  \right)  $ or
$Q\left(  x\right)  $ are \emph{scalar fields.} This must be emphasized
because their four complex (or eight real) components superficially suggest
otherwise. If they \emph{could not be} regarded as scalars, the algebra
$\mathcal{Q}$ would not be a number system. If they \emph{are not} regarded as
scalars, the suggestive value of number systems is lost (they suggest very
specific types of research questions).

Since it will be shown in Part 2 that Dirac's equation is equivalent to a
quantionic field equation, \emph{Dirac's 4-spinor fields are scalar quantionic
fields. }Thus, while the standard interpretation of Dirac's equation%
\[
\left(  \square-m^{2}\right)  ^{-1/2}\phi=0
\]
calls for a transition from scalar fields to the 4-spinorial representation
of the Lorentz group, the field remains a scalar in the quantionic interpretation,
but the ground on which it stands shifts from the nonrelativistic complex
numbers to the quantum relativistic quantions.

\vspace{5pt}%

\emph{(B)\ Physically essential properties.} Associativity is essential for
any practical number system.

The Leibniz identity is essential for the existence of structurally sound
differential equations.

Symmetry, which refers to the linear isomorphism of the real and imaginary
parts of the number system, is essential because it supports the
well-established quantum mechanical equivalence of observables and generators.

We note that the three physically relevant number systems are the only ones
which enjoy these three essential properties (though the last one is not
applicable to real numbers).

\vspace{5pt}%

\emph{(C)\ Inessential properties.} The field of complex numbers is both
commutative and a division algebra, while the algebra of quantions is neither.

\vspace{5pt}%

As a general observation, let us point out that commutativity is a trivializing property. Indeed,
many generalizations stem from the elimination of commutativity from some
relevant structure. (Incidentally, a major current approach to the unification
problem is by way of a non-commutative geometry.)

While unrestricted division is often considered desirable (whence the
popularity of division algebras), it plays no role whatsoever in quantum mechanics, where
dividing by wave functions is meaningless and the need for it never arises.%

\begin{center}%

\setlength{\unitlength}{0.01in}
\begin{picture}(590,275)
\put(0,0){\line(1,0){590}}
\put(0,270){\line(1,0){590}}
\put(0,212){\line(1,0){590}}
\put(0,210){\line(1,0){590}}
\put(0,49){\line(1,0){590}}
\put(0,121){\line(1,0){590}}
\put(0,0){\line(0,1){270}}
\put(28,0){\line(0,1){210}}
\put(140,0){\line(0,1){270}}
\put(142,0){\line(0,1){270}}
\put(230,0){\line(0,1){270}}
\put(320,0){\line(0,1){270}}
\put(410,0){\line(0,1){270}}
\put(500,0){\line(0,1){270}}
\put(590,0){\line(0,1){270}}
\thicklines\put(31,51){\line(0,1){68}}
\put(407,51){\line(0,1){68}}
\put(31,51){\line(1,0){376}}
\put(31,119){\line(1,0){376}}
\thinlines\put(70,240){\makebox(0,0){Properties}}
\put(14,165){\makebox(0,0){A}}
\put(14,85){\makebox(0,0){B}}
\put(14,25){\makebox(0,0){C}}
\put(185,255){\makebox(0,0){$\mathbb{R}$}}
\put(185,237){\makebox(0,0){Real}}
\put(185,225){\makebox(0,0){numbers}}
\put(185,195){\makebox(0,0){1}}
\put(185,175){\makebox(0,0){0}}
\put(185,155){\makebox(0,0){0}}
\put(185,135){\makebox(0,0){$E^1$}}
\put(185,105){\makebox(0,0){Y}}
\put(185,85){\makebox(0,0){Y}}
\put(185,65){\makebox(0,0){n.a.}}
\put(185,35){\makebox(0,0){Y}}
\put(185,15){\makebox(0,0){Y}}
\put(275,255){\makebox(0,0){$\mathbb{C}$}}
\put(275,237){\makebox(0,0){Complex}}
\put(275,225){\makebox(0,0){numbers}}
\put(275,195){\makebox(0,0){2}}
\put(275,155){\makebox(0,0){0}}
\put(275,175){\makebox(0,0){1}}
\put(275,135){\makebox(0,0){$E^2$}}
\put(275,105){\makebox(0,0){Y}}
\put(275,85){\makebox(0,0){Y}}
\put(275,65){\makebox(0,0){Y}}
\put(275,35){\makebox(0,0){Y}}
\put(275,15){\makebox(0,0){Y}}
\put(365,255){\makebox(0,0){$\mathcal{Q}$}}
\put(365,231){\makebox(0,0){Quantions}}
\put(365,195){\makebox(0,0){8}}
\put(365,175){\makebox(0,0){4}}
\put(365,155){\makebox(0,0){1}}
\put(365,135){\makebox(0,0){$M^4$}}
\put(365,105){\makebox(0,0){Y}}
\put(365,85){\makebox(0,0){Y}}
\put(365,65){\makebox(0,0){Y}}
\put(365,35){\makebox(0,0){N}}
\put(365,15){\makebox(0,0){N}}
\put(455,255){\makebox(0,0){$\mathbb{H}$}}
\put(455,231){\makebox(0,0){Quaternions}}
\put(455,195){\makebox(0,0){4}}
\put(455,175){\makebox(0,0){2}}
\put(455,155){\makebox(0,0){n.a.}}
\put(455,135){\makebox(0,0){$E^4$}}
\put(455,105){\makebox(0,0){Y}}
\put(455,85){\makebox(0,0){N}}
\put(455,65){\makebox(0,0){N}}
\put(455,35){\makebox(0,0){N}}
\put(455,15){\makebox(0,0){Y}}
\put(545,255){\makebox(0,0){$\mathbb{O}$}}
\put(545,231){\makebox(0,0){Octonions}}
\put(545,195){\makebox(0,0){8}}
\put(545,175){\makebox(0,0){4}}
\put(545,155){\makebox(0,0){n.a.}}
\put(545,135){\makebox(0,0){$E^8$}}
\put(545,105){\makebox(0,0){N}}
\put(545,85){\makebox(0,0){N}}
\put(545,65){\makebox(0,0){N}}
\put(545,35){\makebox(0,0){N}}
\put(545,15){\makebox(0,0){Y}}
\put(136,195){\makebox(0,0)[r]{N(real)}}
\put(136,175){\makebox(0,0)[r]{N(complex)}}
\put(136,155){\makebox(0,0)[r]{N(quantion)}}
\put(136,135){\makebox(0,0)[r]{Real Gauss space}}
\put(136,105){\makebox(0,0)[r]{Associativity}}
\put(136,85){\makebox(0,0)[r]{Leibniz}}
\put(136,65){\makebox(0,0)[r]{Symmetry}}
\put(136,35){\makebox(0,0)[r]{Commutativity}}
\put(136,15){\makebox(0,0)[r]{Division}}
\end{picture}%

\vspace{0.1in}

Table \ref{S21}.1: A comparative table of quantionic properties.
\end{center}

\section*{Appendix}

Referring to Figure \ref{S2}.2, the algebra of quantions was obtained by the
generalization $\mathbb{H}\rightarrow\mathcal{Q},$ which is very mild, for it
takes place within the same level of the Cayley-Dickson construction. By
contrast, the first version of the present paper, discarded following
Lozi\'{c}'s objections, was based on the inter-level generalization
$\mathbb{C}\rightarrow\mathcal{Q}.$ It thus required two steps: Identifying in
the complex numbers the concepts that will admit a
generalization, and then actually generalizing these concepts to arrive at the quantions.

The secons step was rather straightforward, but the first was objectionable (or rather, objected to) for it required recognizing as essential some deeply hidden and normally irrelevant properties of the complex numbers.

The same properties must also be recognized as essential in the new approach
by way of quaternions, but they are not hidden.

Since the original approach may nevertheless be interesting in its own right,
it is reproduced in the present appendix with only minor modifications of the text.

\subsubsection*{The complex numbers}

The complex numbers, $z=x+iy,$ admit the four \emph{arithmetic binary
operations} ($+,-,\times,\div$) and an \emph{additional unary operation}
(complex conjugation) denoted by the \textbf{star operator. }This operation is
obviously an\emph{\ }\textbf{involution:}%
\[
z^{\ast\ast}=z
\]

The positive definite function%
\[
A\left(  z\right)  =z^{\ast}z=\left(  x-iy\right)  \left(  x+iy\right)
=x^{2}+y^{2}%
\]
will be referred to as the \textbf{algebraic norm} of $z,$ and its positive
square root%
\[
r=\sqrt{z^{\ast}z}%
\]
as the \textbf{modulus} of $z.$

For $z\neq0,$ the inverse of $z$ is always written in the form%
\[
\frac{1}{z}\equiv\frac{1}{z}\frac{z^{\ast}}{z^{\ast}}\equiv\frac{z^{\ast}%
}{zz^{\ast}}=\frac{z^{\ast}}{x^{2}+y^{2}}%
\]
While correct numerically, this expression is wrong conceptually because it
depends on the star operator. Let's emphasize this objection, which is crucial
to the present work:

\emph{Division being, by definition, a purely arithmetic operation, the star
operator is out of place in the expression for the inverse.}

While the conceptual error in the above expression is inconsequential within
the field of complex numbers, it is misleading: Assigning a central importance
to the positive definiteness of the algebraic norm is analogous to assigning a
central importance to the positive definiteness of the Pythagorean norm. The
latter leads to the conclusion that the only generalizations of the Euclidian
space $E^{3}$ are the spaces $E^{n},$ thus missing relativity. The former
leads to the conclusion (by Hurwitz's theorem) that the only generalizations
of the complex numbers are the division algebras, thus missing (if the author
is not deeply mistaken) relativistic quantum mechanics.

\subsubsection*{The geometric representation of complex numbers}

While it is self-evident to us that a complex number $z=x+iy$ can be
represented by the point $\left(  x,y\right)  $ in a plane, this geometric
interpretation is of relatively recent origin. Independently discovered by
Wentzel and Argand around 1800, it was accepted by mathematicians only after
Gauss published it in 1831 --- which was almost three hundred years after the
complex numbers saw the light of day in Cardano's \textit{Ars Magna} (1545).

Since we are seeking an algebraic generalization of the complex numbers, and
since the origin of an algebra is fixed, the Gaussian plane is to be
interpreted as a real two-dimensional linear space, $M_{0}^{2},$ not as a
sheet of a Riemann surface. Complex numbers are thus to be viewed as vectors
in $M_{0}^{2},$ where the label \textquotedblleft2\textquotedblright\ refers
to the dimension, and the label \textquotedblleft0\textquotedblright\ to the
fact that the origin is fixed (in other words, $M_{0}^{2}$ is not to be viewed
as affine but as linear).

In preparation for a generalization to more than two dimensions, let us denote
the basis vectors in $M_{0}^{2}$ by $\mathbf{e}$ and $\mathbf{e}_{1},$ where
$\mathbf{e}$ represents the real unit and $\mathbf{e}_{1}$ the imaginary unit:%
\[
z=x\mathbf{e}+y\mathbf{e}_{1}%
\]

The addition of complex numbers naturally corresponds to vector addition in
$M_{0}^{2},$
\[
\left(  x\mathbf{e}+y\mathbf{e}_{1}\right)  +\left(  u\mathbf{e}%
+v\mathbf{e}_{1}\right)  =\left(  x+u\right)  \mathbf{e}+\left(  y+v\right)
\mathbf{e}_{1}%
\]
while multiplication is given by the second order relation%
\[
\left(  x\mathbf{e}+y\mathbf{e}_{1}\right)  \left(  u\mathbf{e}+v\mathbf{e}%
_{1}\right)  =\left(  xu-yv\right)  \mathbf{e}+\left(  yu+xv\right)
\mathbf{e}_{1}%
\]

The vectors representing complex numbers in $M_{0}^{2}$ can be rotated, but
the basis vectors $\mathbf{e}$ and $\mathbf{e}_{1}$ are fixed. This is because
no non-trivial linear combination $\mathbf{e}^{\prime}=\alpha\mathbf{e}%
+\beta\mathbf{e}_{1}$ enjoys the property $\mathbf{e}^{\prime}\mathbf{e}%
^{\prime}=\mathbf{e}^{\prime}.$ There are therefore no non-trivial
automorphisms%
\[
T\left(  z_{1}z_{2}\right)  =T\left(  z_{1}\right)  T\left(  z_{2}\right)
\]
in the field of complex numbers (the trivial one is $T=I$).

Let us now compute the inverse of an arbitrary non-vanishing complex number
without invoking the star operation. Writing%
\[
z^{-1}=u\mathbf{e}+v\mathbf{e}_{1}%
\]
the real unknowns $u$ and $v$ are subject to the condition%
\[
\mathbf{e}\equiv zz^{-1}=\left(  x\mathbf{e}+y\mathbf{e}_{1}\right)  \left(
u\mathbf{e}+v\mathbf{e}_{1}\right)
\]
which is equivalent to a system of two real linear equations:%
\begin{align*}
xu-yv  &  =1\\
yu+xv  &  =0
\end{align*}
The system's determinant,%
\[
M\left(  z\right)  =x^{2}+y^{2}%
\]
is obviously the Euclidean norm in $M_{0}^{2}.$ We shall refer to this
function as the \textbf{metric norm} of $z.$

\vspace{5pt}%

The solutions of the equations
are thus%
\begin{align*}
u  &  =\frac{1}{M\left(  z\right)  }x\\
v  &  =-\frac{1}{M\left(  z\right)  }y
\end{align*}

Let us introduce the notation%
\begin{equation}
z^{\#}=x\mathbf{e}-y\mathbf{e}_{1} \label{4}%
\end{equation}
and refer to $z^{\#}$ as the \textbf{metric dual} of $z,$ and to \# as the
\textbf{sharp operator.} The metric norm now assumes the form%
\[
M\left(  z\right)  =z^{\#}z
\]
Clearly, the metric dual is an involution:%
\[
z^{\#\#}=z
\]

The conceptually correct expression for the inverse is therefore%
\[
z^{-1}=\frac{z^{\#}}{M\left(  z\right)  }%
\]
It is to be contrasted with the conceptually misleading expression%
\[
z^{-1}=\frac{z^{\ast}}{A\left(  z\right)  }%
\]

\vspace{5pt}%

\subsubsection*{The complex numbers in polar form}

In 1691 Jacques Bernoulli introduced the polar factorization of the complex
numbers%
\[
z=r~e^{i\phi}%
\]
The factors $r$ and $e^{i\phi}$ are referred to, respectively, as the modulus
and the \textbf{phase factor.}

The star and sharp operations affect only the phase factor,%
\[
z^{\ast}=z^{\#}=r~e^{-i\phi}%
\]
while both norms equal the modulus squared,%
\[
M\left(  z\right)  =A\left(  z\right)  =r^{2}%
\]

To extend the polar factorization beyond the complex numbers, we shall regard
it as a manifestation of some general property. Such a property is the
invariance of the two fundamental norms. It gives rise to two invariance groups:

\vspace{5pt}%

1. The\ \textbf{orthogonal group} $SO\left(  2\right)  $ is defined as
\emph{the group of mappings of the linear Euclidean space }$M_{0}^{2}%
$\emph{\ onto itself which preserve the metric norm:}%
\[
\left(
\begin{array}
[c]{c}%
x\\
y
\end{array}
\right)  \longmapsto\left(
\begin{array}
[c]{c}%
u\\
v
\end{array}
\right)  =%
\begin{pmatrix}
\cos\alpha & -\sin\alpha\\
\sin\alpha & \cos\alpha
\end{pmatrix}
\left(
\begin{array}
[c]{c}%
x\\
y
\end{array}
\right)
\]

2. The\ \textbf{gauge group} $U\left(  1\right)  $ is defined as \emph{the
group of mappings of the field }$\mathbb{C}$\emph{\ onto itself which preserve
the algebraic norm:}%
\[
z\longmapsto w=e^{i\alpha}z
\]

These two groups are (locally) isomorphic,%
\[
U\left(  1\right)  \sim SO\left(  2\right)
\]

\subsubsection*{The degeneracy of the complex numbers}

The following table brings out the distinction between the algebraic and
geometric concepts:

\begin{center}%
\setlength{\unitlength}{0.01in}
\begin{picture}(510,140)
\thicklines\put(0,0){\line(1,0){500}}
\put(0,110){\line(1,0){500}}
\put(0,130){\line(1,0){500}}
\put(0,0){\line(0,1){130}}
\put(80,0){\line(0,1){130}}
\put(500,0){\line(0,1){130}}
\thinlines\put(0,55){\line(1,0){500}}
\put(220,0){\line(0,1){130}}
\put(360,0){\line(0,1){130}}
\put(75,82){\makebox(0,0)[r]{Algebra}}
\put(75,27){\makebox(0,0)[r]{Geometry}}
\put(430,40){\makebox(0,0){The orthogonal group}}
\put(430,15){\makebox(0,0){$SO(2)$}}
\put(430,95){\makebox(0,0){The unitary group}}
\put(430,70){\makebox(0,0){$U(1)$}}
\put(430,120){\makebox(0,0){Invariance groups}}
\put(290,40){\makebox(0,0){The metric norm}}
\put(290,15){\makebox(0,0){$M(z)=z^\# z$}}
\put(290,95){\makebox(0,0){The algebraic norm}}
\put(290,70){\makebox(0,0){$A(z)=z^* z$}}
\put(290,120){\makebox(0,0){Norm functions}}
\put(150,40){\makebox(0,0){The sharp operation}}
\put(150,15){\makebox(0,0){$z \rightarrow z^\#$}}
\put(150,95){\makebox(0,0){The star operation}}
\put(150,70){\makebox(0,0){$z \rightarrow z^*$}}
\put(150,120){\makebox(0,0){Involutions}}
\end{picture}%
\vspace{4pt}%

The dissociation of algebra and geometry.
\end{center}

While conceptually very different, the two involutions and the two norms are
numerically indistinguishable:%
\begin{align*}
z^{\ast}  &  =z^{\#}\\
A\left(  z\right)   &  =M\left(  z\right)
\end{align*}

It is sometimes justified to refer to such a \emph{loss of generality}\ as a
\textbf{degeneracy.}\ In the opposite direction, a \emph{generalization} may
be based on the elimination of a degeneracy only if the mathematical context
justifies it.

The above two relations warrant asking whether we are in the presence of a
degeneracy. If we are, there exists a generalization of the complex numbers in
which these relations are false, that is,
\begin{align*}
z^{\ast}  &  \neq z^{\#}\\
A\left(  z\right)   &  \neq M\left(  z\right)
\end{align*}
We are thus to solve the following problem:

\emph{Does a generalization of the complex numbers exist, in which a star and
a sharp operator defined as above are not numerically equal? }If they are
equal, the solutions are\ the quaternions and the octonions.

If this problem has a unique solution (it will be shown that it does), it will
be a number system arrived at by a generalization procedure never considered
before for number systems. If the new number system happens to be relativistic
(it will be shown that it is), it might be the `unifying number system' we are
seeking. The results obtained in this paper support this expectation.

\subsubsection*{The matrix representation of complex numbers}

The orthogonal and the polar representations of complex numbers are
complementary, in the sense that the former is natural for addition,%
\[
z_{1}+z_{2}=\left(  x_{1}+x_{2}\right)  +i\left(  y_{1}+y_{2}\right)
\]
and the latter for multiplication,
\[
z_{1}z_{2}=\left(  r_{1}r_{2}\right)  ~e^{i\left(  \phi_{1}+\phi_{2}\right)  }%
\]

\vspace{5pt}%

For the purpose of generalization, it would be desirable that these two
operations be on the same footing. The representation of complex numbers by
matrices has this property.

Using upper-case letters for matrices, the mappings%
\begin{align*}
1  &  \longmapsto I=%
\begin{pmatrix}
1 & 0\\
0 & 1
\end{pmatrix}
\\
i  &  \longmapsto J=%
\begin{pmatrix}
0 & -1\\
1 & 0
\end{pmatrix}
\end{align*}
yield%
\[
z=x+iy\longmapsto Z=xI+yJ=%
\begin{pmatrix}
x & -y\\
y & x
\end{pmatrix}
\]

\vspace{7pt}%

The algebra of these real $2\times2$ matrices is therefore isomorphic with the
field of complex numbers. In this representation, addition and multiplication
are both natural matrix operations.

The complex conjugate may be viewed as a matrix transposition,
\[
i^{\ast}\longmapsto-J=J^{\top}=%
\begin{pmatrix}
0 & 1\\
-1 & 0
\end{pmatrix}
\]
or as a Hermitian conjugation,%
\[
i^{\ast}\longmapsto-J=J^{\dag}=%
\begin{pmatrix}
0 & 1\\
-1 & 0
\end{pmatrix}
\]
Adopting the latter, we have%
\[
A\left(  Z\right)  =Z^{\dag}Z=%
\begin{pmatrix}
x & y\\
-y & x
\end{pmatrix}%
\begin{pmatrix}
x & -y\\
y & x
\end{pmatrix}
=\left(  x^{2}+y^{2}\right)  I
\]

\vspace{5pt}%

The inverse matrix%
\[
Z^{-1}=%
\begin{pmatrix}
x & -y\\
y & x
\end{pmatrix}
^{-1}=\frac{1}{x^{2}+y^{2}}%
\begin{pmatrix}
x & y\\
-y & x
\end{pmatrix}
\]
confirms that the dual of $Z$ is%
\[
Z^{\#}=%
\begin{pmatrix}
x & y\\
-y & x
\end{pmatrix}
\]
and that the metric norm is%

\vspace{5pt}%

\[
M\left(  Z\right)  =Z^{\#}Z=%
\begin{pmatrix}
x & y\\
-y & x
\end{pmatrix}%
\begin{pmatrix}
x & -y\\
y & x
\end{pmatrix}
=\left(  x^{2}+y^{2}\right)  I
\]

The inverse matrix obtained by solving the algebraic equations defining it
assumes the conceptually correct form%
\[
Z^{-1}=\frac{1}{M\left(  Z\right)  }Z^{\#}%
\]
while its standard form corresponds to the\ conceptually wrong expression%
\[
Z^{-1}=\frac{1}{A\left(  Z\right)  }Z^{\dag}%
\]
which is numerically correct for the complex numbers, for the quaternions, and
for the octonions --- but only owing to their degeneracy.

\subsubsection*{The left regular representation}

To avoid confusion, we shall use the symbol $\mathcal{L}$ to denote the matrix representation of
the field $\mathbb{C}$ of complex numbers.

\vspace{5pt}%

Let us interpret the matrices $Z\in\mathcal{L}$ as operators acting on a
two-dimensional real linear space (by convention from the left). We shall
denote this \textbf{representation space} by $\mathcal{H}_{c},$ and the column
vectors in it by the ket symbol (written as $\left\vert \ast\right)  $ instead
of $\left\vert \ast\right\rangle $ to avoid association with quantum
mechanical state vectors). Thus,%
\[
\left\vert w\right)  =\left(
\begin{array}
[c]{c}%
u\\
v
\end{array}
\right)  \in\mathcal{H}_{c}%
\]
and%
\[
Z\left\vert w\right)  =%
\begin{pmatrix}
x & -y\\
y & x
\end{pmatrix}
\left(
\begin{array}
[c]{c}%
u\\
v
\end{array}
\right)  =\left(
\begin{array}
[c]{c}%
xu-yv\\
yu+xv
\end{array}
\right)  =\left\vert zw\right)  \in\mathcal{H}_{c}%
\]

This matrix representation of the field of complex numbers is thus the
\emph{left regular representation of the complex numbers. }We see that a
natural one-to-one correspondence%
\[
\mathcal{L}\ni%
\begin{pmatrix}
x & -y\\
y & x
\end{pmatrix}
\rightleftarrows\left(
\begin{array}
[c]{c}%
x\\
y
\end{array}
\right)  \in\mathcal{H}_{c}%
\]
exists between the representations of complex numbers by matrices and by kets.

The Hermitian conjugate of the above mapping is evidently%
\[
\left(  w\right\vert Z^{\dag}=%
\begin{pmatrix}
u & v
\end{pmatrix}%
\begin{pmatrix}
x & y\\
-y & x
\end{pmatrix}
=%
\begin{pmatrix}
ux-vy & uy+vx
\end{pmatrix}
=\left(  wz\right\vert
\]

The inner product%
\[
\left(  w|w\right)  =%
\begin{pmatrix}
u & v
\end{pmatrix}
\left(
\begin{array}
[c]{c}%
u\\
v
\end{array}
\right)  =u^{2}+v^{2}%
\]
is positive definite. 

The space $\mathcal{H}_{c}$ is therefore equipped with a
positive definite metric. It is thus a real two-dimensional inner product
space. 

\vspace{5pt}%

While metrically indistinguishable from $M_{0}^{2},$ the space
$\mathcal{H}_{c}$ is conceptually very different from it. It must therefore be
treated differently in the expectation of different generalizations.

Let us introduce a fixed ket
\[
\left\vert \omega\right)  =\left(
\begin{array}
[c]{c}%
\alpha\\
\beta
\end{array}
\right)
\]
meant to formalize the one-to-one correspondence between matrices and kets by
imposing the condition%
\[
\left\vert z\right)  =Z\left\vert \omega\right)
\]
Expanded, this relation reads%
\[
\left(
\begin{array}
[c]{c}%
x\\
y
\end{array}
\right)  =%
\begin{pmatrix}
x & -y\\
y & x
\end{pmatrix}
\left(
\begin{array}
[c]{c}%
\alpha\\
\beta
\end{array}
\right)  =\left(
\begin{array}
[c]{c}%
x\alpha-y\beta\\
y\alpha+x\beta
\end{array}
\right)
\]
and yields $\alpha=1,$ $\beta=0.$ Hence%
\[
\left\vert \omega\right)  =\left(
\begin{array}
[c]{c}%
1\\
0
\end{array}
\right)
\]
This vector is automatically normalized:%
\[
\left(  \omega|\omega\right)  =1
\]

We refer to $\left\vert \omega\right)  $ as the \textbf{linking vector}, for
it establishes a one-to-one correspondence between the vectors $\left\vert
z\right)  \in$ $\mathcal{H}_{c}$ and the matrices $Z\in\mathcal{L}.$

Since $\left\vert \omega\right)  =I\left\vert \omega\right)  ,$ it follows
that $\left\vert \omega\right)  \in$ $\mathcal{H}_{c}$ corresponds to the unit
matrix $I\in\mathcal{L}.$

Another important identity is%
\[
\left(  \omega\left\vert Z\right\vert \omega\right)  =\Re z=x=\frac{1}%
{2}Tr\ Z
\]
where $Tr$ is the trace operator. This identity makes it possible to work with
the trace concept abstractly, that is, without reference to a matrix representation.

\subsubsection{Derivatives of complex functions}

The most general complex derivation operator is a linear combination of two
real derivations:%
\[
\frac{\partial}{\partial z}=\alpha\partial_{x}+\beta\partial_{y}%
\]
Two conditions are needed to determine the coefficients $\alpha$ and $\beta,$
one of them being%
\[
\frac{\partial}{\partial z}z=1
\]

For the other condition, we may say that either $z^{\ast}$ or $z^{\#}$ is
independent of $z$ (because $x$ and $y$ are independent variables). Since the
two options are equivalent owing to the degeneracy of the complex numbers, we
have%
\[
\frac{\partial}{\partial z}z^{\ast}\equiv\frac{\partial}{\partial z}z^{\#}=0
\]
or, after expansion,%
\begin{align*}
\left(  \alpha\partial_{x}+\beta\partial_{y}\right)  \left(  x+iy\right)   &
=\alpha+i\beta=1\\
\left(  \alpha\partial_{x}+\beta\partial_{y}\right)  \left(  x-iy\right)   &
=\alpha-i\beta=0
\end{align*}
The solutions $\alpha=1/2$ and $\beta=-i/2$ yield%
\begin{align*}
\frac{\partial}{\partial z}  &  =\frac{1}{2}\left(  \partial_{x}-i\partial
_{y}\right) \\
\frac{\partial}{\partial z^{\ast}}  &  =\frac{1}{2}\left(  \partial
_{x}+i\partial_{y}\right)
\end{align*}

A general complex function is an arbitrary function of the real variables $x$
and $y,$ or, equivalently, of $z$ and $z^{\ast}.$ An \textbf{analytic}
function is a function of $z$ only. The analyticity condition is therefore%
\[
\frac{\partial}{\partial z^{\ast}}w\left(  z\right)  =0
\]
Expansion of this equation yields%
\[
\frac{\partial}{\partial z^{\ast}}w=\frac{1}{2}\left(  \partial_{x}%
+i\partial_{y}\right)  \left(  u+iv\right)  =\frac{1}{2}\left(  \partial
_{x}u-\partial_{y}v\right)  +i\frac{1}{2}\left(  \partial_{x}v+\partial
_{y}u\right)  =0
\]
or, in the real domain,%
\begin{align*}
\partial_{x}u-\partial_{y}v  &  =0\\
\partial_{x}v+\partial_{y}u  &  =0
\end{align*}
These are the Cauchy-Riemann analyticity conditions.

From%
\[
\frac{\partial}{\partial z}\frac{\partial}{\partial z^{\ast}}=\frac{1}%
{4}\left(  \partial_{x}-i\partial_{y}\right)  \left(  \partial_{x}%
+i\partial_{y}\right)  =\frac{1}{4}\left(  \partial_{x}^{2}+\partial_{y}%
^{2}\right)  =\frac{1}{4}\Delta
\]
follows that every analytic function $w\left(  z\right)  $ is a harmonic
function, $\Delta w=0,$ and since the Laplaceian is a real operator, $u$ and
$v$ are real harmonic functions.

The matrix form of the operators $\partial_{z}$ and $\partial_{z}^{\ast}$ is
thus%
\[
\frac{\partial}{\partial z}\longrightarrow\mathcal{D}_{z}=%
\begin{pmatrix}
\partial_{x} & \partial_{y}\\
-\partial_{y} & \partial_{x}%
\end{pmatrix}
\]%
\[
\frac{\partial}{\partial z^{\ast}}=\frac{\partial}{\partial z^{\#}%
}\longrightarrow\mathcal{D}_{z}^{\#}=%
\begin{pmatrix}
\partial_{x} & -\partial_{y}\\
\partial_{y} & \partial_{x}%
\end{pmatrix}
\]

It follows from%
\[
\mathcal{D}_{z}^{\#}W=%
\begin{pmatrix}
\partial_{x} & -\partial_{y}\\
\partial_{y} & \partial_{x}%
\end{pmatrix}%
\begin{pmatrix}
u & -v\\
v & u
\end{pmatrix}
=%
\begin{pmatrix}
u_{x}-v_{y} & -\left(  u_{y}+v_{x}\right) \\
u_{y}+v_{x} & u_{x}-v_{y}%
\end{pmatrix}
\]
that the matrix form of the Cauchy-Riemann equations is $\mathcal{D}_{z}%
^{\#}W=0.$

The purpose of listing these well-known relations in matrix form is to show
that the matrix formalism lends itself to generalization by increasing the
number of dimensions.%

\end{document}